\documentclass[%
twocolumn,
superscriptaddress,
 amsmath,amssymb,
 aps,
 prx,
 showkeys,
longbibliography
]{revtex4-2}

\pdfoutput=1
\usepackage{graphicx}
\usepackage{dcolumn}
\usepackage{bm}
\usepackage{bbm}
\usepackage[dvipsnames]{xcolor}

\usepackage[outdir=./]{epstopdf}

\makeatletter
\providecommand*{\diff}%
{\@ifnextchar^{\DIfF}{\DIfF^{}}}
\def\DIfF^#1{%
	\mathop{\mathrm{\mathstrut d}}%
	\nolimits^{#1}\gobblespace}
\def\gobblespace{%
	\futurelet\diffarg\opspace}
\def\opspace{%
	\let\DiffSpace\!%
	\ifx\diffarg(%
	\let\DiffSpace\relax
	\else
	\ifx\diffarg[%
	\let\DiffSpace\relax
	\else
	\ifx\diffarg\{%
	\let\DiffSpace\relax
	\fi\fi\fi\DiffSpace}

\newcommand{\ExpPei}[1]{\mathbb{E}_{\mathrm{ei}}\left[{#1}\right]}
\newcommand{\ExpPe}[1]{\mathbb{E}_{\mathrm{e}}\left[{#1}\right]}
\newcommand{\ExpPi}[1]{\mathbb{E}_{\mathrm{i}}\left[{#1}\right]}

\newcommand{\Cov}[1]{\mathbb{C}\left[#1\right]}
\newcommand{\one}[1]{\mathbbm{1}_{\left\{{#1}\right\}}}
\newcommand{\Exp}[1]{\mathbb{E}\left[{#1}\right]}
\newcommand{\Var}[1]{\mathbb{V}\left[{#1}\right]}

\newcommand{\Prob}[1]{\mathbb{P}\left[{#1}\right]}
\newcommand{\ProbP}[1]{\mathbb{P}_0\left[{#1}\right]}

\newcommand{\dd}{\textrm{d}}

\newcommand{\eeii}{\mathrm{e/i}}
\newcommand{\ee}{\mathrm{e}}
\newcommand{\ii}{\mathrm{i}}
\newcommand{\leak}{\mathrm{L}}
\newcommand{\Th}{\mathrm{T}}

\newcommand{\fref}[1]{Fig.~\ref{#1}}
\newcommand{\eref}[1]{Eq.~\eqref{#1}}

\begin{document}

\preprint{APS/123-QED}

\title{Exact analysis of the subthreshold variability for conductance-based neuronal models with synchronous synaptic inputs}


\author{Logan A. Becker}
 	\affiliation{%
 	Center for Theoretical and Computational Neuroscience, The University of Texas at Austin
 	}%
  	\affiliation{%
 	Department of Neuroscience, The University of Texas at Austin
 	}%
 
 \author{Baowang Li}
  	\affiliation{%
 	Center for Theoretical and Computational Neuroscience, The University of Texas at Austin
 	}%
	\affiliation{%
	Department of Neuroscience, The University of Texas at Austin
	}%
 	\affiliation{%
 	Center for Perceptual Systems, The University of Texas at Austin
	}%
	\affiliation{%
	Center for Learning and Memory, The University of Texas at Austin
	}%
	\affiliation{%
	Department of Psychology, The University of Texas at Austin
 	}%
	
 \author{Nicholas J. Priebe}
	 \affiliation{%
	Center for Theoretical and Computational Neuroscience, The University of Texas at Austin
	 }%
	 \affiliation{%
	Department of Neuroscience, The University of Texas at Austin
 	}%
	 \affiliation{%
 	Center for Learning and Memory, The University of Texas at Austin
	}%
 
 \author{Eyal Seidemann}
	 \affiliation{%
 	Center for Theoretical and Computational Neuroscience, The University of Texas at Austin
 	}%
	\affiliation{%
	Department of Neuroscience, The University of Texas at Austin
	 }%
	 \affiliation{%
 	Center for Perceptual Systems, The University of Texas at Austin
	}%
	\affiliation{%
	Department of Psychology, The University of Texas at Austin
	 }%

\author{Thibaud Taillefumier}%
 \altaffiliation{%
 	Corresponding author}%
\email{ttaillef@austin.utexas.edu}
	\affiliation{%
	Center for Theoretical and Computational Neuroscience, The University of Texas at Austin
	 }%
	 \affiliation{%
 	Department of Neuroscience, The University of Texas at Austin
	 }%
	  \affiliation{%
 	Department of Mathematics, The University of Texas at Austin
	 }%

%



\begin{abstract}
The spiking activity of neocortical neurons exhibits a striking level of variability, even when these networks are driven by identical stimuli. 
The approximately Poisson firing of neurons has led to the hypothesis that these neural networks operate in the asynchronous state. 
In the asynchronous state neurons fire independently from one another, so that the probability that a neuron experience synchronous synaptic inputs is exceedingly low. 
While the models of asynchronous neurons lead to observed spiking variability, it is not clear whether the asynchronous state can also account for the level of subthreshold membrane potential variability. 
We propose a new analytical framework to rigorously quantify the subthreshold variability of a single conductance-based neuron in response to synaptic inputs with prescribed degrees of synchrony. 
Technically we leverage the theory of exchangeability to model input synchrony via jump-process-based synaptic drives; we then perform a moment analysis of the stationary response of a neuronal model with all-or-none conductances that neglects post-spiking reset. As a result, we produce exact, interpretable closed forms for the first two stationary moments of the membrane voltage, with explicit dependence on the input synaptic numbers, strengths, and synchrony. 
For biophysically relevant parameters, we find that the asynchronous regime only yields realistic subthreshold variability (voltage variance $\simeq 4-9\mathrm{mV^2}$) when driven by a restricted number of large synapses, compatible with strong thalamic drive. 
By contrast, we find that achieving realistic subthreshold variability with dense cortico-cortical inputs requires including weak but nonzero input synchrony, consistent with measured pairwise spiking correlations. 
We also show that without synchrony, the neural variability averages out to zero for all scaling limits with vanishing synaptic weights, independent of any balanced state hypothesis. This result challenges the theoretical basis for mean-field theories of the asynchronous state.
\end{abstract}

\maketitle


\section{Introduction}

A common and striking feature of cortical activity is the high degree of neuronal spiking variability~\cite{churchland:2010aa}.  
This high variability is notably present in sensory cortex and motor cortex, as well as in regions with intermediate representations~\cite{tolhurst:1981,tolhurst:1983,churchland:2006,rickert:2009}.  
The prevalence of this variability has led to it being a major constraint for modeling cortical networks. 
Cortical networks may operate in distinct regimes depending on species, cortical area, and brain states.
In asleep or anesthetized state, neurons tend to fire synchronously with strong correlations between the firing of distinct neurons~\cite{stevens1998input,lampl:1999,ecker2014state}.
In the awake state, although synchrony has been reported as well, stimulus drive, arousal, or attention tend to promote an irregular firing regime whereby neurons spike in a seemingly random manner, with decreased or little correlation~\cite{poulet2008internal,churchland:2010aa,ecker2014state}. 
This has lead to the hypothesis that cortex primarily operates asynchronously~\cite{renart:2010aa,ecker:2010,cohen2011measuring}.
In the asynchronous state, neurons fire independently from one another, so that the probability that a neuron experiences synchronous synaptic inputs is exceedingly low. 
That said, the asynchronous state hypothesis appears at odds with the high-degree of observed spiking variability in cortex.
Cortical neurons are thought to receive a large number of synaptic inputs ($\simeq 10^4$)~\cite{braitenberg2013cortex}. 
Although the impact of these inputs may vary across synapses, the law of large numbers implies that variability should average out when integrated at the soma.
In principle, this would lead to clock-like spiking responses, contrary to experimental observations~\cite{koch:1992}.

A number of mechanisms have been proposed to explain how high spiking variability emerges in cortical networks~\cite{bell:1995}. 
The prevailing approach posits that excitatory and inhibitory inputs converge on cortical neurons in a balanced manner.  
In balanced models, the overall excitatory and inhibitory drives cancel each other so that transient imbalances in the drive can bring the neuron's membrane voltage across the spike-initiation threshold.  
Such balanced models result in spiking statistics that match those found in the neocortex~\cite{amit:1997,brunel:2000}. 
However, these statistics can emerge in distinct dynamical regimes depending on whether the balance between excitation and inhibition is tight or loose~\cite{ahmadian2021dynamical}.
In tightly balanced networks, whereby the net neuronal drive is negligible  compared to the antagonizing components, activity correlation is effectively zero, leading to a strictly asynchronous regime~\cite{sompolinski:1988,vreeswijk:1996,vreeswijk:1998aa}.
By contrast, in loosely-balanced networks, the net neuronal drive remains of the same order as the antagonizing components, which allows for strong neuronal correlations during evoked activity, compatible with a synchronous regime~\cite{ahmadian2013analysis,rubin2015stabilized,hennequin2018dynamical}.

\begin{figure}[htbp]
\begin{center}
\includegraphics[width=8.6cm]{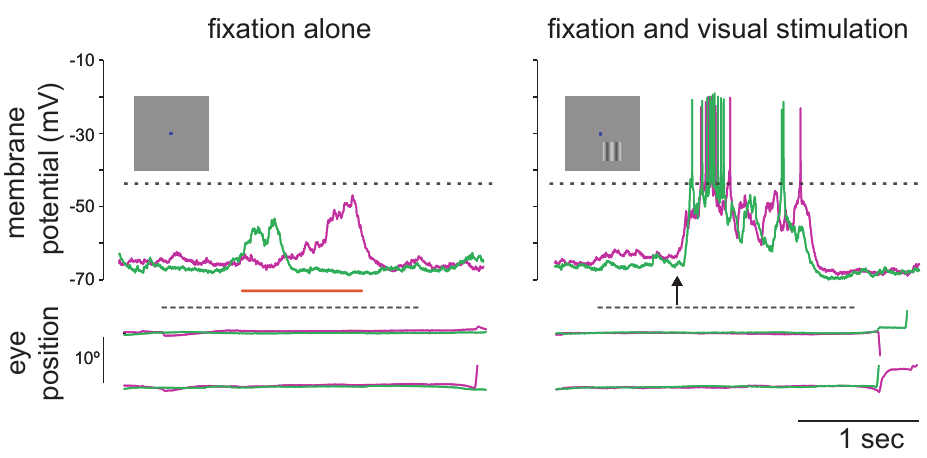}
  \caption{\label{fig:expVar}\textbf{Large trial-by-trial membrane voltage fluctuations.} Membrane voltage responses are shown using whole cell recordings in awake behaving primates for both fixation alone trials (left) and visual stimulation trials (right). A drifting grating was presented for 1 second beginning at the arrow.  Below the membrane voltage traces are records of horizontal and vertical eye movements, illustrating that the animal was fixating during the stimulus.  Red and green traces indicate different trials under the same conditions. Adapted from~\cite{tan:2014aa}. 
  }
\end{center}
\end{figure}

While the high spiking variability is an important constraint for cortical network modeling, there are other biophysical signatures that may be employed. 
We now have access to the subthreshold membrane voltage fluctuations that underlie spikes in awake, behaving animals (see Fig.~\ref{fig:expVar}).  
Membrane voltage recordings reveal two main deviations from the asynchronous hypothesis: 
first, membrane voltage does not hover near spiking threshold and is modulated by the synaptic drive; second, it exhibits state- or stimulus dependent non-Gaussian fluctuation statistics with positive skewness~\cite{Haider:2013aa,tan:2013aa,tan:2014aa,okun2015diverse}. 
In this work, we further argue that membrane voltage recordings reveal much larger voltage fluctuations than predicted by balanced cortical models~\cite{hansel:2012,pattadkal:2018}.

How could such large subthreshold variations in membrane voltage emerge?  
One way that fluctuations could emerge, even for large numbers of input, is if there is synchrony in the driving inputs~\cite{shadlen:1998}.  
In practice, input synchrony is revealed by the presence of positive spiking correlations, which quantify the propensity of distinct synaptic inputs to coactivate.
Measurements of spiking correlations between pairs of neurons vary across reports, but have generally been shown to be weak~\cite{renart:2010aa,ecker:2010,cohen2011measuring}. 
That said, even weak correlations can have a large impact when the population of correlated  inputs is large~\cite{chen:2006,polk:2012}. 
Further, the existence of input synchrony, supported by weak but persistent spiking correlations, is consistent with at least two other experimental observations. 
First, intracellular recordings from pairs of neurons in both anesthetized and awake animals reveal a high degree of membrane voltage correlations~\cite{lampl:1999,yu:2010,arroyo:2018}.  
Second, excitatory and inhibitory conductance inputs are highly correlated with each other within the same neuron~\cite{okun:2008,arroyo:2018}.  
These observations suggest that input synchrony could explain the observed level of subthreshold variability.

While our focus is on achieving realistic subthreshold variability, other challenges to asynchronous networks have been described. 
In particular, real neural networks exhibit distinct regimes of activity depending on the strength of their afferent drives.
In that respect, Zerlaut {\it et al.}~\cite{zerlaut:2019} showed that asynchronous networks can exhibit a spectrum of realistic regimes of activity if they have moderate recurrent connections and are driven by strong thalamic projections (see also \cite{brunel:2000}).
Furthermore, it has been a challenge to identify the scaling rule that should apply to synaptic strengths for asynchrony to hold stably in idealized networks.
Recently, Sanzeni {\it et al.}~\cite{sanzeni:2022} proposed that a realistic asynchronous regime is achieved for a particular large-coupling rule, whereby synaptic strengths scale in keeping with the logarithmic size of the network.
Both studies consider balanced networks with conductance-based neuronal models but neither focuses on the role of synchrony, consistent with the asynchronous state hypothesis.  
The asynchronous state hypothesis is theoretically attractive because it represents a naturally stable regime of activity in infinite-size, balanced networks of current-based neuronal models~\cite{vreeswijk:1996,amit:1997,vreeswijk:1998aa,brunel:2000}.
Such neuronal models, however, neglect the voltage dependence of conductances and it remains unclear whether the asynchronous regime is asymptotically stable for infinite-size, conductance-based network models.

Here, independent of the constraint of network stability, we ask whether biophysically relevant neuronal models can achieve the observed subthreshold variability under realistic levels of input synchrony.
To answer this question, we derive exact analytical expressions for the stationary voltage variance of a single conductance-based neuron in response to synchronous shot-noise drives~\cite{stein:1965,tuckwell:1988}.
 A benefit of shot-noise models compared to diffusion models is to allow for individual synaptic inputs to be temporally separated in distinct impulses, each corresponding to a transient positive conductance fluctuations ~\cite{richardson:2004,richardson:2005, richardson:2006}.
We develop our shot-noise analysis for a variant of classically considered neuronal models.
We call this variant the all-or-none-conductance-based (AONCB) model for which synaptic activation occurs as an all-or-none process rather than as an exponentially relaxing process.
To perform an exact treatment of these models, we develop original probabilistic techniques inspired from Marcus' work about shot-noise driven dynamics~\cite{marcus:1978, marcus:1981}.
To model shot-noise drives with synchrony, we develop a statistical framework based on the property of input exchangeability, which assumes that no synaptic inputs play a particular role. 
In this framework, we show that input drives with varying degree of synchrony can be rigorously modeled via jump processes, while synchrony can be quantitatively related to measures of pairwise spiking correlations.

Our main results are biophysically interpretable formulas for the voltage mean and variance in the limit of instantaneous synapses.
Crucially, these formulas explicitly depend on the input numbers, weights, and synchrony,  and hold without any forms of diffusion approximation.
This is in contrast with analytical treatments which elaborate on the diffusion and effective-time-constant approximations~\cite{destexhe:2001,meffin:2004,zerlaut:2019,sanzeni:2022}.
We leverage these exact, explicit formulas to determine under which synchrony conditions a neuron can achieve the experimentally observed subthreshold variability.
For biophysically relevant synaptic numbers and weights, we find that achieving realistic variability is possible in response to a restricted number of large asynchronous connections, compatible with the dominance of thalamo-cortical projections in the input layers of the visual cortex.
However, we find that achieving realistic variability in response to a large number of moderate cortical inputs, as in superficial cortical visual layers, necessitates nonzero input synchrony in amounts that are consistent with the weak levels of measured spiking correlations observed {\it in vivo}.

In practice, persistent synchrony may spontaneously emerge in large but finite neural networks, as nonzero correlations are the hallmark of finite-dimensional interacting dynamics.
The network structural features responsible for the magnitude of such correlations remains unclear, and we do not address this question here (see~\cite{doiron:2016,ocker:2017} for review).
The persistence of synchrony is also problematic for theoretical approaches that consider networks in the infinite-size limits.
Indeed, our analysis supports that in the absence of synchrony and for all scaling of the synaptic weights, subthreshold variability must vanish in the limit of arbitrary large numbers of synapses.
This suggests that independent of any balanced condition, the mean-field dynamics that emerge in infinite-size networks of conductance-based neurons will not exhibit Poisson-like spiking variability, at least in the absence of additional constraints on the network structure or on the biophysical properties of the neurons.
In current-based neuronal models, however, variability is not dampened by a conductance-dependent effective time constant.
These findings therefore challenge the theoretical basis for the asynchronous state in conductance-based neuronal networks.

Our exact analysis, as well as its biophysical interpretations, is only possible at the cost of several caveats:
First, we neglect the impact of the spike-generating mechanism (and of the post-spiking reset) in shaping the subthreshold variability.
Second, we quantify synchrony under the assumption of input exchangeability, that is, for synapses having a typical strength as opposed to being heterogeneous.
Third, we consider input drives that implement an instantaneous form of synchrony with temporally precise synaptic coactivations.
Fourth, we do not consider slow temporal fluctations in the mean synaptic drive.
Fifth, and perhaps most concerning, we do not account for the stable emergence of a synchronous regime in network models.
We argue in the discussion that all the above caveats but the last one can be addressed without impacting our findings.
Addressing the last caveat remains an open problem.

 For reference, we list in Table \ref{tab:notations} the main notations used in this work.
These notations utilize the subscript $\{\}_\ee$, $\{\}_\ii$ to refer to excitation or inhibition, respectively. 
The notation $\{\}_\eeii$ means that the subscript can be either $\{\}_e$ or $\{\}_i$.
The notation $\{\}_{\ee\ii}$ is used to emphasize that a quantity depends jointly on excitation and inhibition.


\section{Stochastic modeling and analysis}\label{sec:model}



\subsection{All-or-none-conductance-based neurons}\label{sec:AONCB}

\begin{table}
\caption{\label{tab:notations}Main notations.}
\begin{ruledtabular}
\begin{tabular}{lp{6.5cm}}
   $a_{\eeii,1}$  & first-order synaptic efficacies\\
   $a_{\eeii,2}$  & second-order synaptic efficacies \\
   $a_{\eeii,12}$  & auxiliary second-order synaptic efficacies\\
   $b$,  & rate of the driving Poisson process $N$\\
   $b_{\eeii}$ & rate of the excitatory/inhibitory Poisson process $N_\eeii$\\
   $C$ & membrane capacitance \\
   $c_{\ee\ii}$,  & crosscorrelation synaptic efficacy\\
   $\Cov{\cdot,\cdot}$ & stationary covariance \\
   $\Exp{\cdot}$ & stationary expectation \\
   $\ExpPei{\cdot}$ & expectation w.r.t. the joint distribution $p_{\ee\ii}$ or $p_{\ee\ii,kl}$ \\
   $\mathrm{E}_\eeii \left[ \cdot \right]$ & expectation w.r.t. the marginal distribution $p_{\eeii}$ or $p_{\eeii,k}$  \\
   $\epsilon=\tau_\mathrm{s}/\tau $ & fast-conductance small parameter\\
   $G$ &  passive leak conductance\\
   $g_{\eeii}$ &  overall excitatory/inhibitory conductance\\
   $h_{\eeii}=g_{\eeii}/C$ &  reduced excitatory/inhibitory conductance\\
   $k_{\eeii}$ &  number of coactivating excitatory/inhibitory synaptic inputs\\
   $K_\eeii$ & total number of excitatory/inhibitory synaptic inputs \\
   $N$ & driving Poisson process with rate $b$\\
   $N_{\eeii}$ & excitatory/inhibitory driving Poisson process with rate $b_\eeii$\\
   $p_{\ee\ii}$ & bivariate jump distribution of $(W_\ee,W_\ii)$\\
   $p_\eeii$ & marginal jump distribution of $W_\eeii$\\
   $p_{\ee\ii,kl}$ & bivariate distribution for the numbers of coactivating synapses $(k_\ee,k_\ii)$\\
   $p_{\eeii,k}$ & marginal synaptic count distribution $k_\eeii$\\
   $r_{\eeii}$ & individual excitatory/inhibitory synaptic rate\\
   $\rho_{\ee\ii}$ & spiking correlation between excitatory and inhibitory inputs\\
   $\rho_\eeii$ & spiking correlation within  excitatory/inhibi\-tory inputs\\
   $\tau$ & passive membrane time constant \\
   $\tau_\mathrm{s}$ & synaptic time constant\\
   $\Var{\cdot}$ & stationary variance \\
   $W_{\eeii}$ & excitatory/inhibitory random jumps\\
   $V_{\eeii}$ & excitatory/inhibitory reversal potentials\\
   $w_{\eeii}$ & typical value for excitatory/inhibitory synaptic weights\\
   $X_k$  & binary variable indicating the activation of excitatory synapse $k$ \\
   $Y_l$  & binary variable indicating the activation of inhibitory synapse $l$ \\
   $Z$  & driving compound Poisson process with base rate $b$ and jump distribution $p_{\ee\ii}$ \\
\end{tabular}
\end{ruledtabular}
\end{table}

\begin{figure}[htbp]
\begin{center}
\includegraphics[width=8.6cm]{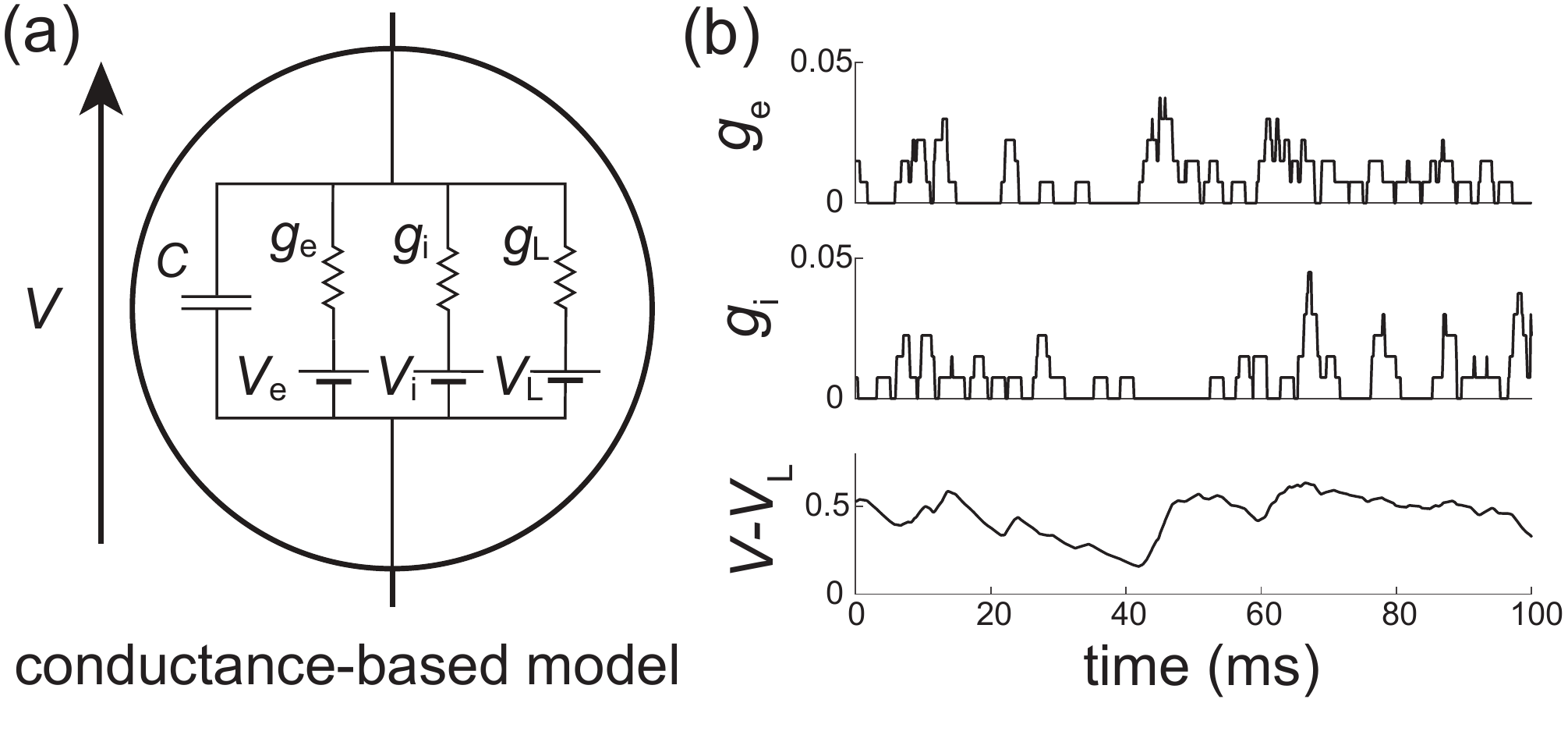}
\caption{\label{fig:AONCB}{\bf All-or-none-conductance-based models.}
(a) Electrical diagram of conductance-based model for which the neuronal voltage $V$ evolves in response to fluctuations of excitatory and inhibitory conductances $g_\ee$ and $g_\ii$.
(b)
In all-or-none models, inputs delivered as Poisson processes transiently activate the excitatory and inhibitory conductances $g_\ee$ and $g_\ii$ during a finite, nonzero synaptic activation time $\tau_\mathrm{s}>0$. 
Simulation parameters: $K_\ee=K_\ii=50$, $r_\ee=r_\ii=10\mathrm{Hz}$, $\tau=15\mathrm{ms}$ $\tau_\mathrm{s}=2\mathrm{ms}>0$.
}
\end{center}
\end{figure}

We consider the subthreshold dynamics of an original neuronal model, which we called the all-or-none-conductance-based (AONCB) model.
In this model, the membrane voltage $V$ obeys the first-order stochastic differential equation
\begin{eqnarray}\label{eq:Vdyn}
C \dot{V} = G(V_\leak-V) + g_\ee(V_\ee-V) +  g_\ii(V_\ii-V) + I\, ,
\end{eqnarray}
where randomness arises from the stochastically activating excitatory and inhibitory conductances, respectively denoted by $g_{\mathrm{e}}$ and $g_\ii$ (see \fref{fig:AONCB}a).
These conductances result from the action of $K_\ee$ excitatory and $K_\ii$ inhibitory synapses: $g_{\mathrm{e}}(t)= \sum_{k=1}^{K_\ee} g_{\ee,k}(t)$ and $g_{\ii}(t)= \sum_{k=1}^{K_\ii}  g_{\ii,k}(t)$.
In the absence of synaptic inputs, i.e., when $g_{\mathrm{e}}=g_\ii=0$, and of external current $I$, the voltage exponentially relaxes toward its leak reversal potential $V_\leak$ with passive time constant $\tau=C/G$, where $C$ denotes the cell's membrane capacitance and  $G$ denotes the cellular passive conductance~\cite{rall:1969}.
In the presence of synaptic inputs, the transient synaptic currents $I_\ee=g_{\ee}(V_\ee-V)$ and $I_\ii=g_{\ii}(V_\ii-V)$ cause the membrane voltage to fluctuate.
Conductance-based models account for the voltage-dependence of synaptic currents via the driving forces $V_\ee-V$ and $V_\ii-V$, where $V_\ee$ and $V_\ii$ denotes the excitatory and inhibitory reversal potential, respectively.
Without loss of generality, we assume in the following that $V_\leak=0$ and that $V_\ii < V_\leak=0<V_\ee$.

We model the spiking activity of the $K_\ee+K_\ii$ upstream neurons as shot noise~\cite{stein:1965,tuckwell:1988}, which can be generically modeled as a $(K_\ee+K_\ii)$-dimensional stochastic point process~\cite{daley2003introduction,daley2007introduction}.
Let us denote by $\{ N_{\ee,k}(t)\}_{1\leq k \leq K_\ee}$ its excitatory component and by $\{ N_{\ii,k}(t)\}_{1\leq k \leq K_\ii}$ its inhibitory component, where $t$ denotes time and $k$ is the neuron index.
For each neuron $k$, the process $N_{\eeii,k}(t)$ is specified as the counting process registering the spiking occurrences of neuron $k$ up to time $t$.
In other words,  $N_{\eeii,k}(t) = \sum_{k} \mathbbm{1}_{\{ T_{\eeii,k,n} \leq t \}}$, where $\{ T_{\eeii,k,n} \}_{n \in \mathbbm{Z}}$ denotes the full sequence of spiking times of neuron $k$ and where $\mathbbm{1}_A$ denotes the indicator function of set $A$. 
Note that by convention, we label spikes so that $T_{\eeii,k,0} \leq 0 < T_{\eeii,k,1} $ for all neuron $k$. 
Given a point-process model for the upstream spiking activity, classical conductance-based models consider that a single  input to a synapse causes an instantaneous increase of its conductance, followed by an exponential decay with typical time scale $\tau_\mathrm{s}>0$.
Here we depart from this assumption and consider that the synaptic conductances $g_{\eeii, k}$ operates all-or-none with a common activation time still referred to as $\tau_\mathrm{s}$.
Specifically,  we assume that the dynamics of the conductance $g_{\eeii, k}$ follows
\begin{eqnarray}\label{eq:synDyn}
\lefteqn{\dot{g}_{\eeii, k}(t)/G =} \\
&& \hspace{30pt} w_{\eeii,k} \sum_n \big( \delta( t- T_{\eeii,k,n})-\delta( t- T_{\eeii,k,n} - \tau_\mathrm{s}) \big) \, ,\nonumber 
\end{eqnarray}
where $w_{\eeii,k} \geq 0$ is the dimensionless synaptic weight.
The above equation prescribes that the $n$-th spike delivery to synapse $k$ at time $T_{\eeii,k,n}$ is followed by an instantaneous increase of that synapse's conductance by an amount $w_{\eeii,k}$ for a period $\tau_\mathrm{s}$.
Thus, the synaptic response prescribed by \eref{eq:synDyn} is all-or-none as opposed to being graded as in classical conductance-based models.
Moreover, just as in classical models, \eref{eq:synDyn} allows synapses to multi-activate, thereby neglecting nonlinear synaptic saturation (see \fref{fig:AONCB}b).

To be complete, AONCB neurons must in principle include a spike-generating mechanism.
A customary choice is the integrate-and-fire mechanism~\cite{Knight:1972,Knight:1972il}: a neuron emits a spike whenever its voltage $V$ exceeds a threshold value $V_\Th$, and reset instantaneously to some value $V_R$ afterwards.
Such a mechanism impacts the neuronal subthreshold voltage dynamics via post-spiking reset, which implements a nonlinear form of feedback.
However, in this work we focus on the variability that is generated by fluctuating, possibly synchronous, synaptic inputs.
For this reason, we will neglect  the influence of the spiking reset in our analysis and actually,  we will ignore the spike-generating mechanism altogether.
 Finally, although our analysis of AONCB neurons will apply to positive synaptic activation time $\tau_\mathrm{s}>0$, we will only discuss our results in the limit of instantaneous synapses.
This corresponds to taking $\tau_\mathrm{s} \to 0^+$ while adopting the scaling $g_\eeii \propto 1/\tau_s$ in order to maintain the charge transfer induced by a synaptic event.
We will see that this limiting process preserves the response variability of AONCB neurons.


\subsection{Quantifying the synchrony of exchangeable synaptic inputs}\label{sec:DM}

\begin{figure}[tbp]
	\begin{center}
		\includegraphics[width=7.6cm]{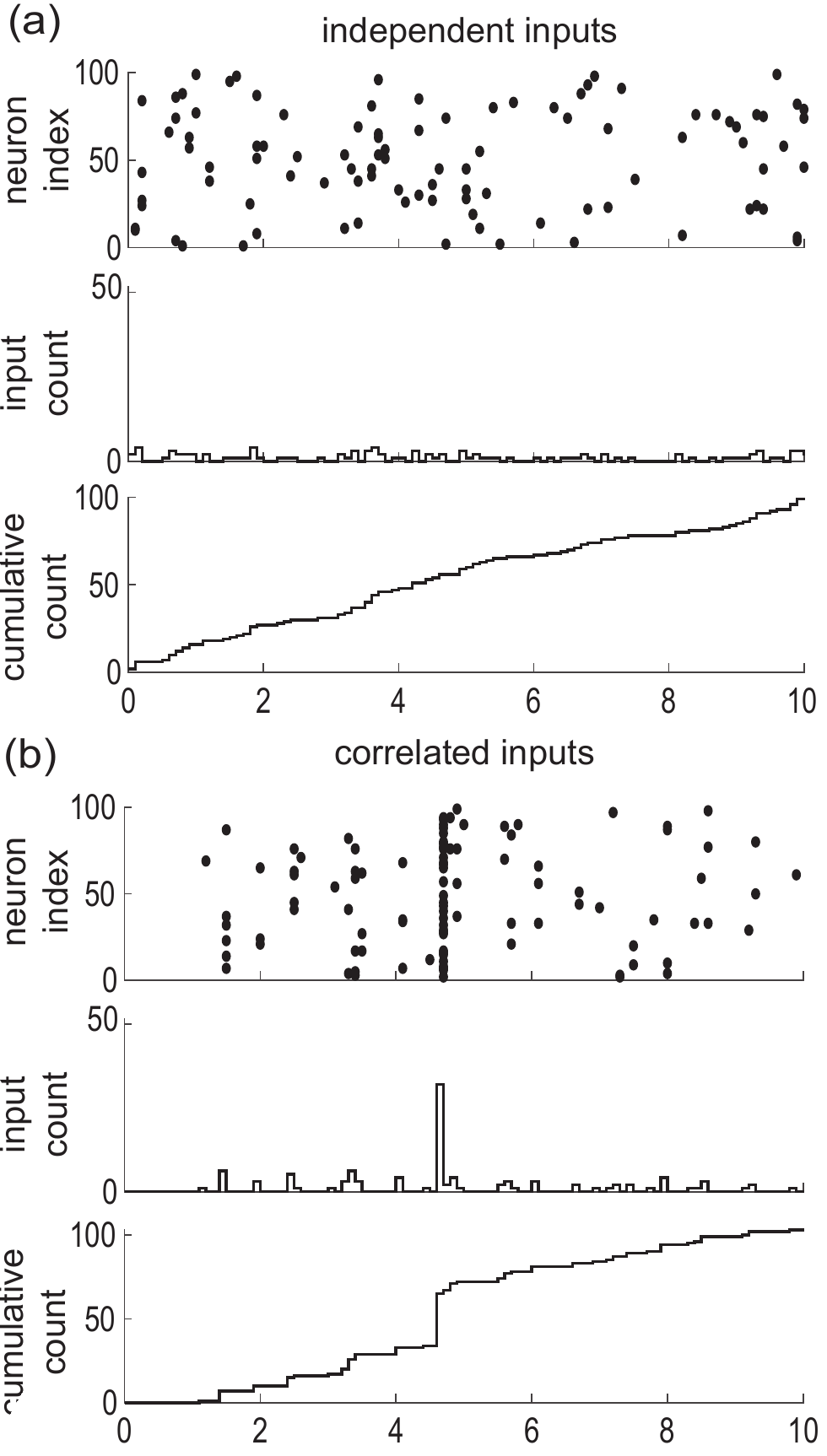}
		\caption{\label{fig:CorrDeFinetti}{\bf Parametrizing correlations via exchangeability.}
		The activity of $K_\ee=100$ exchangeable synaptic inputs collected over $N$ consecutive time bins can be represented as $\{0,1\}$-valued array $\{X_{k,i}\}_{1\leq k \leq K_\ee, 1\leq i \leq N}$, where $X_{k,i}=1$ if input $k$ activates in time bin $i$. 
		Under assumptions of exchangeability, the input spiking correlation is entirely captured by the count statistics of how many inputs coactivate within a given time bin.
		In the limit $K_\ee \to \infty$, the distribution of the fraction of coactivating inputs coincides with the directing de Finetti measure, which we consider as a parametric choice in our approach.
		 In the absence of correlation, synapses tend to activate in isolation: $\rho_\ee=0$ in (a).
		 In the presence of correlation, synapses tend to coactivate yielding disproportionately large synaptic activation event: $\rho_\ee=0.1$ in (b).
		 Considering the associated cumulative counts specifies discrete-time jump processes that can be generalized to the continuous-time limit, i.e., for time bins of vanishing duration $\Delta t \to 0^+$.
		}
	\end{center}
\end{figure}

Our goal here is to introduce a discrete model for synaptic inputs, whereby synchrony can be rigorously quantified.
To this end, let us suppose that the neuron under consideration receives inputs from $K_\ee$ excitatory neurons and $K_\ii$ inhibitory neurons, chosen from arbitrary large pools of $N_\ee \gg K_\ee$ excitatory neurons and $N_\ii \gg K_\ii$ inhibitory neurons.
Adopting a discrete-time representation with elementary bin size $\Delta t$, we denote by $\{\{ x_{1,n}, \ldots, x_{K_\ee,n}\} , \{y_{1,n}, \ldots, y_{K_\ii,n}\} \}$ in $\{0,1\}^{K_\ee} \times \{0,1\}^{K_\ii}$ the input state within the $n$-th bin.
Our main simplifying assumption consists in modeling the $N_\ee$ excitatory inputs and the $N_\ii$ inhibitory inputs as separately exchangeable random variables $\{ X_{1,n}, \ldots, X_{K_\ee,n} \}$ and $\{Y_{1,n}, \ldots, Y_{K_\ii,n}\}$ that are distributed identically over $\{0,1\}^{N_\ee}$ and $\{0,1\}^{N_\ii}$, respectively, and independently across time. 
This warrants dropping the dependence on time index $n$.
By separately exchangeable, we mean that no subset of excitatory inputs or inhibitory inputs plays a distinct role so that at all time, the respective distributions of  $\{ X_{1,n}, \ldots, X_{K_\ee,n} \}$ and $\{Y_{1,n}, \ldots, Y_{K_\ii,n}\}$ are independent of the inputs labelling.
In other words, for all permutations $\sigma_\ee$ of $\{ 1, \ldots, N_\ee\}$ and $\sigma_\ii$ of $\{ 1, \ldots, N_\ii\}$,  the joint distribution of $\{ X_{\sigma_\ee(1)}, \ldots, X_{\sigma_\ee(N_\ee)} \}$ and $\{ Y_{\sigma_\ii(1)}, \ldots, Y_{\sigma_\ii(N_\ii)} \}$ is identical to that of  $\{ X_1, \ldots, X_{N_\ee} \}$ and $\{ Y_1, \ldots, Y_{N_\ii} \}$~\cite{kingman:1978,aldous1985exchangeability}.
By contrast with independent random spiking variables, exchangeable ones can exhibit nonzero correlation structure.
By symmetry, this structure is specified by three correlation coefficients
\begin{eqnarray}
\rho_\ee = \frac{\Cov{X_k,X_l}}{\Var{X_k}}  \, , \,
\rho_\ii = \frac{\Cov{Y_k,Y_l}}{\Var{Y_k}}  \, , \,
\rho_{\ee\ii} = \frac{\Cov{X_k,Y_l}}{\sqrt{\Var{X_k} \Var{Y_l}}}  \, , \nonumber
\end{eqnarray}
where $\Cov{X,Y}$ and $\Var{X}$ denote the covariance and the variance of the binary variables $X$ and $Z$, respectively.

Interestingly, a more explicit form for $\rho_\ee$, $\rho_\ee$, and $\rho_{\ee\ii}$ can be obtained in the limit of an infinite-size pool $N_\ee,N_\ii \to \infty$.
This follows from de Finetti's theorem~\cite{definetti:1929}, which states that the probability of observing a given input configuration for $K_\ee$ excitatory neurons and $K_\ii$ inhibitory neurons is given  by
\begin{eqnarray}\label{eq:deFinetti}
\lefteqn{\Prob{X_1, \ldots, X_{K_\ee},Y_1, \ldots, Y_{K_\ii}} 
= } \nonumber\\
&&
 \int \prod_{k=1}^{K_\ee} \theta_\ee^{X_k} (1-\theta_\ee)^{1-X_k}  \prod_{l=1}^{K_\ii} \theta_\ii^{X_l} (1-\theta_\ii)^{1-X_l} \, \dd F_{\ee \ii}(\theta_\ee, \theta_\ii) \, . \nonumber
\end{eqnarray}
where $F_{\ee\ii}$ is the directing de Finetti measure, defined as a bivariate distribution over the unit square $[0,1]\times [0,1]$.
In the equation above, the number $\theta_\ee$ and $\theta_\ii$ represent the (jointly fluctuating) probabilities that an excitatory neuron and an inhibitory neuron spike in a given time bin, respectively.
The core message of de Finetti theorem is that the spiking activity of neurons from infinite exchangeable pools is obtained as a mixture of conditionally independent binomial laws.
This mixture is specified by the directing measure $F_{\ee\ii}$, which fully parametrizes our synchronous input model.
Independent spiking corresponds to choosing $F_{\ee\ii}$ as a point-mass measure concentrated on some probabilities $\pi_\eeii=r_\eeii \Delta t$, where $r_{\eeii}$ denotes the individual spiking rate of a neuron:  $\dd F_{\ee\ii}(\theta)= \delta(\theta_\ee-\pi_\ee) \delta(\theta_\ii-\pi_\ii)  \dd \theta_\ee \dd \theta_\ii $ (see Fig~\ref{fig:CorrDeFinetti}a).
By contrast, a dispersed directing measure $F_{\ee\ii}$ corresponds to the existence of correlations among the inputs (see Fig~\ref{fig:CorrDeFinetti}b).
Accordingly, we show in Appendix \ref{app:CorrDisc} that the spiking pairwise correlation $\rho_\eeii$ take the explicit form
\begin{eqnarray}\label{eq:corr}
\rho_\eeii = \frac{\Var{\theta_\eeii}}{\Exp{\theta_\eeii} (1-\Exp{\theta_\eeii} )} \, , 
\end{eqnarray}
whereas $\rho_{\ee\ii}$, the correlation between excitation and inhibition, is given by
\begin{eqnarray}\label{eq:corrEI}
\rho_{\ee \ii} = \frac{\Cov{\theta_\ee, \theta_\ii}}{\sqrt{\Exp{\theta_\ee}\Exp{\theta_\ii}(1-\Exp{\theta_\ee})(1-\Exp{\theta_\ii})}}  \, .
\end{eqnarray}
In the above formulas, $\Exp{\theta_{\eeii}}$, $\Var{\theta_{\eeii}}$, and $\Cov{\theta_\ee,\theta_\ii}$ denote expectation, variance, and covariance of $(\theta_\ee,\theta_i )\sim F_{\ee\ii}$, respectively.
Note that these formulas show that nonzero correlations $\rho_\eeii$  correspond to nonzero variance, as is always the case for dispersed distribution.
Independence between excitation and inhibition for which $\rho_{\ee \ii} =0$ corresponds to directing measure $F_{\ee \ii}$ with product form, i.e., $F_{\ee \ii}(\theta_\ee,\theta_\ii)=F_{\ee}(\theta_\ee) F_{\ii}(\theta_\ii)$, where $F_\ee$ and $F_\ii$ denote the marginal distributions. 
Alternative forms of the directed measure $F_{\ee \ii}$ generally lead to nonzero cross correlation $\rho_{\ee \ii}$, which necessarily satisfies $0<\vert \rho_{\ee \ii} \vert \leq \sqrt{\rho_\ee \rho_\ii}$.

In this exchangeable setting, a reasonable parametric choice for the marginals $F_\ee$ and $F_\ii$ is given by beta distributions $\mathrm{Beta}(\alpha,\beta)$, where $\alpha$ and $\beta$ denote shape parameters~\cite{gupta2004handbook}.
Practically, this choice is motivated by the ability of beta distributions to efficiently fit correlated spiking data generated by existing algorithms~\cite{macke:2009}.
Formally, this choice is motivated by the fact that beta distributions are conjugate priors for the binomial likelihood functions, so that the resulting probabilistic models can be studied analytically~\cite{hjort:1990,thibaux:2007,broderick2012beta}.
For instance, for $F_\ee \sim \mathrm{Beta}(\alpha_\ee,\beta_\ee)$, the probability that $k_\ee$ synapses among the $K_\ee$ inputs are jointly active within the same time bin follows the beta-binomial distribution
\begin{eqnarray}\label{eq:PkDef}
P_{\ee,k} = \binom{K_\ee}{k} \frac{B(\alpha_\ee + k,\beta_\ee+K_\ee -k )}{B(\alpha_\ee,\beta_\ee)} \, . 
\end{eqnarray}
Accordingly, the mean number of active excitatory inputs is $\Exp{k_\ee} = K_\ee \alpha_\ee/(\alpha_\ee+\beta_\ee)=K_\ee r_\ee \Delta t$.
Utilizing \eref{eq:corr}, we also find that $\rho_\ee = 1/(1+\alpha_\ee+\beta_\ee)$.
 Note that the above results show that by changing de Finetti's measure, one can not only modify the spiking correlation  but also the mean spiking rate. 

In the following, we will exploit the above analytical results to illustrate that taking the continuous-time limit $\Delta t \to 0^+$ specifies synchronous input drives as compound Poisson processes~\cite{daley2003introduction,daley2007introduction}.
To do so, we will consider both excitation and inhibition, which in a discrete setting corresponds to considering bivariate probability distributions $P_{\ee\ii,kl}$ defined over $\{ 0, \ldots , K_\ee\} \times \{ 0, \ldots,K_\ii \}$.
Ideally, these distributions $P_{\ee\ii,kl}$ should be such that its conditional marginals $P_{\ee,k}$ and $P_{\ii,l}$, with distributions given by \eref{eq:PkDef}.
Unfortunately, there does not seem to be a simple low-dimensional parametrization for such distributions $P_{\ee\ii,kl}$, except in particular cases.
To address this point, at least numerically, one can resort to a variety of methods including copulas~\cite{berkes:2008,balakrishnan2009continuous}.
For analytical calculations, we will only consider consider two particular cases for which the marginals of $F_{\ee\ii}$ are given by the beta distributions: $(i)$ the case of maximum positive correlation for which $\theta_\ee=\theta_\ii$, i.e., $\dd F_{\ee\ii}(\theta_\ee , \theta_\ii) =\delta(\theta_\ee - \theta_\ii) F(\theta_\ee) \, \dd \theta_\ee \dd \theta_\ii$ with $F_{\ee}= F_{\ii}=F$ and $(ii)$ the case of zero correlation for which $\theta_\ee$ and $\theta_\ii$ are independent, i.e., $ F_{\ee\ii}(\theta_\ee , \theta_\ii) =F_{\ee}(\theta_\ee)F_{\ii}(\theta_\ii)$.


\subsection{Synchronous synaptic drives as compound Poisson processes}\label{sec:CPPE}

\begin{figure*}[tbp]
\begin{center}
\includegraphics[width=0.95\textwidth]{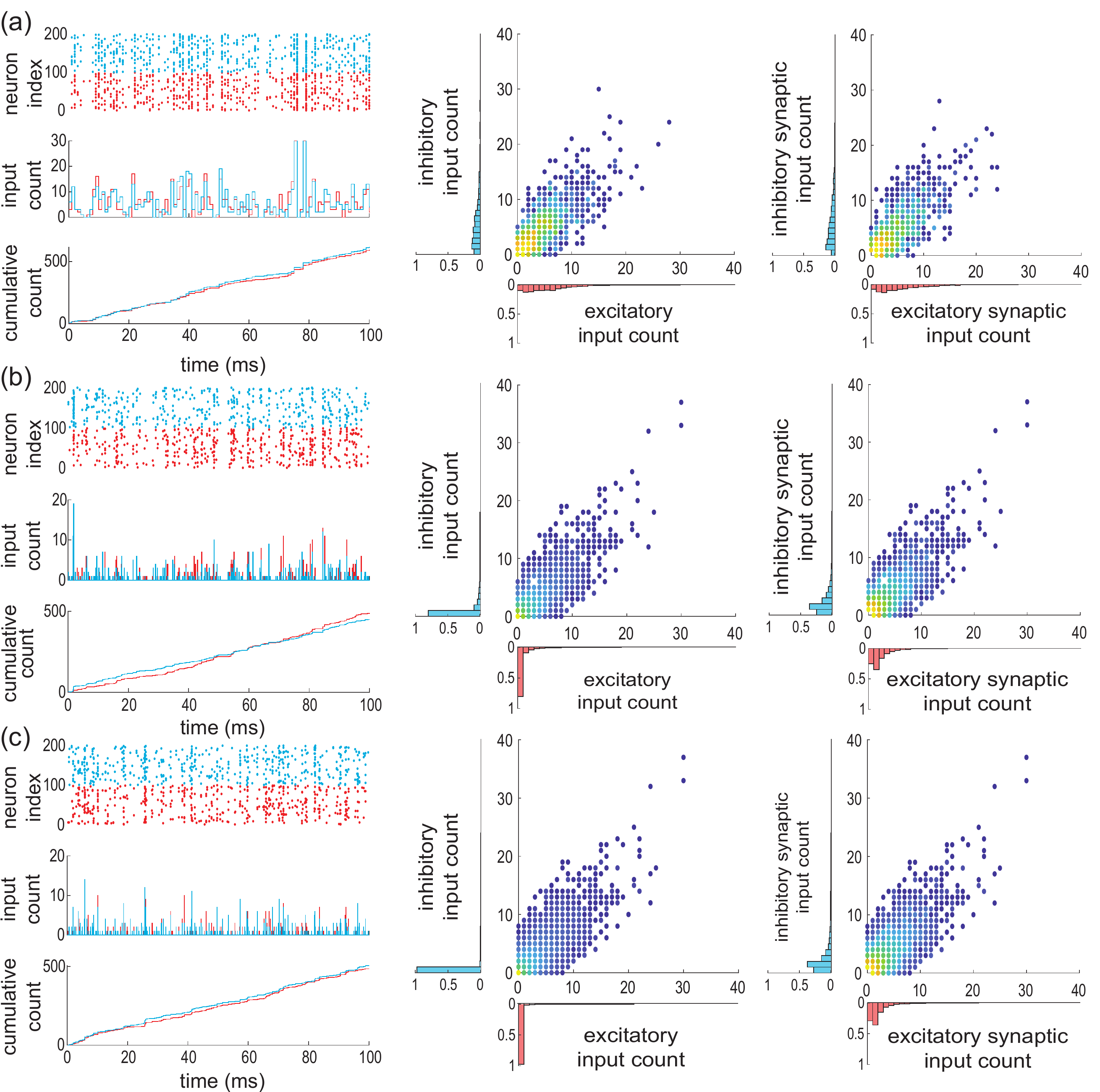}
\caption{\label{fig:ContLim2}{\bf Limit compound Poisson process with excitation and inhibition.}
(a)
Under assumption of partial exchangeability, synaptic inputs can only be distinguished by the fact that they are either excitatory or inhibitory, which is marked by being colored in red or blue in the discrete representation of correlated synaptic inputs with bin size $\Delta t=1\mathrm{ms}$.
Accordingly, considering excitation and inhibition separately specifies two associated input-count processes and two cumulative counting processes.
For nonzero spiking correlation $\rho=0.03$, these processes are themselves correlated as captured by the joint distribution of excitatory and inhibitory input counts $P_{\ee\ii,kl}$ (center) and by the  joint distribution of excitatory and inhibitory jumps $P_{\ee\ii,kl}/(1-P_{00})$ (right).
(b)
The input count distribution $P_{\ee\ii,kl}$ is a finite-size approximation of the bivariate directing de Finetti measure $F_{\ee\ii}$, which we consider as a parameter as usual.
For a smaller bin size $\Delta t =0.1\mathrm{ms}$, this distribution concentrates in $(0,0)$, as an increasing proportion of time bins does not register any synaptic events, be they excitatory or inhibitory.
In the presence of correlation however, the conditioned jump distribution remains correlated but also dispersed.
(c)
In the limit $\Delta t \to 0$, the input-count distribution is concentrated in $(0,0)$, consistent with the fact that the average number of synaptic activations remains constant while the number of bins diverges.
By contrast, the distribution of synaptic event size conditioned to distinct from $(0,0)$ converges toward a well-defined distribution: $p_{\ee\ii,kl}=\lim_{\Delta t \to 0^+}P_{\ee\ii,kl}/(1-P_{\ee\ii,00})$.
This distribution characterizes the jumps of a bivariate compound Poisson process, obtained as the limit of the cumulative count process when considering $\Delta t \to 0^+$.
}
\end{center}
\end{figure*}

 Under assumption of input exchangeability and given typical excitatory and inhibitory synaptic weights $w_\eeii$, the overall synaptic drive to a neuron is determined by $(k_\ee,k_\ii)$, the numbers of active excitatory and inhibitory inputs at each discrete time step.
As AONCB dynamics unfolds in continuous time, we  need to consider this discrete drive in the continuous-time limit as well, i.e., for vanishing time bins $\Delta t \to 0^+$.
When $\Delta t \to 0^+$, we show in Appendix \ref{app:CorrProc1} that the overall synaptic drive specifies  a compound Poisson process $Z$ with bivariate jumps $(W_\ee,W_\ii)$.
Specifically, we have
\begin{eqnarray}\label{def:biPoiss}
Z(t)= \left( \sum_{n}^{N(t)} W_{\ee,n},  \sum_{n}^{N(t)} W_{\ii,n} \right) \, , 
\end{eqnarray}
where $(W_{\ee,n}, W_{\ii,n})$ are i.i.d. samples with bivariate distribution denoted by $p_{\ee\ii}$ and where the overall driving Poisson process $N$ registers the number of synaptic events without mulitple counts (see \fref{fig:ContLim2}).
By synaptic events, we mean these times for which at least one excitatory synapse or one inhibitory synapse activates.
We  say that $N$ registers these events without multiple count as it counts one event independent of the number of possibly coactivating synapses.
Similarly, we denote by $N_\ee$ and $N_\ii$ the counting processes registering synaptic excitatory events and synaptic  inhibitory events alone, respectively.
These processes $N_\ee$ and $N_\ii$ are Poisson processes that are correlated in the presence of synchrony, as both $N_\ee$ and $N_\ii$ may register the same event.
Note that this implies that $\max(N_\ee(t),N_\ii(t)) \leq N(t) \leq N_\ee(t)+N_\ii(t)$.
More generally, denoting by $b$ and $b_\eeii$ the rates of $N$ and $N_\eeii$, respectively, the presence of synchrony implies that $\max(b_\ee,b_\ii) \leq b \leq b_\ee+b_\ii$ and $r_\eeii \leq b_\eeii \leq K_\eeii r_\eeii$, where $r_\eeii$ is the typical activation rates of a single synapse.

For simplicity, we explain how to obtain such limit compound Poisson processes by reasoning on the excitatory inputs alone.
To this end, let us denote the marginal jump distribution of $W_\ee$ as $p_\ee$.
Given a fixed typical synaptic weight $w_\ee$, the jumps are quantized as $W_\ee = k w_\ee$, with $k$ distributed on $\{1, \ldots, K_\ee\}$, as by convention jumps cannot have zero size.
These jumps are naturally defined in the discrete setting, i.e., with $\Delta t>0$, and their discrete distribution is given via conditioning as $P_{\ee,k}/ (1-P_{\ee,0})$.
For beta distributed marginals $F_\ee$, we show in Appendix \ref{app:CorrProc1} that considering $\Delta t \to 0^+$ yields the jump distribution
\begin{eqnarray}\label{eq:betaParam}
p_{\ee, k} \!= \!\! \lim_{\Delta t \to 0^+} \frac{P_{\ee,k}}{1-P_{\ee,0}} = \binom{K_\ee}{k} \frac{B(k,\beta_\ee+K_\ee-k)}{\psi(\beta_\ee+K_\ee)-\psi(\beta_\ee)} \, , 
\end{eqnarray}
where $\psi$ denotes the digamma function.
In the following, we will explicitly index discrete count distributions, e.g. $p_{\ee, k}$, to distinguish them form the corresponding jump distributions, i.e., $p_\ee$.
\eref{eq:betaParam} follows from observing that the probability to find a spike within a bin is $\Exp{X_i} = \alpha_\ee / (\alpha_\ee+\beta_\ee)= r_\ee \Delta t$, so that for fixed excitatory spiking rate $r_\ee$, $\alpha_\ee \to 0^+$ when $\Delta t \to 0^+$.
As a result, the continuous-time spiking correlation is $\rho_\ee=1/(1+\beta_\ee)$, so that we can interpret $\beta_\ee$ as a parameter controlling correlations.
More generally, we show in Appendix \ref{app:CorrCont} that the limit correlation $\rho_\ee$ only depends on the  count distribution $p_{\ee, k}$  via
\begin{eqnarray}\label{eq:CorrRhoe}
\rho_\ee = 
\frac{
\ExpPe{k(k-1)}
}
{
\ExpPe{k}(K_\ee -1)
} \, ,
\end{eqnarray}
where $\ExpPe{\cdot}$ denotes expectations with respect to $p_{\ee,k}$.
This shows that zero spiking correlation corresponds to single synaptic activations, i.e., to an input drive modeled as a Poisson process, as opposed to a compound Poisson process.
For Poisson-process models, the overall rate of synaptic events is necessarily equal to the sum of the individual spiking rate: $b_\ee=K_\ee r_\ee$.
This is no longer the case in the presence of synchronous spiking, when nonzero input correlation $\rho_\ee>0$ arises from coincidental synaptic activations.
Indeed, as the population spiking rate is conserved when  $\Delta t \to 0^+$, the rate of excitatory synaptic events $b_\ee$ governing $N_\ee$ satisfies
$K_\ee r_\ee=b_\ee \ExpPe{k}$ so that
\begin{eqnarray}\label{eq:bDef}
b_\ee = \frac{K_\ee r_\ee}{\ExpPe{k}} = r_\ee \beta_\ee \left(\psi(\beta_\ee+K_\ee)-\psi(\beta_\ee) \right)\, .
\end{eqnarray}
Let us reiterate for clarity, that if $k_\ee$ synapses activate synchronously, this only counts as one synaptic event, which can come in variable size $k$. 
Consistently, we have in general $r_\ee \leq b_\ee \leq  K_\ee r_\ee$.
When $\beta_\ee \to 0$, we have perfect synchrony with $\rho_\ee=1$ and  $b_\ee \to r_\ee$, whereas the independent spiking regime with $\rho_\ee=0$ is attained for $\beta_\ee \to \infty$, for which we have  $b_\ee \to K_\ee r_\ee$.

It is possible to generalize the above construction to mixed excitation and inhibition but a closed-from treatment is only possible for special cases.
For the independent case $(i)$, in the limit $\Delta t \to 0^+$, jumps are either excitatory alone or inhibitory alone, i.e., the jump distribution $p_{\ee\ii}$ has support on 
$\{ 1, \ldots, K_\ee \} \times \{ 0 \} \cup \{ 0 \} \times \{1 , \ldots, K_\ii \}$.
Accordingly, we show in Appendix \ref{app:CorrProc2} that
\begin{eqnarray}\label{eq:peiDefInd}
p_{\ee \ii, kl} 
&=&
\lim_{\Delta t \to 0^+} \frac{P_{k}P_{l}}{1-P_{\ee,0}P_{\ii,0}}   \, , \\
&=&
\frac{b_\ee}{b_\ee+b_\ii} p_{\ee, k} \one{l=0}  + \frac{b_\ee}{b_\ee+b_\ii}  p_{\ii, l} \one{k=0}  \, , \nonumber
\end{eqnarray}
where $p_{\eeii, k}$ and $b_\eeii$ are specified by \eref{eq:betaParam} and \eref{eq:bDef} the parameters $\beta_{\eeii}$ and $K_{\eeii}$, respectively.
This shows that as expected, in the absence of synchrony the driving compound poisson process $Z$ with bidimensional jump is obtained as the direct sum of two independent compound Poisson processes.
In particular, the driving processes are such that $N=N_\ee+N_\ii$, with rates satisfying $b=b_\ee+b_\ii$.
By contrast, for the maximally correlated case with $r_\ee=r_\ii=r$ $(ii)$, we show in Appendix \ref{app:CorrProc2} that the jumps are given as $(W_\ee,W_\ii) = (k w_\ee, l w_\ii)$, with $(k,l)$ distributed on $\{0, \ldots, K_\ee \}  \times \{0, \ldots, K_\ii \} \setminus \{0,0\}$ (see \fref{fig:ContLim2}b and c) according to
\begin{eqnarray}\label{eq:peiDef}
p_{\ee \ii, kl} &=& \lim_{\Delta t \to 0^+} \frac{P_{\ee\ii,kl}}{1-P_{\ee\ii,00}}  \\
&=& \binom{K_\ee}{k} \binom{K_\ii}{l} \frac{B(k+l,\beta+K_\ee+K_\ii-k-l)}{\psi(\beta+K_\ee+K_\ii)-\psi(\beta)} \nonumber \, .
\end{eqnarray}
Incidentally, the driving Poisson process $N$ has a rate $b$ determined by adapting \eref{eq:bDef}
\begin{eqnarray}\label{eq:bDef2}
b = r \beta \left(\psi(\beta+K_\ee+K_\ii)-\psi(\beta) \right)\, ,
\nonumber
\end{eqnarray}
for which one can check that $r \leq b \leq (K_\ee + K_\ii )r$.

All the closed-form results so far have been derived for synchrony parametrization in terms of beta distribution.
There are other possible parametrizations and these would lead to different count distributions $p_{\ee\ii,kl}$, but without known closed-form.
To address this limitation in the following, all our results will hold for arbitrary distributions $p_{\ee\ii}$ of the jump sizes $(W_\ee,W_\ii)$ on the positive orthant $(0,\infty) \times(0,\infty)$.
In particular, our results will be given in terms of expectations with respect to  $p_{\ee\ii}$, still denoted by $\ExpPei{\cdot}$.
Nonzero correlation between excitation and inhibition will correspond to those choices of $p_{\ee\ii}$ for which $W_\ee W_\ii>0$ with nonzero probability, which indicates the presence of synchronous excitatory and inhibitory inputs.
Note that this modeling setting restricts nonzero correlations to be positive, which is an inherent limitation of our synchrony-based approach.
When considering an arbitrary $p_{\ee\ii}$, the main caveat is understanding how such a distribution may correspond to a given input numbers $K_\eeii$ and spiking correlations $\rho_\eeii$ and $\rho_{\ee\ii}$.
For this reason, we will always consider that  $k_\ee=W_\ee/w_\ee$ and $k_\ii=W_\ii/w_\ii$ follows beta distributed marginal distributions when discussing the roles of $w_\eeii$, $K_\eeii$, $\rho_\eeii$, and $\rho_{\ee\ii}$ in shaping the voltage response of a neuron.
In that respect, we show in Appendix \ref{app:CorrCont} that the coefficient $\rho_{\ee\ii}$ can always be deduced from the knowledge of a discrete count distribution $p_{\ee\ii,kl}$ on $\{0, \ldots, K_\ee \}  \times \{0, \ldots, K_\ii \} \setminus \{0,0\}$ via 
\begin{eqnarray}
\rho_{\ee\ii} = \frac{\ExpPei{k_\ee k_\ii}}{\sqrt{K_\ee  \ExpPei{k_\ee}K_\ii  \ExpPei{k_\ii}}} \geq 0 \, ,
\nonumber
\end{eqnarray}
where the expectations are with respect to $p_{\ee\ii,kl}$.


\subsection{Instantaneous synapses and Marcus integrals}\label{sec:Marcus}

We are now in a position to formulate the mathematical problem at stake within the framework developed by Marcus to study shot-noise driven systems~\cite{marcus:1978, marcus:1981}.
Our goal is quantifying the subthreshold variability of an AONCB neuron subjected to synchronous inputs.
Mathematically, this amounts to computing the first two moments of the stationary process solving the following stochastic dynamics
\begin{eqnarray}\label{eq:modMath}
 \dot{V} = -V /\tau +  h_\ee(V_\ee-V) +  h_\ii(V_\ii-V) + I/C\, ,
\end{eqnarray}
where $V_\ii<0<V_\ee$ are constants and where the reduced conductances $h_\ee=g_\ee/C$ and $h_\ii=g_\ii/C$ follows stochastic processes defined in terms of a compound Poisson process $Z$ with bivariate jumps.
Formally, the compound Poisson process $Z$ is specified by $b$, the rate of its governing Poisson process $N$, and by the joint distribution  of its jumps  $p_{\ee \ii}$.
Each point of the Poisson process $N$ represents a synaptic activation time $T_n$, where $n$ is in $\mathbbm{Z}$ with the convention that $T_0 \leq 0 \leq T_1$.
At all these times, the synaptic input sizes are drawn as i.i.d. random variables $(W_{\ee,n},W_{\ii,n})$ in $\mathbbm{R}^+ \times \mathbbm{R}^+$ with probability distribution $p_{\ee\ii}$.

\begin{figure}[tbp]
\begin{center}
\includegraphics[width=0.45\textwidth]{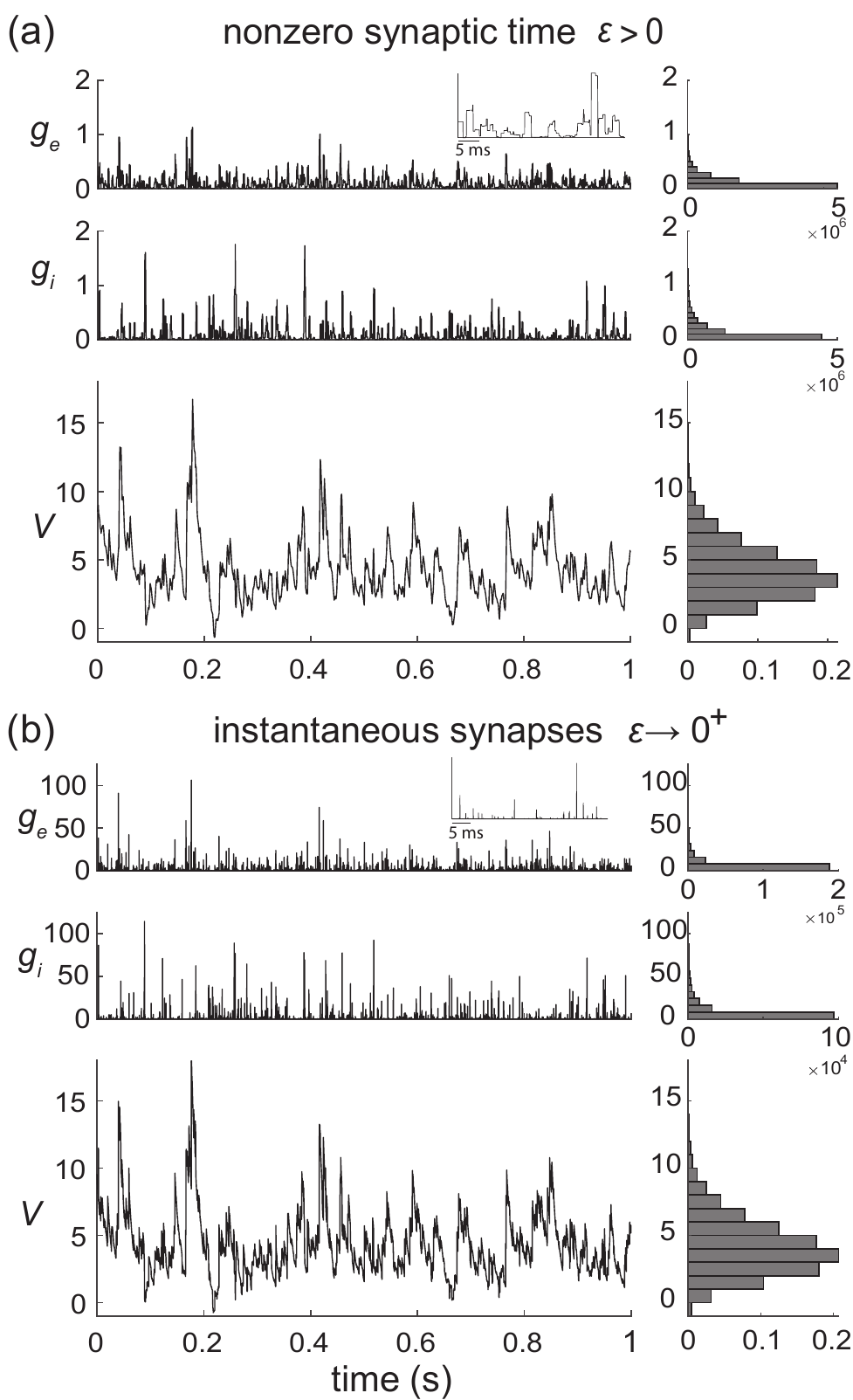}
\caption{\label{fig:InstSyn}{\bf Limit of instantaneous synapses.}
The voltage trace and the empirical voltage distribution are only marginally altered by taking the limit $\epsilon \to 0^+$ for short synaptic time constant: $\tau_\mathrm{s}=2\mathrm{ms}$ in (a) and $\tau_\mathrm{s}=0.02\mathrm{ms}$ in (b).
In both (a) and (b), we consider the same compound Poisson-process drive with  $\rho_\ee=0.03$, $\rho_\ii=0.06$, and $\rho_{\ee\ii}=0$, and the resulting fluctuating voltage $V$ is simulated via a standard Euler discretization scheme.
The corresponding empirical conductance and voltage distributions are shown on the right.
The later voltage distribution asymptotically determines the stationary moments of $V$.  
}
\end{center}
\end{figure}

At this point, it is important to observe that the driving process $Z$ is distinct from the conductance process $h=(h_\ee,h_\ii)$.
The latter process is formally defined for AONCB neurons as
\begin{eqnarray}
h(t) &=& \frac{Z(t)-Z(t-\epsilon \tau)}{\epsilon \tau} \, , \nonumber\\
&=&  \frac{1}{\epsilon \tau } \left( \sum^{N(t)}_{n=N(t-\epsilon \tau)+1} \hspace{-10pt} W_{\ee, n} \; ,\sum^{N(t)}_{n=N(t-\epsilon \tau)+1} \hspace{-10pt} W_{\ii,n}  \right)\, , 
\nonumber
\end{eqnarray}
where the dimensionless parameter $\epsilon=\tau_\mathrm{s}/\tau>0$ is the ratio of the duration of synaptic activation relative to the passive membrane time constant.
 Note that the amplitude of $h$ scales in inverse proportion to $\epsilon$ in order to maintain the overall charge transfer during synaptic events of varying durations.
Such a scaling ensures that the voltage response of AONCB neurons has finite, nonzero variability for small or vanishing synaptic time constant, i.e., for $\epsilon \ll 1$ (see Fig~\ref{fig:InstSyn}).
The simplifying limit of instantaneous synapses is obtained for $\epsilon=\tau_\mathrm{s}/\tau \to 0^+$, which corresponds to infinitely fast synaptic activation.
By virtue of its construction, the conductance process $h$ becomes a shot noise in the limit $\epsilon \to 0^+$, which can be formally identified to $\dd Z/\dd t$.
This is consistent with the definition of shot-noise processes as temporal derivative of compound Poisson processes, i.e., as collections of randomly weighted Dirac-delta masses.


 Due to their high degree of idealization, shot-noise models are often amenable to exact stochastic analysis, albeit with some caveats.
For equations akin to \eref{eq:modMath} in the limit of instantaneous synapses, such a caveat follows from the multiplicative nature of the conductance shot noise $h$.
In principle, one might expect to solve \eref{eq:modMath} with shot-noise drive via stochastic calculus, as for diffusion-based drive.
This would involve interpreting  the stochastic integral representations of solutions in terms of Stratonovich representations~\cite{stratonovich:1966}.
However, Stratonovich calculus is not well-defined for shot-noise drives~\cite{chechkin:2014}.
To remedy this point, Marcus has proposed to study stochastic  equations subjected to regularized versions of shot noises, whose regularity is controlled by a nonnegative parameter $\epsilon$~\cite{marcus:1978, marcus:1981}.
For $\epsilon>0$, the dynamical equations admit classical solutions, whereas the shot-noise-driven regime is recovered in the limit $\epsilon \to 0^+$.
The hope is to be able to characterize analytically the shot-noise-driven solution, or at least some of its moments, by considering regular solutions in the limit $\epsilon \to 0^+$.
We choose to refer to the control parameter as $\epsilon$ by design in the above.
This is because AONCB models represent Marcus-type regularizations that are amenable to analysis in the limit of instantaneous synapses, i.e., when $\epsilon=\tau_\mathrm{s}/\tau \to 0^+$, for which the conductance process $h$ converges toward a form of shot noise.

Marcus interpretation of stochastic integration has practical implications for numerical simulations with shot noise~\cite{richardson:2004}.
According to this interpretation, shot-noise-driven solutions shall be conceived as limits of regularized solutions for which standard numerical scheme applies.
Correspondingly, shot-noise-driven solutions to \eref{eq:modMath} can be simulated via a limit numerical scheme.
We derive such a limit scheme in Appendix \ref{app:Marcus}.
Specifically, we show that the voltage of shot-noise-driven AONCB neurons exponentially relaxes toward the leak reversal potential $V_\leak=0$, except when subjected to synaptic impulses at times $\{T_n\}_{n \in \mathbbm{Z}}$.
At these times, the voltage $V$ updates discontinuously according to $V(T_n)=V(T_n^-)+J_n$, where the jumps are given in Appendix \ref{app:Marcus} via the Marcus rule
\begin{eqnarray}\label{eq:MarcusJump}
\lefteqn{J_n =} \\ 
&& \left( \frac{W_{\ee,n} V_\ee + W_{\ii,n} V_\ii }{W_{\ee,n} +W_{\ii,n} } - V(T^-_n) \right) \! \left( 1- e^{-(W_{\ee,n}+W_{\ii,n})}\right) \nonumber\, .
\end{eqnarray}
Observe that the above Marcus rule directly implies that no jump can cause the voltage to exit $(V_\ii,V_\ee)$, the allowed range of variation for $V$.
Moreover, note that this rule specifies an exact even-driven simulation scheme given knowledge of the synaptic activation times and sizes $\{T_n,W_{\ee,n},W_{\ii,n}\}_{n \in \mathbbm{Z}}$~\cite{Mathes:1964}.
We adopt the above Marcus-type numerical scheme in all the simulations that involve instantaneous synapses.


\subsection{Moment calculations}\label{sec:moment}

\begin{figure}[htbp]
\begin{center}
\includegraphics[width=8.6cm]{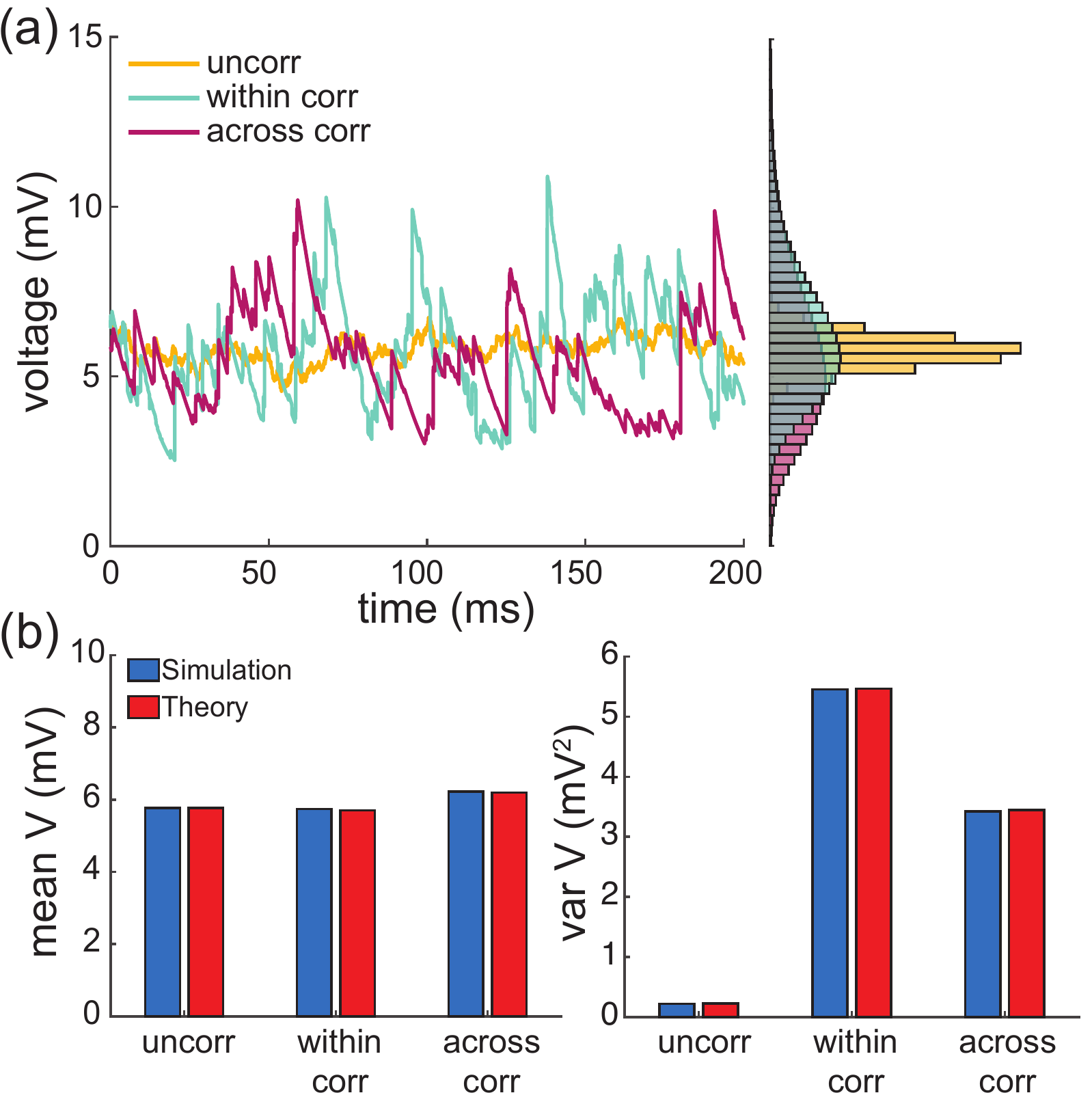}
\caption{\label{fig:SimTheory}{\bf Comparison of simulation and theory.}
(a) 
Examples of voltage traces obtained via Monte-Carlo simulations of an AONCB neuron for various type of synchrony-based input correlations: 
uncorrelated $\rho_\ee = \rho_\ii = \rho_{\ee\ii}=0$ (uncorr, yellow),
within correlation $\rho_\ee, \rho_\ii>0$ and $\rho_{\ee\ii}=0$ (within corr, cyan),
within and across correlation $\rho_\ee, \rho_\ii, \rho_{\ee\ii}>0$ (across corr, magenta).
(b)
Comparison of the analytically derived expressions \eqref{eq:statmean} and \eqref{eq:statvar} with numerical estimates obtained via Monte-Carlo simulations for the synchrony conditions considered in (a).
}
\end{center}
\end{figure}

When driven by stationary compound-Poisson processes,  AONCB neurons exhibit ergodic voltage dynamics.
As a result, the typical voltage state, obtained by sampling the voltage at random time, is captured by a unique stationary distribution.
Our main analytical results, which we give here, consists in exact formulas for the first two voltage moment with respect to that stationary distribution.
Specifically, we derive the stationary mean voltage \eref{eq:statmean}  in Appendix \ref{app:mean} and  the stationary voltage variance \eref{eq:statvar} in Appendix \ref{app:var}.
These results are obtained by a probabilistic treatment exploiting the properties of compound Poisson processes within Marcus' framework.
This treatment yields compact, interpretable formulas in the limit of instantaneous synapses $\epsilon = \tau_\mathrm{s}/\tau \to 0^+$.
Readers who are interested in the method of derivation for these results are encouraged to go over the calculations presented in Appendices \ref{app:mean}, \ref{app:var}, \ref{app:QE}, \ref{app:RO}, \ref{app:ID}, \ref{app:VE1}, and \ref{app:VE2}.

In the limit of instantaneous synapses, $\epsilon \to 0^+$, we find that the stationary voltage mean is
\begin{eqnarray}\label{eq:statmean}
\Exp{V} 
=
\lim_{\epsilon \to 0^+}\Exp{V_\epsilon}
=
\frac{
 a_{\ee,1} V_\ee+a_{\ii,1}  V_\ii  + I/G
}{
1+ a_{\ee,1} +a_{\ii,1}
} \, ,
\end{eqnarray}
where we define the first-order synaptic efficacies as
\begin{eqnarray}
\begin{array}{ccc}
a_{\ee,1} &=& \displaystyle b \tau \ExpPei{\frac{W_\ee}{W_\ee+W_\ii}\left(1-e^{-(W_\ee+W_\ii)} \right)}  \, , \vspace{5pt} \\
a_{\ii,1} &=&  \displaystyle b \tau \ExpPei{\frac{W_\ii}{W_\ee+W_\ii}\left(1-e^{-(W_\ee+W_\ii)} \right)} \, .
\end{array}
 \label{eq:a1} 
\end{eqnarray}
Note the $\ExpPei{\cdot}$ refers to the expectation with respect to the jump distribution $p_{\ee\ii}$ in \eref{eq:a1}, whereas $\Exp{\cdot}$ refers to the stationary expectation in \eref{eq:statmean}.
\eref{eq:statmean} has the same form as for deterministic dynamics with constant conductances, in the sense that the mean voltage is a weighted sum of the reversal potentials $V_\ee$, $V_\ii$ and $V_\leak=0$.
One can check that for such deterministic dynamics, the synaptic efficacies involved in the stationary mean simply read $a_{\eeii,1}=K_\eeii r_\eeii w_\eeii$.
Thus, the impact of synaptic variability, and in particular of synchrony, entirely lies in the definition of the efficacies in \eref{eq:a1}.
In the absence of synchrony, one can check that accounting for the shot-noise nature of the synaptic conductances  leads to synaptic efficacies under exponential form: $a_{\eeii,1}=K_\eeii r_\eeii (1-e^{-w_\eeii})$.
In turn, accounting for input synchrony leads to synaptic efficacies expressed as expectation of these exponential forms in \eref{eq:a1}, consistent with the stochastic nature of the conductance jumps $(W_\ee,W_\ii)$.
Our other main result, the formula for the stationary voltage variance, involves synaptic efficacies of similar form.
Specifically, we find that
\begin{eqnarray}\label{eq:statvar}
\lefteqn{\Var{V} = \frac{1}{1+ a_{\ee,2}+a_{\ii,2}}\times } \\
&& \left( a_{\ee,12} (V_\ee \!-\! \Exp{V})^2 + a_{\ii,12} (V_\ii \!-\! \Exp{V})^2 - c_{\ee \ii} (V_\ee \!-\! V_\ii)^2 \right) \nonumber \, ,
\end{eqnarray}
where we define the second-order synaptic efficacies  as
\begin{eqnarray}
\begin{array}{ccc}
a_{\ee,2} &=&   \displaystyle \frac{b \tau}{2} \ExpPei{ \frac{W_\ee}{W_\ee+W_\ii} \left(1- e^{-2(W_\ee+W_\ii)}\right)} \, ,  \vspace{5pt}\\
a_{\ii,2} &=&  \displaystyle \frac{b \tau}{2} \ExpPei{ \frac{W_\ii}{W_\ee+W_\ii} \left(1- e^{-2(W_\ee+W_\ii)}\right)} \, .
\end{array}
\label{eq:ae2}
\end{eqnarray}
\eref{eq:statvar} also prominently features auxiliary second-order efficacies defined by $a_{\eeii,12} = a_{\eeii,1} - a_{\eeii,2}$.
Owing to their prominent role, we also mention their explicit form:
\begin{eqnarray}
\label{eq:ae12}
\begin{array}{ccc}
a_{\ee,12} &=&  \displaystyle \frac{b\tau}{2}  \ExpPei{\frac{W_\ee}{W_\ee+W_\ii}\left(1- e^{-(W_\ee+W_\ii)}\right)^2} \, , \vspace{5pt}\\
a_{\ii,12} &=& \displaystyle \frac{b\tau}{2}  \ExpPei{\frac{W_\ii}{W_\ee+W_\ii}\left(1- e^{-(W_\ee+W_\ii)}\right)^2} \, .
\end{array}
\label{eq:a12}
\end{eqnarray}
The other quantity of interest featuring in \eref{eq:statvar} is the crosscorrelation coefficient 
\begin{eqnarray}
\label{eq:cei}
c_{\ee \ii} = \frac{b \tau}{2}  \ExpPei{\frac{W_\ee W_\ii}{(W_\ee +W_\ii)^2}\left(1- e^{-(W_\ee+W_\ii)}\right)^2} \, ,
\end{eqnarray}
which entirely captures the (nonnegative) correlation between excitatory and inhibitory inputs and shall be seen as an efficacy as well.

To conclude, let us stress that for AONCB models, establishing the above exact expressions does not require any approximation other than taking the limit of instantaneous synapses.
In particular, we neither resort to any diffusion approximations~\cite{zerlaut:2019,sanzeni:2022} nor invoke the effective-time-constant approximation~\cite{richardson:2004,richardson:2005,richardson:2006}. 
We give in Appendix \ref{app:VE2} an alternative factorized form for $\Var{V}$ to justify the nonnegativity of  expression \eqref{eq:statvar}.
In Fig.~\ref{fig:SimTheory}, we illustrate the excellent agreement of the analytically derived expressions \eqref{eq:statmean} and \eqref{eq:statvar} with numerical estimates obtained via Monte-Carlo simulations of the AONCB dynamics for various input synchrony conditions.
Discussing and interpreting quantitatively \eref{eq:statmean} and \eref{eq:statvar} within a biophysically relevant context will be the main focus of the remaining of this work.


\section{Comparison with experimental data}\label{sec:result}



\subsection{Experimental measurements and parameter estimations}\label{sec:expData}

 Cortical activity typically exhibits a high degree of variability in response to identical stimuli~\cite{arieli:1996aa,mainen:1995}, with individual neuronal spiking exhibiting Poissonian characteristics~\cite{tolhurst:1983, bialek:1999}.
Such variability is striking because neurons are thought to typically receive a large number ($\simeq 10^4$) of synaptic contacts~\cite{braitenberg2013cortex}.
As a result, in the absence of correlations, neuronal variability should average out, leading to quasi-deterministic neuronal voltage dynamics~\cite{softky:1993}.
To explain how variability seemingly defeats averaging in large neural networks, it has been proposed that neurons operate in a special regime, whereby inhibitory and excitatory drive nearly cancel one another~\cite{sompolinski:1988,vreeswijk:1996,amit:1997,vreeswijk:1998aa,brunel:2000}.
In such balanced networks, the voltage fluctuations become the main determinant of the dynamics, yielding a Poisson-like spiking activity~\cite{sompolinski:1988,vreeswijk:1996,amit:1997,vreeswijk:1998aa,brunel:2000}. 
 However, depending upon the tightness of this balance, networks can exhibit distinct dynamical regimes with varying degree of synchrony \cite{ahmadian2021dynamical}.

In the following, we will exploit the analytical framework of AONCB neurons to argue that the asynchronous picture predicts voltage fluctuations are an order of magnitude smaller than experimental observations~\cite{churchland:2010aa,tan:2013aa,tan:2014aa,okun2015diverse}.
Such observations indicate that the variability of the neuronal membrane voltage exhibits typical variance values of  $\simeq 4-9\mathrm{mV}^2$.
Then we will claim that achieving such variability requires input synchrony within the setting of AONCB neurons.
Experimental estimates of the spiking correlations are typically thought as weak with coefficients ranging from $0.01$ to $0.04$~\cite{renart:2010aa,ecker:2010,cohen2011measuring}.
Such weak values do not warrant the neglect of correlations owing to the typically high number of synaptic connections.
Actually, if $K$ denotes the number of  inputs, all assumed to play exchangeable roles, an empirical criterion to decide whether a correlation coefficient $\rho$ is weak is that $\rho<1/K$~\cite{chen:2006,polk:2012}.
Assuming the lower estimate of $\rho \simeq 0.01$, this criterion is achieved for  $K \simeq 100$ inputs, which is well below the typical number of excitatory synapses for cortical neurons.
In the following, we will only consider the response of AONCB neurons to synchronous drive with biophysically realistic spiking correlations ($0 \leq \rho \leq 0.03$).

Two key parameters for our argument will be the excitatory and inhibitory synaptic weights denoted by $w_\ee$ and $w_\ii$, respectively.
Typical values for these weights can be estimated via biophysical considerations within the framework of AONCB neurons.
In order to develop these considerations, we assume the values $V_\ii=-10\mathrm{mV} < V_\leak=0<V_\ee=60\mathrm{mV}$ for reversal potentials and $\tau=15\mathrm{ms}$ for the passive membrane time constant.
Given these  assumptions, we set the upper range of excitatory synaptic weights so that when delivered to a neuron close to its resting state, unitary excitatory inputs cause peak membrane fluctuations of $\simeq 0.5 \mathrm{mV}$ at the soma, attained after a peak time of  $\simeq 5 \mathrm{ms}$.
Such fluctuations correspond to typically large \emph{in-vivo} synaptic activations of thalamo-cortical projections in rats~\cite{bruno2006cortex}.
Although activations of similar amplitude have been reported for cortico-cortical connections~\cite{jouhanneau2015vivo,pala:2015vivo}, recent large-scale \emph{in vivo} studies have revealed that cortico-cortical excitatory connections are typically much weaker~\cite{seeman:2018sparse,campagnola2022local}.
At the same time, these studies have shown that inhibitory synaptic conductances are about fourfold larger than excitatory ones, but with similar timescales. 
Fitting these values within the framework of AONCB neurons for $\epsilon=\tau_\mathrm{s}/\tau\simeq 1/4$ reveals that the largest possible synaptic inputs correspond to dimensionless weights $w_\ee \simeq 0.01$ and $w_\ii \simeq 0.04$.
Following on~\cite{seeman:2018sparse,campagnola2022local}, we consider that the comparatively moderate cortico-cortical recurrent connections are an order of magnitude weaker than typical thalamo-cortical projections, i.e.,  $w_\ee \simeq 0.001$ and $w_\ii \simeq 0.004$.
Such a range is in keeping with estimates used in~\cite{sanzeni:2022}.

\begin{figure}[tbp]
\begin{center}
\includegraphics[width=8.6cm]{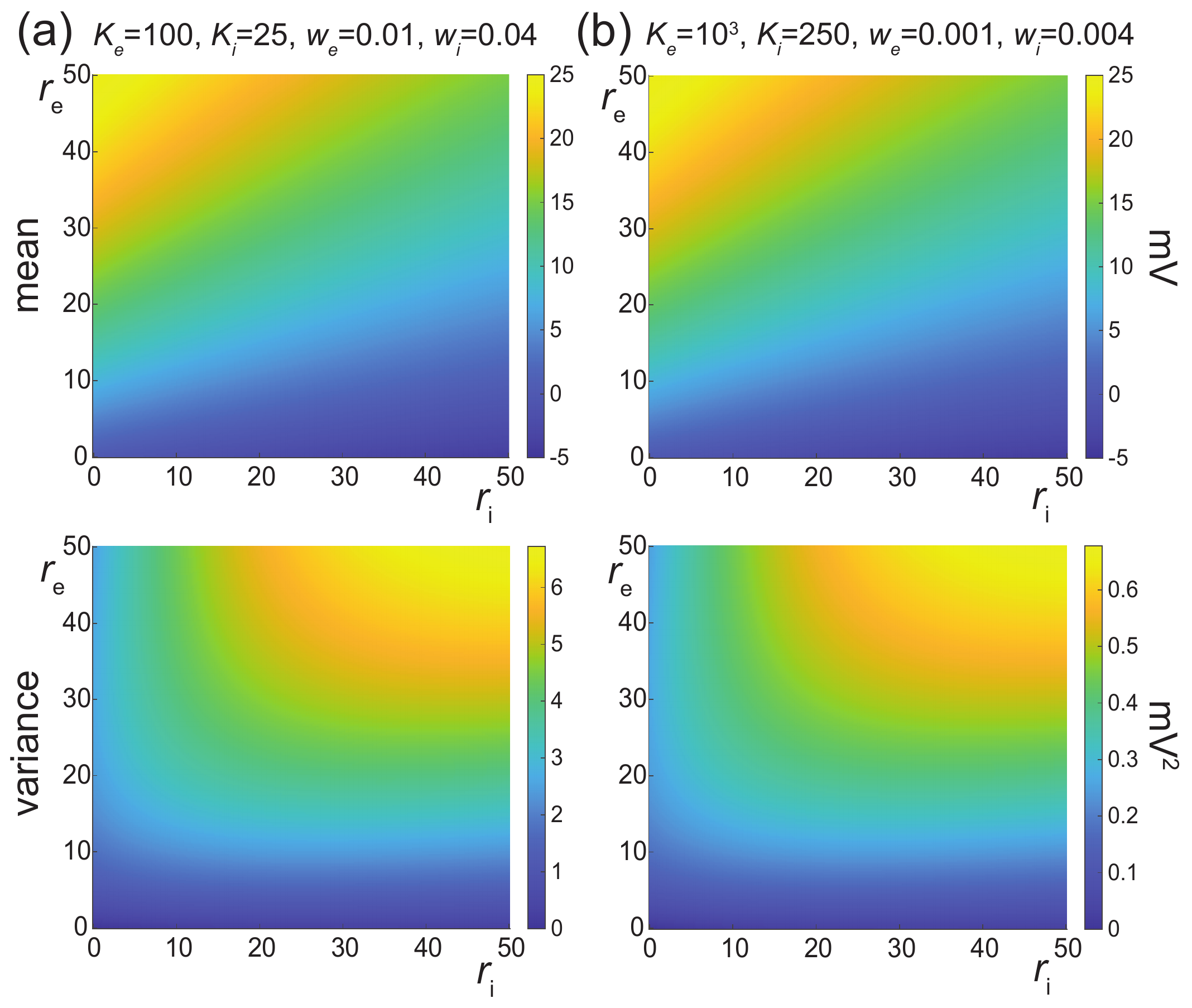}
\caption{\label{fig:Corr0}{\bf Voltage mean and variance in the absence of input correlations.}
Column (a) depicts the stationary subthreshold response of an AONCB neurons driven by $K_\ee=100$ and $K_\ii=25$ synapses with large weights $w_\ee=0.01$ and $w_\ii=0.04$.
Column (b) depicts the stationary subthreshold response of an AONCB neurons driven by $K_\ee=10^3$ and $K_\ii=250$ synapses with moderate weights $w_\ee=0.001$ and $w_\ii=0.004$.
For synaptic weights $w_\ee,w_\ii \ll 1$, the mean  response is identical as $K_\ee w_\ee=K_\ii w_\ii=1$ for (a) and (b).
By contrast, for $\rho_\ee=\rho_\ii=\rho_{\ee\ii}=0$, the variance is at least an order-of-magnitude smaller than that experimentally observed ($4-9\mathrm{mV}^2$) for moderate weights as shown in (a).
Reaching the lower range of realistic neural variability requires driving the cell via large weights as shown in (b).
}
\end{center}
\end{figure}


\subsection{The effective-time constant approximation holds in the asynchronous regime}\label{sec:EffectTime}

Let us consider that neuronal inputs have zero (or negligible) correlation structure, which corresponds to assuming that all synapses are driven by independent Poisson processes.
Incidentally, excitation and inhibition act independently.
Within the framework of AONCB neurons, this latter assumption corresponds to choosing a joint jump distribution of the form 
\begin{eqnarray}\label{eq:indProb}
p_{\ee\ii}(W_\ee,W_\ii) =  \frac{b_\ee}{b} p_\ee(W_\ee) \delta(W_\ii) + \frac{b_\ii}{b} p_\ii(W_\ii) \delta(W_\ee) \, . \nonumber
\end{eqnarray}
where $\delta(\cdot)$ denotes the Dirac delta function so that $W_\ee W_\ii=0$ with probability one.
Moreover, $b_\ee$ and $b_\ii$ are independently specified via \eref{eq:bDef} and the overall rate of synaptic events is purely additive: $b=b_\ee+b_\ii$.
Consequently, the crosscorrelation efficacy $c_{\ee\ii}$ in \eref{eq:statvar} vanishes and the dimensionless  efficacies simplify to 
\begin{eqnarray}
a_{\ee,1} = b_\ee\tau \ExpPe{1-e^{-W_\ee}} \quad  \mathrm{and} \quad a_{\ii,1} = b_\ii\tau \ExpPi{1-e^{-W_\ii}} \, . \nonumber
\end{eqnarray}
Further assuming that individual excitatory and inhibitory synapses act independently leads to considering that $p_\ee$ and $p_\ii$ depict the size of individual synaptic inputs, as opposed to aggregate events.
This corresponds to taking $\beta_\ee \to \infty$ and $\beta_\ii \to \infty$ in our parametric model based on beta distributions.
Then, as intuition suggests, the overall rates of excitation and inhibition activation are recovered as $b_\ee=K_\ee r_\ee$ and $b_\ii=K_\ii r_\ii$, where $r_{\ee}$ and $r_{\ii}$ are the individual spiking rates.

Individual synaptic weights are small in the sense that $w_{\ee},w_{\ii} \ll 1$, which warrants neglecting exponential corrections for the evaluation of the synaptic efficacies, at least in the absence of synchrony-based correlations.
Accordingly, we have
\begin{eqnarray}
a_{\ee,1} \simeq K_\ee r_\ee\tau w_\ee \quad \mathrm{and} \quad a_{\ee,12} \simeq  K_\ee r_\ee\tau w_\ee^2/2  \, , \nonumber
\end{eqnarray}
as well as symmetric expressions for inhibitory efficacies.
Plugging these values in \eref{eq:statvar} yields the classical mean-field estimate for the stationary variance
\begin{eqnarray}\label{eq:indVar}
\Var{V} \simeq
 \frac{K_\ee r_\ee w_{\ee}^2 (V_\ee - \Exp{V})^2 + K_\ii r_\ii w_{\ii}^2 (V_\ii - \Exp{V})^2
}{ 2(1/\tau + K_\ee r_\ee w_{\ee}+ K_\ii r_\ii w_\ii)} \, , \nonumber
 \end{eqnarray}
which is exactly the same expression as that derived via the diffusion and effective-time-constant approximations in \cite{destexhe:2001,meffin:2004}.
However, observe that the only approximation we made in obtaining the above expression is to neglect exponential corrections due to the relative weakness of biophysically relevant synaptic weights, which we hereafter refer to as the small-weight approximation.


\subsection{Asynchronous inputs yield exceedingly small neural variability}\label{sec:async}

In  \fref{fig:Corr0}, we represent the stationary mean $\Exp{V}$ and variance $\Var{V}$ as a function of the neuronal spiking input rates $r_e$ and $r_i$, but for distinct values of synaptic weights $w_\ee$ and $w_\ii$.
In \fref{fig:Corr0}a, we consider  synaptic weights as large as biophysically admissible based on recent {\it in-vivo} studies~\cite{seeman:2018sparse,campagnola2022local}, i.e., $w_\ee=0.01$ and $w_\ii=0.04$.
By contrast, in \fref{fig:Corr0}b, we consider moderate synaptic weights $w_\ee=0.001$ and $w_\ii=0.004$, which yield somatic post-synaptic deflections of typical amplitudes.
In both cases, we consider input numbers $K_\ee$ and $K_\ii$ such that the mean voltage $\Exp{V}$ covers the same biophysical range of values as $r_\ee$ and $r_\ii$ varies between $0\mathrm{Hz}$ and $50\mathrm{Hz}$.
Given a zero resting potential, we set this biophysical range to be bounded by $\Delta \Exp{V} \leq 20\mathrm{mV}$ as typically observed experimentally in electrophysiological recordings.
These conditions correspond to constant aggregate weights set to  $K_\ee w_\ee=K_\ii w_\ii=1$ so that 
\begin{eqnarray*}
K_\ee r_\ee w_\ee = K_\ii r_\ii w_\ii \leq  50 \mathrm{Hz} \simeq 1/\tau \, .
\end{eqnarray*}
This implies that the AONCB neurons under consideration do not reach the high-conductance regime for which the passive conductance can be neglected, i.e.,  $K_\ee r_\ee w_\ee + K_\ee r_\ee w_\ii \gg  1/\tau$~\cite{destexhe:2003}.
Away from the high-conductance regime, the variance magnitude is controlled by the denominator of \eref{eq:indVar}.
Accordingly, the variance in both cases is primarily dependent on the excitatory rate $r_\ee$ since for $K_\ee w_\ee=K_\ii w_\ii=1$, the effective excitatory driving force $F_\ee=K_\ee w_\ee^2 (V_\ee-\Exp{V})^2$ dominates the effective inhibitory driving force $F_\ii=K_\ii w_\ii^2 (V_\ii-\Exp{V})^2$.
This is because the neuronal voltage typically sits close to the inhibitory reversal potential but far from the excitatory reversal potential $V_\ee -\Exp{V} > \Exp{V}-V_\ii$.
For instance, when close to rest $\Exp{V}\simeq 0$, the ratio of the effective driving forces is $(K_\ee w_\ee^2 V_\ee^2)/(K_\ii w_\ii^2 V_\ii^2)\simeq 9$ fold in favor of excitation.
Importantly, the magnitude of the variance is distinct for moderate synapses and for large synapses.
This is because for constant aggregate weights $K_\ee w_\ee=K_\ii w_\ii=1$, the ratio of effective driving forces for large and moderate synapses scales in keeping with the ratio 
of the weights, and so does the ratio of variances away from the high conductance regime.
Thus we have 
\begin{eqnarray*}
F_\ee\vert_{w_\ee=10^{-2}}/F_\ee\vert_{w_\ee=10^{-3}}=F_\ii\vert_{w_\ii=10^{-2}}/F_\ii\vert_{w_\ii=10^{-3}}=10 \, ,
\end{eqnarray*}
and the variance decreases by one order of magnitude from large  weights in \fref{fig:Corr0}a to moderate weights in \fref{fig:Corr0}b.

\begin{figure}[tbp]
\begin{center}
\includegraphics[width=8.6cm]{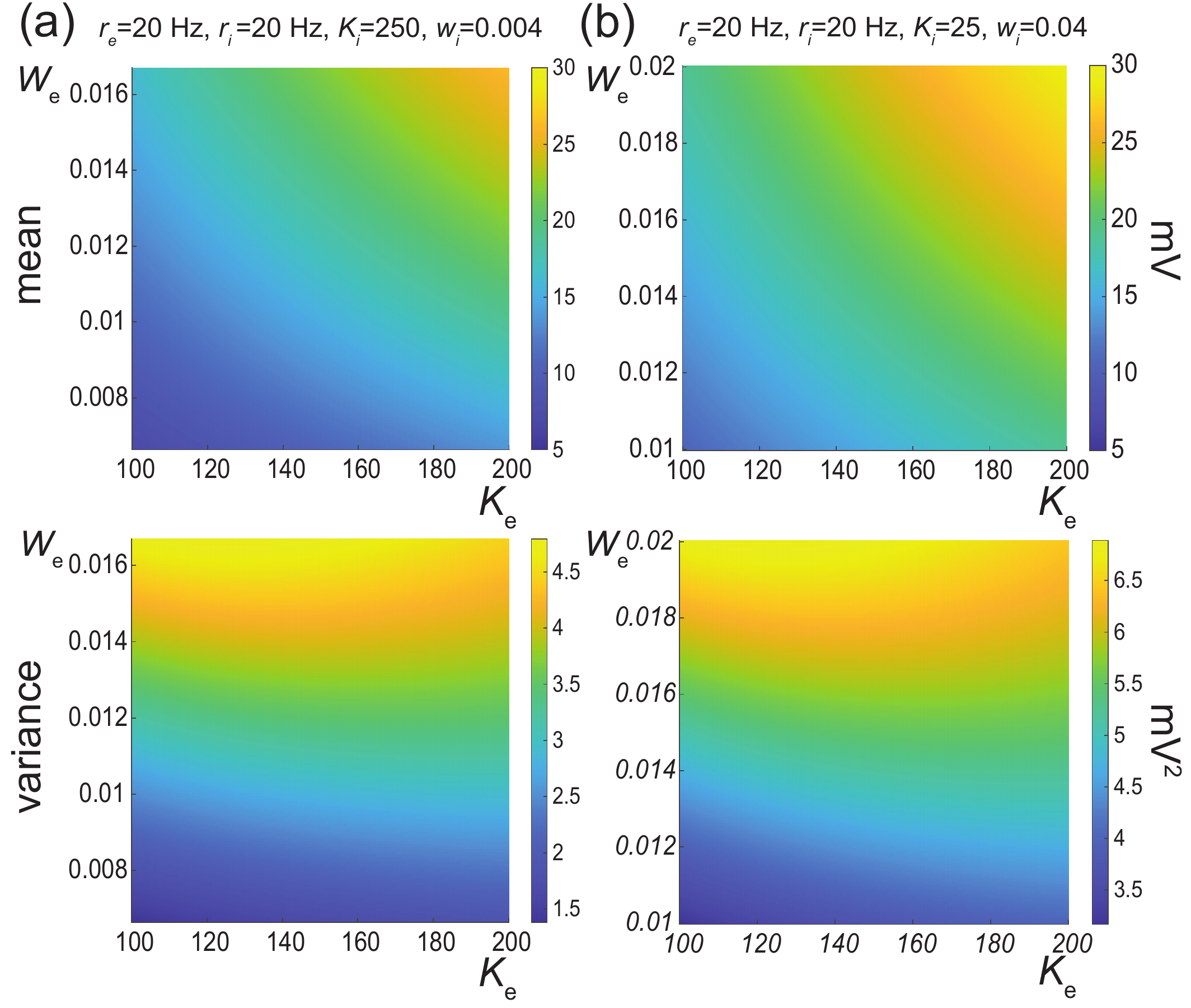}
\caption{\label{fig:Kw}{\bf Dependence on the number of inputs and the synaptic weights in the absence of correlations.}
Column (a) depicts the stationary subthreshold response of an AONCB neurons driven by a varying number of excitatory synapses $K_\ee$ with varying weight $w_\ee$ at rate $r_\ee=20\mathrm{Hz}$, with background inhibitory drive given by $K_\ii=250$ with moderate weights $w_\ii=0.004$ and $r_\ii=20\mathrm{Hz}$. 
Column (b) depicts the same as in column (a) but for a background inhibitory drive given by $K_\ii=25$ with large weights $w_\ii=0.04$ and $r_\ii=20\mathrm{Hz}$.
For both conditions, achieving realistic level of variance, i.e., $\Var{V}\simeq4-9\mathrm{mV}^2$, while ensuring a biophysically relevant mean range of variation, i.e., $\Delta \Exp{V}\simeq10\text{-}20\mathrm{mV}$, is only possible for large weights: $w_\ee \geq 0.015$ for moderate inhibitory weights in \fref{fig:Kw}a and  $w_\ee \geq 0.01$ for large weights.
}
\end{center}
\end{figure}

The above numerical analysis reveals that achieving realistic levels of subthreshold variability for a biophysical mean range of variation requires AONCB neurons to be exclusively driven by large synaptic weights.
This is confirmed by considering the voltage mean $\Exp{V}$ and variance $\Var{V}$ in \fref{fig:Kw} as a function of the number of inputs $K_\ee$ and of the synaptic weights $w_\ee$ for a given level of inhibition.
We choose this level of inhibition to be set by $K_\ii=250$ moderate synapses $w_\ii=0.004$ with $r_\ii=20\mathrm{Hz}$ in \fref{fig:Kw}a and by $K_\ii=25$ large synapses $w_\ii=0.04$ with $r_\ii=20\mathrm{Hz}$ in \fref{fig:Kw}b.
As expected, assuming that $r_\ee=20\mathrm{Hz}$ in the absence of input correlations, the voltage mean $\Exp{V}$ only depends on the product $K_\ee w_\ee$, which yields similar mean range of variations for $K_\ee$ varying up to $2000$ in \fref{fig:Kw}a and up to $200$ in \fref{fig:Kw}b.
Thus, it is possible to achieve the same range of variations as with moderate synaptic with a fewer number of larger synaptic weights. 
By contrast, the voltage variance $\Var{V}$ only achieves realistic levels for large synaptic weights in both conditions, with $w_\ee \geq 0.015$ for moderate inhibitory background synapses in \fref{fig:Kw}a and  $w_\ee \geq 0.01$ for large inhibitory background synapses in \fref{fig:Kw}b.


\subsection{Including input correlations yields realistic subthreshold variability}\label{sec:syncForm1}

Without synchrony, achieving the experimentally observed variability necessitates an excitatory drive mediated via synaptic weights $w_\ee \simeq 0.01$, which corresponds to the upper bounds of the biophysically admissible range and is in agreement with numerical results presented in~\cite{sanzeni:2022}.
Albeit possible, this is unrealistic given the wide distribution of amplitudes observed experimentally, whereby the vast majority of synaptic events are small to moderate, at least for cortico-cortical connections~\cite{seeman:2018sparse,campagnola2022local}.
In principle, one can remedy this issue by allowing for synchronous activation of, say, $k_\ee=10$ synapses with moderate weight $w_\ee=0.001$, as it amounts to the activation of a single synapse with large weight $k_\ee w_\ee=0.01$.
A weaker assumption that yields a similar increase in neural variability is to only ask for synapses to tend to synchronize probabilistically, which amounts to require $k_\ee$ to be a random variable with some distribution mass on $\{k_\ee>1 \}$.
This exactly amounts to model the input drive via a jump process as presented in Section \ref{sec:model}, with a jump distribution  $p_\ee$ that probabilistically captures this degree of input synchrony.
In turn, this distribution $p_\ee$ corresponds to a precise input correlation $\rho_\ee$ via \eref{eq:CorrRhoe}.

\begin{figure}[tbp]
\begin{center}
\includegraphics[width=8.6cm]{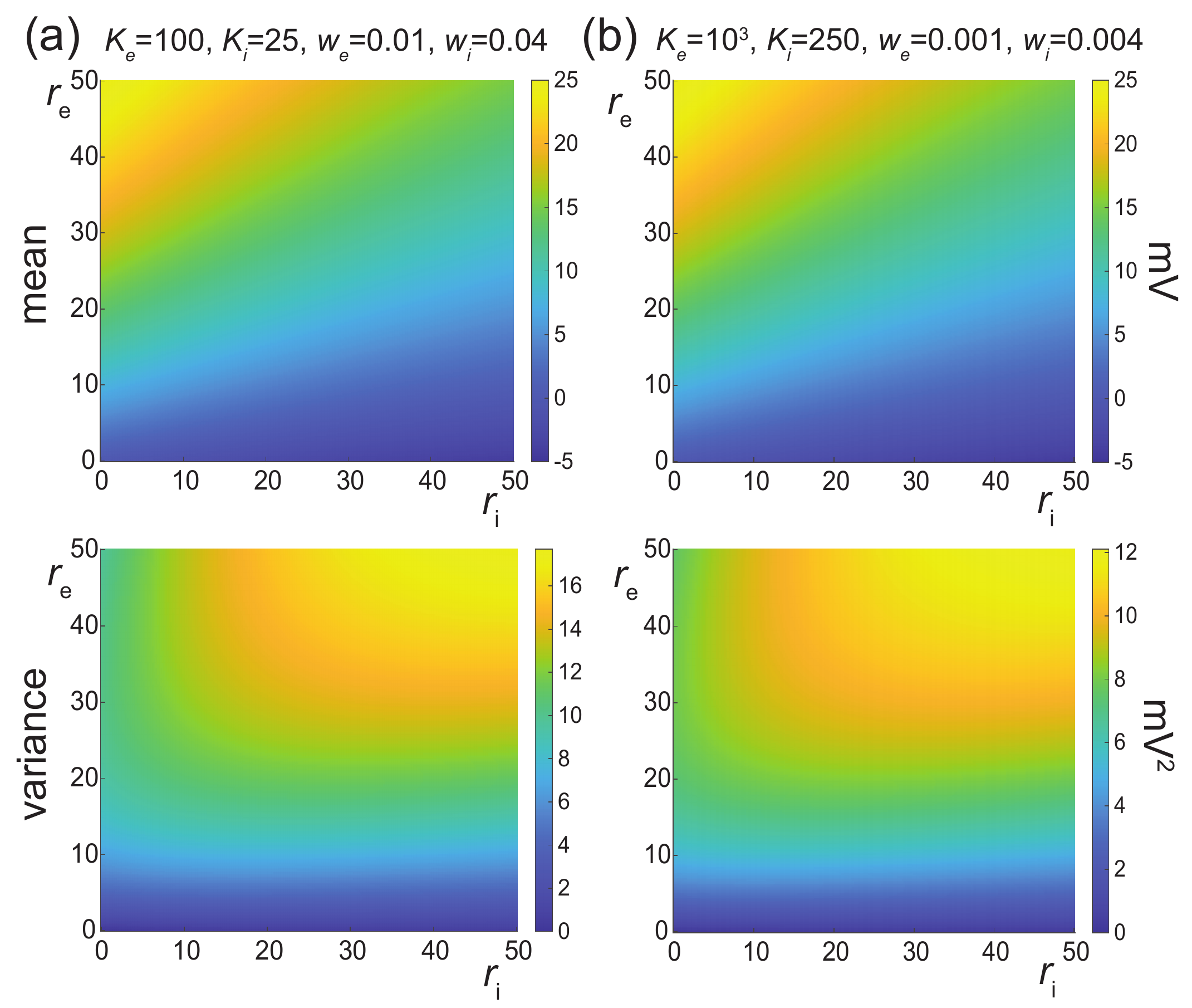}
\caption{\label{fig:Corr1}{\bf Voltage mean and variance in the presence of  excitatory and inhibitory input correlations but without correlation across excitation and inhibition: $\rho_\ee=\rho_\ii>\rho_{\ee\ii}=0$.}
Column (a) depicts the stationary subthreshold response of an AONCB neurons driven by $K_\ee=100$ and $K_\ii=25$ synapses with large weights $w_\ee=0.01$ and $w_\ii=0.04$.
Column (b) depicts the stationary subthreshold response of an AONCB neurons driven by $K_\ee=10^3$ and $K_\ii=250$ synapses with moderate dimensionless weights $w_\ee=0.001$ and $w_\ii=0.004$.
For synaptic weights $w_\ee,w_\ii \ll 1$, the mean response is identical as $K_\ee w_\ee=K_\ii w_\ii=1$ for (a) and (b).
By contrast with the case of no correlation in \fref{fig:Corr0}, for $\rho_\ee=\rho_\ii=0.03$ and $\rho_{\ee\ii}=0$, the variance achieves similar levels as experimentally observed ($4-9\mathrm{mV}^2$) for  moderate weights as shown in (b), but slightly larger levels for large weights as shown in (a).}
\end{center}
\end{figure}

We quantify the impact of nonzero correlation in \fref{fig:Corr1} where we consider the cases of moderate weights $w_\ee=0.001$ and $w_\ee=0.004$ and large weights $w_\ee=0.01$ and $w_\ii=0.04$ as in \fref{fig:Corr0} but for $\rho_\ee=\rho_\ii=0.03$.
Specifically, we consider an AONCB neuron subjected to two independent beta-binomial-derived compound Poisson process drives with rate $b_\ee$ and $b_\ii$, respectively.
These rates $b_\ee$ and $b_\ii$ are obtained via \eref{eq:bDef} by setting $\beta_\ee=\beta_\ii=1/\rho_\ee-1=1/\rho_\ii-1$ and for given input numbers $K_\ee$ and $K_\ii$ and spiking rates $r_\ee$ and $r_\ii$.
This ensures that the mean number of synaptic activations $b_\ee \ExpPei{k_\ee}=K_\ee r_\ee$ and $b_\ii \Exp{k_\ii}=K_\ii r_\ii$ remains constant when compared with \fref{fig:Corr0}.
As a result, the mean response of the AONCB neuron is  essentially left unchanged by the presence of correlations, with virtually identical biophysical range of variations $\Delta\ExpPei{V} \simeq 10\text{-}20\mathrm{mV}$.
This is because for correlation $\rho_\ee=\rho_\ii \simeq 0.03$, the aggregate weights still satisfy $k_\ee w_\ee, k_\ii w_\ii < 1$ with probability close to one given that $K_\ee w_\ee = K_\ii w_\ii=1$.
Then, in the absence of crosscorrelation, i.e., $\rho_{\ee\ii}=0$, we still have
\begin{eqnarray*}
a_{\ee,1}= b_\ee \tau \ExpPe{1-e^{-k_\ee w_\ee}} \simeq  b_\ee \tau w_\ee  \ExpPe{k_\ee}  = K_\ee r_\ee \tau w_\ee \, , \nonumber
\end{eqnarray*}
as well as $a_{\ii,1} \simeq K_\ii r_\ii \tau w_\ii$ by symmetry.
However, for both moderate and large synaptic weights, the voltage variance $\Var{V}$ now exhibits slightly larger magnitudes than observed experimentally.
This is because we show in Appendix \ref{app:smallApprox1} that in the small-weight approximation
\begin{eqnarray*}
a_{\ee,12}
&=&
\frac{b_\ee \tau}{2} \ExpPe{\left(1-e^{-k_\ee w_\ee}\right)^2} \, , \\
&\simeq&
\left(1+\rho_\ee(K_\ee-1) \right) \frac{K_\ee r_\ee \tau  w_\ee^2}{2}\, ,
\end{eqnarray*}
where we recognize $K_\ee r_\ee \tau  w_\ee^2 /2=a_{\ee,12} \vert_{\rho_\ee=0}$ as the second-order efficacy in the absence of correlations from \fref{fig:Corr0}.
A similar statement holds for  $a_{\ii,12}$.
This shows that correlations increase neural variability whenever $\rho_\ee > 1/K_\ee$ or $\rho_\ii > 1/K_\ii$, which coincides with our previously given criterion  to assess the relative weakness of correlations.
Accordingly, when excitation and inhibition act independently, i.e., $\rho_{\ee\ii}=0$, we find that the increase in variability due to input synchrony $\Delta_{\rho_{\eeii}} =\Var{V} \vert_{\rho_{\ee\ii}=0}-\Var{V} \vert_{\rho_\eeii=\rho_{\ee\ii}=0}$ satisfies
\begin{eqnarray}\label{eq:statvar2}
\Delta_{\rho_{\eeii}}
&\simeq&
\frac{ \rho_\ee(K_\ee-1) K_\ee r_\ee w_\ee^2 (V_\ee - \Exp{V})^2 }{ 2(1/ \tau+ K_\ee r_\ee w_\ee + K_\ii r_\ii w_\ii) +}  \\
&& \hspace{20pt} \frac{  \rho_\ii(K_\ii-1) K_\ii r_\ii w_\ii^2 (V_\ii - \Exp{V})^2}{ 2(1/ \tau+ K_\ee r_\ee w_\ee + K_\ii r_\ii w_\ii) }  \nonumber  \, ,
\end{eqnarray}
The above relation follows from the fact that the small-weight approximation for $\Exp{V}$ is independent of correlations and from neglecting the exponential corrections due to the nonzero size of the synaptic weights.
The above formula remains valid as long as the correlations $\rho_\ee$ and $\rho_\ii$ are weak enough so that the aggregate weights satisfy $k_\ee w_\ee, k_\ii w_\ii<1$ with probability close to one.
To inspect the relevance of exponential corrections, we estimate in Appendix \ref{app:smallApprox2} the error incurred by neglecting exponential corrections.
Focusing on the case of excitatory inputs, we find that for correlation coefficients $\rho_\ee \leq 0.05$, neglecting exponential corrections incurs less than a $3\%$ error if the number of inputs is smaller than $K_\ee \leq 1000$ for moderate synaptic weight $w_\ee=0.001$ or than $K_\ee \leq 100$ for large synaptic weight $w_\ee=0.01$.


\subsection{Including correlations between excitation and inhibition reduces subthreshold variability}\label{sec:sync}

\begin{figure}[tbp]
\begin{center}
\includegraphics[width=8.6cm]{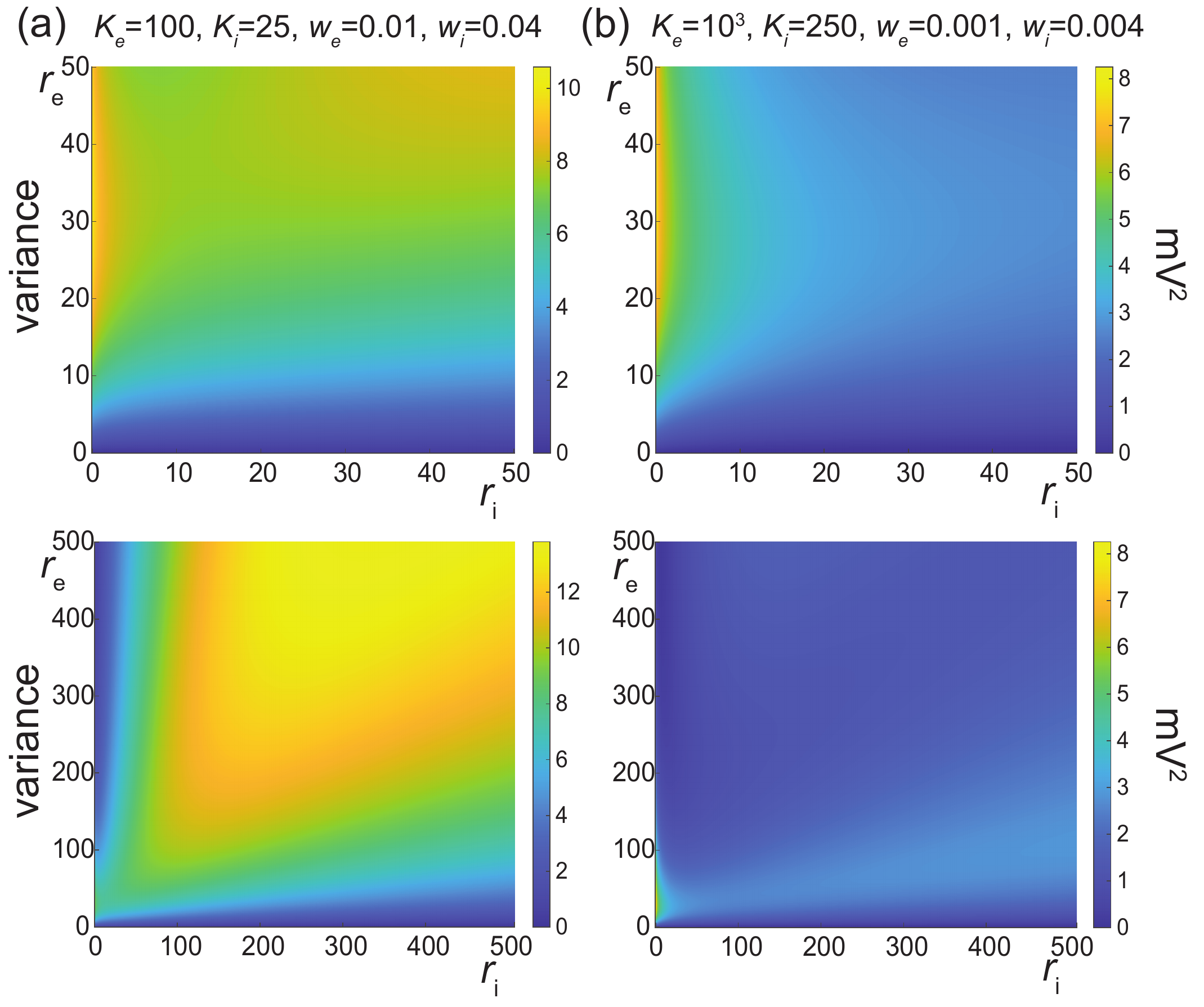}
\caption{\label{fig:Corr2}{\bf Voltage mean and variance in the presence of  excitatory and inhibitory input correlations and with correlation across excitation and inhibition:  $\rho_\ee=\rho_\ii=\rho_{\ee\ii}>0$.}
Column (a) depicts the stationary subthreshold response of an AONCB neurons driven by $K_\ee=100$ and $K_\ii=25$ synapses with large weights $w_\ee=0.01$ and $w_\ii=0.04$.
Column (b) depicts the stationary subthreshold response of an AONCB neurons driven by $K_\ee=10^3$ and $K_\ii=250$ synapses with moderate dimensionless weights $w_\ee=0.001$ and $w_\ii=0.004$.
For synaptic weights $w_\ee,w_\ii \ll 1$, the mean response is identical as $K_\ee w_\ee=K_\ii w_\ii=1$ for (a) and (b).
Compared with the case of no crosscorrelation in \fref{fig:Corr1}, for $\rho_\ee=\rho_\ii=\rho_{\ee\ii}=0.03$, the variance is reduced to a biophysical range similar to that experimentally observed ($4-9\mathrm{mV}^2$) for moderate weights as shown in (a), as well as for large weights as shown in (b).
}
\end{center}
\end{figure}

The voltage variance estimated for realistic excitatory and inhibitory correlations, e.g., $\rho_\ee=\rho_\ii=0.03$ and $\rho_{\ee\ii}=0$, exceeds the typical levels measured {\it in vivo}, i.e., $4-9\mathrm{mV}^2$, for large synaptic weights.
The inclusion of correlations between excitation an inhibition, i.e., $\rho_{\ee\ii}>0$ can reduce the voltage variance to more realistic levels.
We confirm this point in \fref{fig:Corr2} where we consider the cases of moderate weights $w_\ee=0.001$ and $w_\ee=0.004$ and large weights $w_\ee=0.01$ and $w_\ii=0.04$ as in \fref{fig:Corr1} but for $\rho_\ee=\rho_\ii=\rho_{\ee\ii} = 0.03$.
Positive crosscorrelation between excitation and inhibition only marginally impacts the mean voltage response. 
This is due to the fact that exponential corrections become slightly more relevant as the presence of crosscorrelation leads to larger aggregate weights: $W_\ee+W_\ii$ with $W_\ee$ and $W_\ii$ possibly being jointly positive.
By contrast with this marginal impact on the mean response, the voltage variance is significantly reduced when excitation and inhibition are correlated.
This is in keeping with the intuition that the net effect of such crosscorrelation is to cancel excitatory and inhibitory synaptic inputs with one another, before they can cause voltage fluctuations. 
The amount by which the voltage variance is reduced can be quantified in the small-weight approximation.
In this approximation, we show in Appendix \ref{app:smallApprox1} that the efficacy $c_{\ee\ii}$ capturing the impact of crosscorrelations simplifies to
\begin{eqnarray*}
c_{\ee\ii}
&\simeq&
\frac{b \tau}{2} \ExpPei{W_\ee W_\ii} 
=
(\rho_{\ee\ii}  \sqrt{r_\ee r_\ii} \tau/2) (K_\ee w_\ee) (K_\ii w_\ii)  \, .
\end{eqnarray*}
Using the above simplified expression and invoking the fact that the small-weight approximation for $\Exp{V}$ is independent of correlations show a decrease in the amount $\Delta_{\rho_{\ee\ii}} = \Var{V}-\Var{V} \vert_{\rho_{\ee\ii}=0}$ with
\begin{eqnarray}\label{eq:statvar3}
\lefteqn{
\Delta_{\rho_{\ee\ii}} \simeq 
}\\
&&
-\frac{
\rho_{\ee\ii}\sqrt{r_\ee r_\ii}  ( K_\ee w_\ee) (K_\ii w_\ii)  (V_\ee - \Exp{V})(\Exp{V} -V_\ii)
}{ 
1/\tau+ K_\ee r_\ee w_\ee + K_\ii r_\ii w_\ii
}
 \leq 0    \nonumber \, .
\end{eqnarray}
Despite the above reduction in variance, we also show in Appendix \ref{app:smallApprox1} that positive input correlations always cause an overall increase of neural variability:
$$ 0 \leq \Var{V} \vert_{\rho_\eeii=\rho_{\ee\ii}=0} \leq \Var{V}   \leq \Var{V} \vert_{\rho_{\ee\ii}=0} \, . $$
Note that the reduction of variability due to $\rho_{\ee\ii}>0$ crucially depends on the instantaneous nature of correlations between excitation and inhibition.
To see this, observe that Marcus rule \eref{eq:MarcusJump} specifies instantaneous jumps via a weighted average of the reversal potentials $V_\ee$ and $V_\ii$, which represent extreme values for voltage updates.
Thus, perfectly synchronous excitation and inhibition updates the voltage toward an intermediary value rather than extreme ones, leading to smaller jumps on average, 
Such an effect can vanish or even reverse when synchrony breaks down, e.g., when inhibition substantially lags behind excitation.

\subsection{Asynchronous scaling limits require fixed-size synaptic weights}\label{sec:asyncLim}

Our  analysis reveals that the correlations must significantly impact the voltage variability whenever the number of inputs are such that $K_\ee>1/\rho_\ee$ or $K_\ii>1/\rho_\ii$.
Spiking correlations are typically measured {\it in vivo} to be larger than $0.01$.
Therefore, synchrony must shape the response of neurons that are driven by more than $100$ active inputs, which is presumably allowed by the typically high number of synaptic contacts ($\simeq10^4$) in cortex~\cite{braitenberg2013cortex}.
In practice, we find that synchrony can explain the relatively high level of neural variability observed in the subthreshold neuronal responses.
Beyond these practical findings, we predict that input synchrony also have significant theoretical implications with respect to modeling spiking networks.
Analytically tractable models for cortical activity are generally obtained by considering spiking networks in the infinite-size limit.
Such infinite-size networks are tractable because the neurons they comprise only interact via population averages, erasing any role for nonzero correlation structure.
Distinct mean-field models assume that synaptic weights vanish according to distinct scalings with respect to the number of synapses, i.e., $w_{\eeii} \to 0$ as $K_\eeii \to \infty$.
In particular, classical mean-field limits consider the scaling $w_{\eeii}\sim1/K_\eeii$, 
balanced mean-field limits consider the scaling $w_{\eeii}\sim1/\sqrt{K_\eeii}$, with $K_\ee w_\ee - K_\ii w_\ii=O(1)$,
and strong coupling limits consider the scaling $w_{\eeii}\sim1/\ln{K_\eeii}$, with $K_\ee w_\ee - K_\ii w_\ii=O(1)$ as well.

Our analysis of AONCB neurons shows that the neglect of synchrony-based correlations is incompatible with the maintenance of neural variability in the infinite-size limit.
Indeed, \eref{eq:indVar} shows that for any scaling with $1/w_{\ee} =o(K_\ee)$ and $1/w_{\ii} =o(K_\ii)$, as for all the mean-field limits mentioned above, we  have
\begin{eqnarray}
\Var{V} 
= O(w_{\ee}) + O(w_{\ii}) \xrightarrow{K_e, K_i \to \infty} 0 \, . \nonumber
 \end{eqnarray}
Thus, in the absence of correlation and independent of the synaptic weight scaling, the subthreshold voltage variance of AONCB neurons must vanish in the limit of arbitrary large numbers of synapses.
We expect such decay of the voltage variability to be characteristic of conductance-based models in the absence of input correlation.
Indeed, dimensional analysis suggests that voltage variance for both current-based and conductance-based models are generically obtained via normalization by the reciprocal of the membrane time constant.
However, by contrast with current-based models, the reciprocal of the membrane time constant for conductance-based models,  i.e., $1/\tau+K_\ee w_\ee r_\ee+K_\ii w_\ii r_\ii$,  involves contributions from synaptic conductances.
Thus, to ensure nonzero asymptotic variability, the denominator scaling $O(K_\ee w_\ee)+O(K_\ii w_\ii)$ must be balanced by the natural scaling of the Poissonian input drives, i.e., $O(K_\ee w_\ee^2)+O(K_\ii w_\ii^2)$.
In the absence of input correlations, this is only possible for fixed-size weights, which is incompatible with any scaling assumptions.

\subsection{Synchrony allows for variability-preserving scaling limits with vanishing weights}\label{sec:syncLim}

Infinite-size networks with fixed-size synaptic weights are problematic for restricting modeled neurons to operate in the high-conductance regime, whereby the intrinsic conductance properties of the cell play no role.
Such a regime is biophysically unrealistic as it implies that the cell would respond to perturbations infinitely fast.
We propose to address this issue by considering a new type of variability-preserving limit models obtained for the classical scaling but in the presence of synchrony-based correlations. 
For simplicity, let us consider our correlated input model with excitation alone in the limit of an arbitrary large number of inputs $K_\ee \to \infty$.
When $\rho_\ee>0$, the small-weight approximation \eref{eq:statvar2} suggests that adopting the scaling $w_\ee \sim \Omega_\ee/K_\ee$, where $\Omega_\ee$ denotes the aggregate synaptic weight, yields a nonzero contribution when $K_\ee \to \infty$ as the numerator scales as $O(K_\ee^2w_\ee^2)$.
It turns out that this choice can be shown to be valid without resorting to any approximations.
Indeed, under the classical scaling assumption, we show in Appendix \ref{app:InfSize} that the discrete jump distribution $p_{\ee,k}$ weakly converges to the continuous density $\dd \nu_\ee/\dd w$ in the sense that
\begin{eqnarray}\label{eq:Levy}
\lefteqn{
b_\ee \sum_{k=1}^{K_\ee} p_{\ee,k} \delta \left(\frac{w}{\Omega_\ee}-\frac{k}{K_\ee} \right) \, \dd w
\xrightarrow{K_\ee \to \infty}  
} \\
&& \hspace{60pt} \nu_\ee(\dd w) =  \frac{r_\ee \beta_\ee}{w}  \left(1- \frac{W_\ee}{w}\right)^{\beta_\ee-1} \, \dd w  \, .
\nonumber
\end{eqnarray}
The above density has infinite mass over $[0,\Omega_\ee]$ owing to its diverging behavior in zero and is referred to as a degenerate beta distribution.
In spite of its degenerate nature, it is known that densities of the above form define well-posed processes, the so-called beta processes, which have been studied extensively in the field of nonparametric Bayesian inference~\cite{thibaux:2007,broderick2012beta}.
These beta processes represent generalizations of our compound Poisson process drives in so far as they allow for a countable infinity of jumps to occur within a finite time window.
This is a natural requirement to impose when considering an infinite pool of synchronous synaptic inputs, the overwhelming majority of which having nearly zero amplitude.

The above arguments show that one can defined a generalized class of synchronous input models that can serve as the drive of AONCB neurons as well.
Such generalizations are obtained as limits of compound Poisson processes and are specified via their L\'evy-Khintchine measures, which formalize the role of $\nu_\ee$ ~\cite{khintchine:1934,levy:1954}.
Our results naturally extend to this generalized class.
Concretely, for excitation alone, our results extend by replacing all expectations of the form $b_\ee \ExpPe{\cdot}$ by integral with respect to the  measure $\nu_\ee$.
One can easily check that these expectations, which feature prominently in the definition of the various synaptic efficacies, all remain finite for L\'evy-Khintchine measures.
In particular, the voltage mean and variance of AONCB neurons remain finite with
\begin{eqnarray}
&\displaystyle \Exp{V} = \frac{V_\ee \int_0^{\Omega_\ee} (1-e^{-w}) \nu_\ee(\dd w)}{1/\tau +\int_0^{\Omega_\ee} (1-e^{-w}) \nu_\ee(\dd w)} \, , &\nonumber\\
&\displaystyle \Var{V} = \frac{(V_\ee -\Exp{V})^2 \int_0^{\Omega_\ee} (1-e^{-w})^2 \nu_\ee(\dd w)}{2/\tau +\int_0^{\Omega_\ee} (1-e^{-2w}) \nu_\ee(\dd w)}  \, .& \nonumber
\end{eqnarray}
Thus, considering the classical scaling limit $w_\ee \propto 1/K_\ee$ preserves nonzero subthreshold variability in the infinite size limit $K_\ee \to \infty$ as long as $\nu_\ee$ puts mass away from zero, i.e., for $\beta_\ee<\infty \Leftrightarrow \rho_\ee>0$.
Furthermore, we show in Appendix \ref{app:InfSize} that $\Var{V}=O(\rho_\ee)$ so that voltage variability consistently vanishes in the absence of spiking correlation, for which $\nu_\epsilon$ concentrates in zero, i.e.,  when $\beta_\ee \to \infty \Leftrightarrow \rho_\ee = 0$. 


\section{Discussion}\label{sec:discussion}


\subsection{Synchrony modeling}

We have presented a parametric representation of the neuronal drives resulting from a finite number of asynchronous or (weakly) synchronous synaptic inputs.
Several parametric statistical models have been proposed for generating correlated spiking activities in a discrete setting~\cite{qaqish:2003,niebur:2007,macke:2009,macke:2011}.
Such models have been used to analyze the activity of neural populations via Bayesian inference methods~\cite{pillow:2008,park:2013,theis:2013}, as well as maximum entropy methods \cite{schneidman:2006weak,granot:2013}.
Our approach is not to simulate or analyze complex neural dependencies but rather to derive from first principles the synchronous input models that could drive conductance-based neuronal models.
This approach primarily relies on extending the definition of discrete-time correlated spiking models akin to \cite{macke:2009} to the continuous-time setting.
To do so, the main tenet of our approach is to realize that input synchrony and spiking correlation represent equivalent measures under the assumption of input exchangeabilty.

Input exchangeabilty posits that the driving inputs form a subset of an arbitrarily large pool of exchangeable random variables~\cite{kingman:1978, aldous1985exchangeability}. 
In particular, this implies that the main determinant of the neuronal drive is the number of active inputs, as opposed to the magnitude of these synaptic inputs.
Then, de Finetti theorem~\cite{definetti:1929} states that the probability of observing a given input configuration can be represented in the discrete setting under an integral form (see \eref{eq:deFinetti}) involving a directing probability measure $F$.
Intuitively, $F$ represents the probability distribution of the fraction of coactivating inputs at any discrete time. 
Our approach identifies the directing measure $F$ as a free parameter that captures input synchrony. 
The more dispersed the distribution $F$, the more synchronous the inputs, as previously noted in~\cite{bohte:2000,amari:2003}.
Our work elaborates on this observation to develop computationally tractable statistical models for synchronous spiking in the continuous-time limit, i.e., for vanishing discrete time step $\Delta t \to 0^+$.

We  derive our results using a discrete-time directing measure chosen as beta distribution $F \sim B(\alpha,\beta)$, where the parameters $\alpha$ and $\beta$ can be related to the individual spiking rate $r$ and the spiking correlation $\rho$ via $r \Delta t = \alpha/(\alpha+\beta)$ and $\rho=1/(1+\alpha+\beta)$.
For this specific choice of distribution, we are able to construct statistical models of the correlated spiking activity as generalized beta-binomial processes~\cite{hjort:1990}, which play an important role in statistical Bayesian inference~\cite{thibaux:2007,broderick2012beta}. 
This construction allows us to fully parametrize the synchronous activity of a finite number of inputs via the jump distribution of a compound Poisson process, which depends explicitly on the spiking correlation.
For being continuously indexed in time, stationary compound Poisson processes can naturally serve as the drive to biophysically relevant neuronal models.
The idea to utilize compound Poisson processes to model input synchrony was originally proposed in~\cite{kuhn:2003,staude:2010,staude:2010b}, but without constructing these processes as limits of discrete spiking models and without providing explicit functional form for their jump distributions.
 More generally, our synchrony modeling can be interpreted as a limit case of the formalism proposed in~\cite{bauerle2005multivariate,trousdale2013generative} to model correlated spiking activity via multidimensional Poisson processes.


\subsection{Moment analysis}

We analytically characterize the subthreshold variability of a tractable conductance-based neuronal model, the AONCB neurons, when driven by synchronous synaptic inputs.
The analytical characterization of a neuron's voltage fluctuations has been the focus of intense research~\cite{destexhe:2001,shelley:2002,rudolph:2003,meffin:2004,kumar:2008}.
These attempts have considered neuronal models that already incorporate some diffusion scaling hypotheses~\cite{vanKampen:1992,risken:1996}, formally obtained by assuming an infinite number of synaptic inputs.
The primary benefit of these diffusion approximations is that one can treat the corresponding Fokker-Planck equations to quantify neuronal variability in conductance-based integrate-and-fire models, while also including the effect of post-spiking reset~\cite{zerlaut:2019,sanzeni:2022}. 
In practice, subthreshold variability is often estimated in the effective-time-constant approximation, while neglecting the multiplicative noise contributions due to voltage-dependent membrane fluctuations~\cite{destexhe:2001,shelley:2002,rudolph:2003}, although an exact treatment is also possible without this simplifying assumption~\cite{sanzeni:2022}.
By contrast, the analysis of conductance-based models has resisted exact treatments when driven by shot noise, as for compound Poisson input processes, rather than by Gaussian white noise, as in the diffusion approximation~\cite{richardson:2004,richardson:2005, richardson:2006}.

The exact treatment of shot-noise-driven neuronal dynamics is primarily hindered by the limitations of the Ito/Stratonovich integrals~\cite{ito:1944,stratonovich:1966} to capture the effects of point-process-based noise sources, even without including a reset mechanism.
These limitations were originally identified by Marcus, who proposed to approach the problem via a new type of stochastic equation~\cite{marcus:1978,marcus:1981}.
The key to Marcus equation is to define shot noise as limits of regularized, well-behaved approximations of that shot noise, for which classical calculus applies~\cite{chechkin:2014}.
In practice, these approximations are canonically obtained as the solutions of shot-noise-driven Langevin equations with relaxation time scale $\tau_\mathrm{s}$, and shot noise is formally recovered in the limit $\tau_\mathrm{s} \to 0^+$.
Our assertion here is that all-or-none conductances implement such a form of shot-noise regularization for which a natural limiting process can be defined when synapses operate instantaneously, i.e., $\tau_\mathrm{s} \to 0^+$.
The main difference with the canonical Marcus approach is that our regularization is all-or-none, substituting each Dirac delta impulse with a finite step-like impulse of duration $\tau_\mathrm{s}$ and magnitude $1/\tau_\mathrm{s}$, thereby introducing a synaptic timescale but without any relaxation mechanism.

The above assertion is the basis for introducing AONCB neurons, which is supported by our ability to obtain exact formulas for the first two moments of their stationary voltage dynamics (see \eref{eq:statmean} and \eref{eq:statvar}).
For $\tau_\mathrm{s}>0$, these moments can be expressed in terms of synaptic efficacies that takes exact but rather intricate integral forms.
Fortunately, these efficacies drastically simplify in the instantaneous synapse limit $\tau_\mathrm{s} \to 0^+$, for which the canonical shot-noise drive is recovered.
These resulting formulas mirror those obtained in the diffusion and effective-time-constant approximations~\cite{destexhe:2001,meffin:2004}, except that they involve synaptic efficacies whose expressions are oirginal in three ways (see  \eref{eq:a1} \eref{eq:a2}, \eref{eq:a12}, and \eref{eq:cei}):
First, independent of input synchrony, these efficacies all have exponential forms and saturate in the limit of large synaptic weights.
Such saturation is a general characteristic of shot-noise-driven, continuously-relaxing systems~\cite{baccelli:2019,baccelli:2021,yu:2022,yu:2022}.
Second, these efficacies are defined as expectations with respect to the jump distribution $p_{\ee\ii}$ of the driving compound Poisson process (see \eref{eq:peiDef} and Appendix \ref{app:CorrProc1}).
A nonzero dispersion of $p_{\ee\ii}$, indicating that synaptic activation is truly modeled via random variables $W_\ee$ and $W_\ii$, is the hallmark of input synchrony~\cite{staude:2010,staude:2010b}.
Third, these efficacies involve the overall rate of synaptic events $b$ (see \eref{eq:bDef2}), which also depends on input synchrony.
Such dependence can be naturally understood  within the framework of Palm calculus~\cite{baccelli2003palm}, a form of calculus specially developed for stationary point processes.


\begin{figure}[tbp]
\begin{center}
\includegraphics[width=8.6cm]{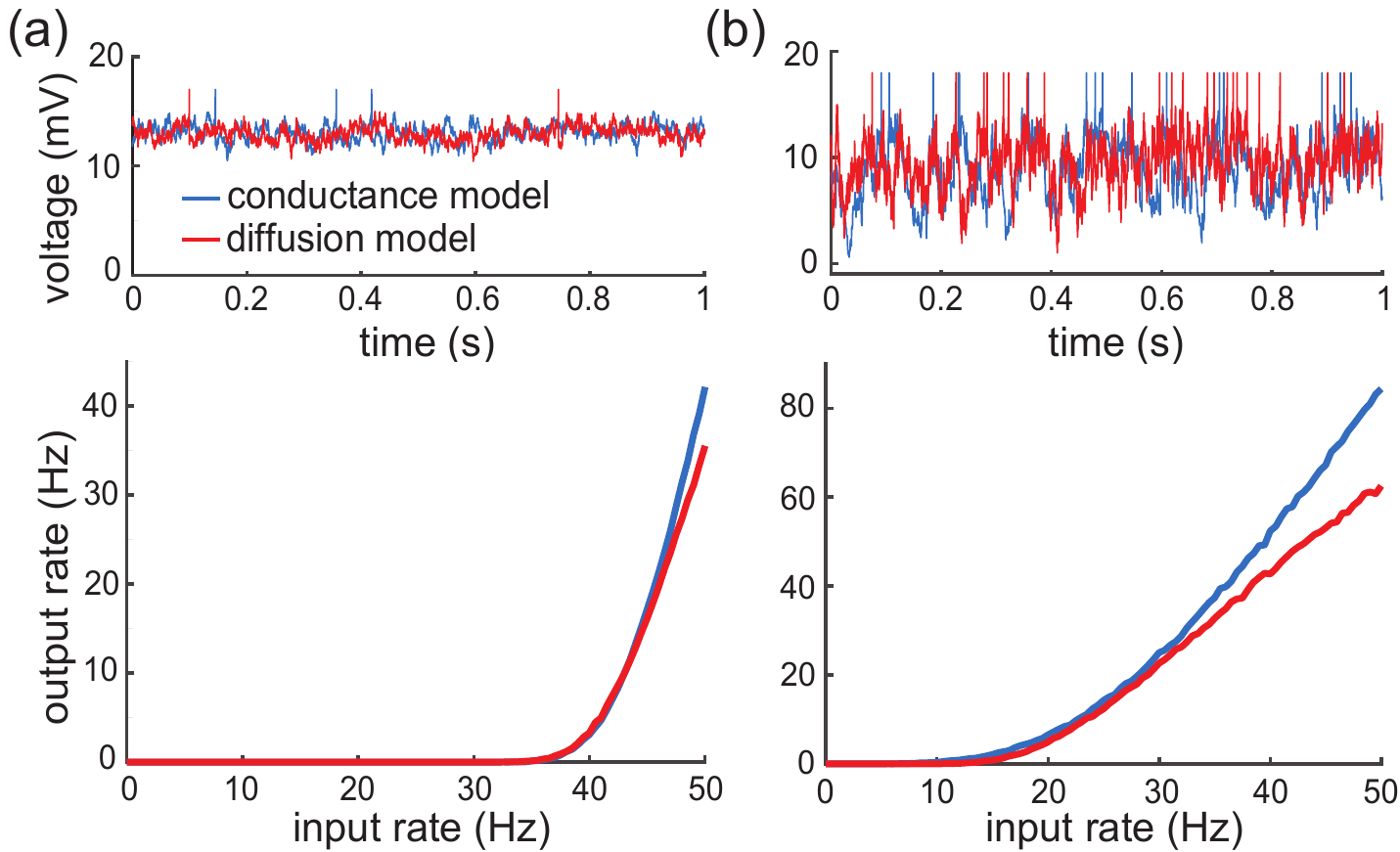}
\caption{\label{fig:diff}{\bf Diffusion approximations in the presence of synchrony.}
(a) Comparison of an asynchronously driven integrate-and-fire AONCB neuron (blue trace) with its diffusion approximation obtained via the effective-time-constant approximation (red trace).
(b) Comparison of a synchronously driven integrate-and-fire AONCB neuron (blue trace) with its diffusion approximation obtained by our exact analysis (red trace).
Parameters:  $K_e = 1000$, $K_i = 350$, $\tau = 15$ ms, $w_e = 0.001$, $w_i = 0.004$, $r_e = r_i = 25$ Hz, $\rho_e = \rho_i = 0.03$, $\rho_{ei} = 0$, $V_\textrm{T} = 15$ mV, and $V_\textrm{R} = 12$ mV. 
}
\end{center}
\end{figure}


\subsection{Biophysical relevance}

Our analysis allows us to investigate quantitatively how subthreshold variability depends on the numbers and strength of the synaptic contacts.
This approach requires that we infer synaptic weights from the typical peak time and peak amplitude of the somatic membrane fluctuations caused by post-synaptic potentials~\cite{bruno2006cortex,seeman:2018sparse,campagnola2022local}.
Within our modeling framework, these weights are dimensionless quantities that we estimate by fitting the AONCB neuronal response to a single all-or-none synaptic activation at rest.
For biophysically relevant parameters, this yields typically small synaptic weights in the sense that $w_\ee , w_\ii  \ll 1$.
These small values warrant adopting the small-weight approximation, for which expressions \eref{eq:statmean} and \eref{eq:statvar} simplify.

In the small-weight approximation, the mean voltage becomes independent of input synchrony, whereas the simplified voltage variance \eref{eq:statvar2} only depends on input synchrony via the spiking correlation coefficients $\rho_\ee$, $\rho_\ii$, and $\rho_{\ee\ii}$, as opposed to depending on a full jump distribution.
Spike-count correlations have been experimentally shown to be weak in cortical circuits~\cite{renart:2010aa,ecker:2010,cohen2011measuring} and for this reason, most theoretical approaches argued for asynchronous activity~\cite{london:2010aa,abbott:1993,brunel:2000,baladron:2012aa,touboul2012noise, robert:2016}. 
A putative role for synchrony in neural computations remains a matter of debate~\cite{averbeck:2006aa,ecker:2011,shea:2014}. 
In modeled networks, although the tight balance regime implies asynchronous activity,~\cite{sompolinski:1988,vreeswijk:1996,vreeswijk:1998aa}, the loosely balance regime is compatible with the establishment of strong neuronal correlations~\cite{ahmadian2013analysis,rubin2015stabilized,hennequin2018dynamical}.
When distributed over large networks, weak correlations can still give rise to precise synchrony, once information is pooled from a large enough number of synaptic inputs~\cite{chen:2006,polk:2012}. 
In this view, and assuming that distinct inputs play comparable roles, correlations measure the propensity of distinct synaptic inputs impinging on a neuron to coactivate, which represents a clear form of synchrony. 
Our analysis shows that considering  synchrony in amounts consistent with the  levels of observed spiking correlation is enough to account for the surprisingly large magnitude of subthreshold neuronal variability~\cite{churchland:2010aa,tan:2013aa,tan:2014aa,okun2015diverse}.
In contrast,  the asynchronous regime yields unrealistically low variability, an observation that challenges the basis for the asynchronous state hypothesis.

Recent theoretical works~\cite{zerlaut:2019,sanzeni:2022} have also noted that the asynchronous state hypothesis seems at odds with certain features of the cortical activity such as the emergence of spontaneous activity or the maintenance of significant average polarization during evoked activity.
Zerlaut {\it et al.} have analyzed under which conditions conductance-based networks can achieve a spectrum of asynchronous states with realistic neural features.
In their work, a key variable to achieve this spectrum is a strong afferent drive that modulates a balanced network with moderate recurrent connections.
Moderate recurrent conductances are inferred from allowing for up to $2\mathrm{mV}$ somatic deflections at rest, whereas the afferent drive is provided via even stronger synaptic conductances that can activate synchronously.
These inferred conductances appear large in light of recent {\it in-vivo} measurements~\cite{bruno2006cortex,seeman:2018sparse,campagnola2022local}, and the corresponding synaptic weights all satisfy $w_\ee,w_\ii \geq 0.01$ within our framework.
Correspondingly, the typical connectivity numbers considered  are small with $K_\ee=200$, $K_\ii=50$ for recurrent connections and $K_\ee=10$ for the coactivating afferent projections.
Thus, results from~\cite{zerlaut:2019} appear consistent with our observation that realistic subthreshold variability can only be achieved asynchronously for a restricted number of large synaptic weights.
Our findings, however, predict that these results follow from connectivity sparseness and will not hold in denser networks, for which the pairwise spiking correlation will exceed the empirical criteria for asynchrony, e.g., $\rho_\ee>1/K_\ee$ ($\rho_\ee<0.005\leq 1/K_\ee$ in~\cite{zerlaut:2019}).
Sanzeni {\it et al.} have pointed out that implementing the effective-time-constant approximation in conductance-based models suppresses subthreshold variability,  especially in the high-conductance state~\cite{destexhe:2003}.
As mentioned here, this suppression causes the voltage variability to decay as $O(w_\ee)+O(w_\ii)$ in any scaling limit with vanishing synaptic weights.
Sanzeni {\it et al.} observe that such decay is too fast to yield realistic variability for the balanced scaling, which assumes $w_\ee \sim 1/\sqrt{K_\ee}$ and $w_\ii \sim 1/\sqrt{K_\ii}$.
To remedy this point, these authors propose to adopt a slower scaling of the weights, i.e., $w_\ee \sim 1/\ln K_\ee$ and $w_\ii \sim 1/\ln K_\ii$, which can be derived from the principle of rate conservation in neural networks.
Such a scaling is sufficiently slow for variability to persist in networks with large connectivity number ($\simeq 10^5$).
However, as any scaling with vanishing weights, our exact analysis shows that such scaling must eventually lead to decaying variability, thereby challenging the basis for the synchronous state hypothesis.

Both of these studies focus on the network dynamics of conductance-based networks under the diffusion approximations. 
Diffusive behaviors only rigorously emerge under some scaling limit with vanishing weights~\cite{vanKampen:1992,risken:1996}.
By focusing on the single-cell level rather than the network level, we are able to demonstrate that the effective-time-constant approximation holds exactly for shot-noise driven, conductance-based neurons, without any diffusive approximations.
Consequently, suppression of variability must occur independent of any scaling choice, except in the presence of input synchrony.
Although this observation poses a serious theoretical challenge to the asynchronous state hypothesis, observe that it does not invalidate the practical usefulness of the diffusion approximation.
For instance, we show in Fig.~\ref{fig:diff} that the mean spiking response of an a shot-noise driven AONCB neuron  with an integrate-and-fire mechanism can be satisfactorily captured via the diffusion approximation.
In addition, our analysis allows one to extend the diffusion approximation to include input synchrony.


\begin{figure}[tbp]
	\begin{center}
		\includegraphics[width=8.6cm]{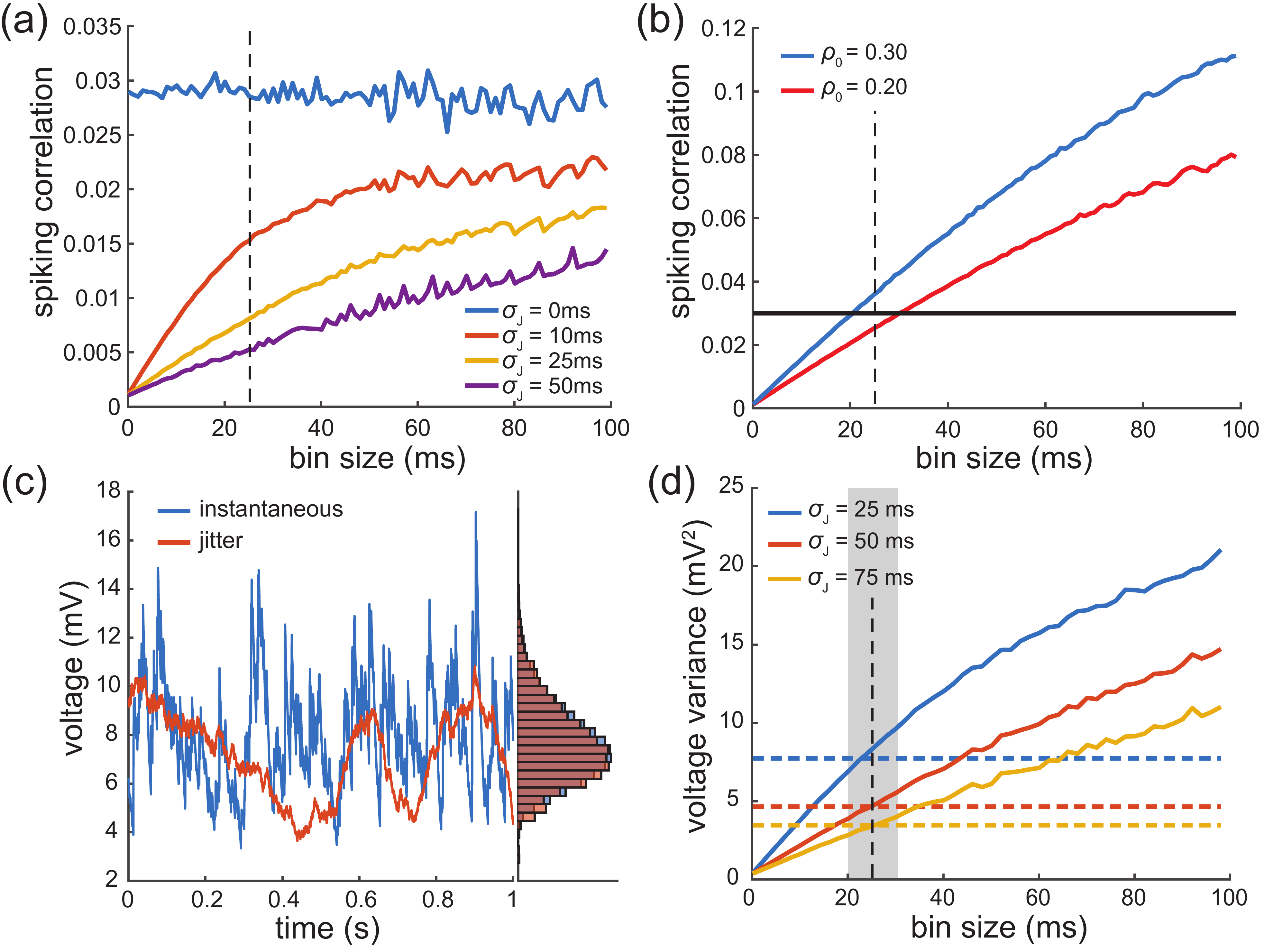}
		\caption{\label{fig:TimeDep}{\bf Impact of jittering synchronous inputs.}
		(a) Effect of jittering synchronous spike times via independent Gaussian centered time shifts with varied standard deviation $\sigma_J$: without jitter, spiking correlation is independent of the size of the time bins used to count spikes (blue trace). Jittering with larger $\sigma_J$ decreases spiking correlation for all bin sizes, with spiking correlation vanishing in the limit of small bin sizes. 
		(b) Given a jitter standard deviation of $\sigma_J=50\mathrm{ms}$, one obtains spike-count correlation of $\rho(\Delta t) = 0.03$ in $\Delta t = 25\mathrm{ms}$ bins by jittering a synchronous input with instantaneous correlation of $\rho_\ee =0.2-0.3$. 
		(c) Comparison of voltage trace obtained with instantaneous synchronous input (blue) and jittered correlated inputs (red) for $\sigma_J=50\mathrm{ms}$. Both types of input are chosen so that they yield the same spiking correlation of $\rho_\ee=\rho(\Delta t) = 0.03$ with bin size of $\Delta t=25\mathrm{ms}$. The stationary distributions are close to identical leading to less than 1\% error in the variance estimates. 
		(d) Comparison between the voltage variances of an AONCB neuron driven by realistic synchronous inputs with various jitter (dashed line) and the voltage variances of  the same AONCB neuron driven by instantaneously synchronous approximations (solid line). For each $\sigma_J$, different instantaneous approximations are obtained for different bin sizes $\Delta t$ by setting $\rho_\ee=\rho(\Delta t)$ for various bin size $\Delta t$. Good approximations are consistently obtained for $\Delta t \simeq 25\mathrm{ms}$  (grey column). Other parameters: $r_e=10\mathrm{Hz}$, $K_e=1000$, $w_e=10^{-3}$.}
	\end{center}
\end{figure}


\subsection{Limitations of the approach}


A first limitation of our analysis is that we neglect the spike-generating mechanism as a source of neural variability.
Most diffusion-based approaches model spike generation via the integrate-and-fire mechanism, whereby the membrane voltages reset to fixed value upon reaching a spike-initiation threshold~\cite{destexhe:2001,shelley:2002,zerlaut:2019,sanzeni:2022,rudolph:2003,meffin:2004,kumar:2008}.
Accounting for such a mechanism can impact our findings in two ways:
$(i)$
By confining voltage below the spiking threshold, the spiking mechanism may suppress the mean response enough for the neuron to operate well in the high-conductance regime for large input drives.
Such a scenario will still produce exceedingly low variability due to variability quenching in the high-conductance regime, consistent with~\cite{churchland:2010aa}.
$(ii)$
The additional variability due to post-spiking resets may dominate the synaptic variability, so that a large overall subthreshold variability can be achieved in spite of low synaptic  variability.
This possibility also seems unlikely as dominant yet stereotypical resets would imply a quasi-deterministic neural response~\cite{softky:1993}.
Addressing the above limitations quantitatively requires extending our exact analysis to include the integrate-and-fire mechanism using technique from queueing theory~\cite{baccelli2003palm}.
This is beyond the scope of this work. 
We note, however, that implementing a post-spiking reset to a fixed voltage level yields simulated trajectories that markedly differ from physiological ones (see \fref{fig:expVar}), for which the post-spiking voltage varies across conditions~\cite{tan:2013aa,tan:2014aa,okun2015diverse}.

A second limitation of our analysis is our assumption of exchangeability, which is the lens through which we operate a link between spiking correlations and input drives.
Taken literally, the exchangeability assumption states that synapses all have a typical strength and that conductance variability primarily stems from the variable numbers of coactivating synapses.
This is certainly an oversimplification as synapses exhibit heterogeneity~\cite{buzsaki2014log}, which likely plays a role in shaping neural variability~\cite{Iyer:2013}. 
Distinguishing between heterogeneity and correlation contributions, however, is a fundamentally ambiguous task~\cite{amarasingham:2015}.
For instance, considering $K_\ee$ synchronous inputs with weight $w_\ee$ at rate $b_\ee$  and with jump probability $p_{\ee}$ (see \eref{eq:PkDef} and \eref{eq:bDef}) is indistinguishable from considering $K_\ee$ independent inputs with heterogeneous weights $\{w_\ee, 2 w_\ee, \ldots, K_\ee w_\ee \}$ and rates $K_\ee r_\ee p_{\ee,k}$.
Within our modeling approach, accounting for synaptic heterogeneity, with dispersed distribution for synaptic weights $q_\ee(w)$, can be done by taking the jump distribution $p_\ee$ as 
\begin{eqnarray*}
p_\ee(w) = \sum_{k=1}^K q_\ee^{(\star k)}(w) p_{\ee,k} \, ,
\end{eqnarray*}
where $q_\ee^{(\star k)}$ refers to the $k$-fold convolution of $q_\ee(w)$.
This leads to an overdispersion of the jump distribution $p_\ee$, and thus increased subthreshold neural variability. 
Therefore, while we have assumed exchangeability, our approach can accommodate weight heterogeneity.
The interpretation of our results in term of synchrony rather than heterogeneity is supported by recent experimental evidence that cortical response selectivity derives from strength in numbers of synapses, rather than difference in synaptic weights~\cite{scholl:2021}.

A third limitation of our analysis is to consider a perfect form of synchrony, with exactly simultaneous synaptic activations.
Although seemingly unrealistic, we argue that perfect input synchrony can still yield biologically relevant estimates of the voltage variability.
For instantaneous synchrony, the empirical spiking correlation is independent of the timescale over which spikes are counted, i.e., $\rho_{\mathrm{emp}}=\rho_\ee$, as shown in \fref{fig:TimeDep}(a) (blue line). 
This is a potential problem because spiking correlations have been measured to vanish on small timescales in experimental recordings~\cite{smith2008spatial,smith2013spatial}. 
More realistic input models can be obtained by jittering instantaneously synchronous spikes.
Such a procedure leads to a general decrease in the empirical spiking correlations $\rho_{\mathrm{emp}}(\Delta t)$ with spiking correlations over all timescales $\Delta t$, including for $\Delta t= 25\mathrm{ms}$ (vertical dashed line in \fref{fig:TimeDep}(a)), which vanish in the limit of small timescales $\Delta t \to 0$ (red, yellow and purple lines in \fref{fig:TimeDep}(a)). 
Analysis of the temporal structure of spiking correlation in~\cite{smith2008spatial,smith2013spatial} suggests that correlations $\rho_{\mathrm{emp}}(\Delta t)$ lie within the range $0.01-0.04$ for $\Delta t \simeq 25\mathrm{ms}$.
We focus on this timescale because it is just larger than the membrane time constant of the neuron. .
Then, to achieve realistic correlations at $\Delta t \simeq25\mathrm{ms}$, the instantaneous spiking correlation of the unjitterred synchronous input model, denoted by $\rho_\infty$, may be increased.
Adopting a jittering timescale of  $\sigma_J=50\mathrm{ms}$, \fref{fig:TimeDep}(b) shows that  $\rho_{\mathrm{emp}}(\Delta t)\simeq 0.03$ with $\Delta t =25\mathrm{ms}$ for instantaneous spiking correlation $\rho_\infty$ within the range $0.2-0.3$. 
Note that for very long timescales $\Delta t \to \infty$, this also implies that the empirical spiking correlation saturates at $\rho_\infty \simeq 0.2-0.3$, as reported in~\cite{smith2008spatial,smith2013spatial}.
To validate that our instantaneous model makes realistic prediction about the subthreshold variability, we simulate AONCB neurons in response to these jittered synchronous inputs. 
\fref{fig:TimeDep}(c) shows that the resulting stationary voltage distribution (red histogram) closely follows the distribution obtained by assuming instantaneous synchrony with $\rho_\ee$ chosen such that $\rho_\ee=\rho_{\mathrm{emp}}(\Delta t=25\mathrm{ms})$ (blue trace and histogram). 
Furthermore, we can justify the choice of the timescale $\Delta t=25\mathrm{ms}$ \emph{a posteriori}. 
Specifically, in \fref{fig:TimeDep}d, we consider temporally structured inputs obtained from the same instantaneous synchrony $\rho_\infty$ but for various jittering timescale $\sigma_J$. Jittering at larger timescale $\sigma_J$ reduces synchrony and voltage variance (vertical dashed lines). 
We then compare the resulting voltage variance with perfectly synchronous approximations obtained by matching spike-count correlation at various time scale (our choice is to match at 25ms).  
\fref{fig:TimeDep}(d) shows that matching at increasing timescale yields higher variance, but matching at $\Delta t \simeq 25\mathrm{ms}$ offers good approximations (grey square where variances are about the same).
Extending our analytic results to include jittering will require modeling spiking correlations via multidimensional Poisson processes rather than via compound Poisson processes~\cite{bauerle2005multivariate,trousdale2013generative}.
However, this is  is beyond the scope of this work.
A remaining limitation of our synchrony modeling is that our analysis can only account for nonnegative, instantaneous correlations between excitation and inhibition, while in reality such correlations may be negative and are expected to peak at a non-zero time lag.

A fourth limitation of our analysis is that it is restricted to a form of synchrony that ignores temporal heterogeneity.
This is a limitation because a leading hypothesis for the emergence of variability is that neurons generate spikes as if through a doubly stochastic process, i.e., as a Poisson process with temporally fluctuating rate~\cite{litwin2012slow}.
To better understand this limitation, let us interpret our exchangeability-based modeling approach within the framework of  doubly stochastic processes~\cite{daley2003introduction,daley2007introduction}.
This can be done  most conveniently by reasoning on the discrete correlated spiking model specified by \eref{eq:deFinetti}.
Specifically, given fixed bin size $\Delta t>0$, one can interpret the collection of {\it i.i.d.} variables $\theta \sim F$ as an instantaneously fluctuating rate.
In this interpretation, nonzero correlations can be seen as emerging from a doubly stochastic process for which the rate fluctuates as uncorrelated noise, i.e., with zero correlation time.
This zero correlation time is potentially a serious limitation as it has been argued that shared variability is best modeled by a low-dimensional latent process evolving with slow, or even smooth, dynamics~\cite{macke:2011}. 
Addressing this limitation will require developing limit spiking model with nonzero correlation time using probabilistic techniques that are beyond the scope of this work~\cite{aldous1985exchangeability}.

A final limitation of our analysis is that it does not explain the consistent emergence of synchrony in network dynamics.
It remains conceptually unclear how synchrony can emerge and persist in neural networks that are fundamentally plagued by noise and exhibit large degrees of temporal and cellular heterogeneity.
It may well be that carefully taking into account the finite-size of networks will be enough to produce the desired level of synchrony-based correlation, which is rather weak after all.
Still, one would have to check wether achieving a given degree of synchrony requires the tuning of certain network features, such as the degree of shared input or the propensity of certain recurrent motifs~\cite{rocha:2007aa} or the relative width of recurrent connections with respect to feedforward projections~\cite{rosenbaum:2017}.
From a theoretical standpoint, the asynchronous state hypothesis answers the consistency problem by assuming no spiking correlations, and thus no synchrony.
One can justify this assumption in idealized mathematical models by demonstrating the so-called ``propagation-of-chaos'' property~\cite{sznitman:1989}, which rigorously holds for certain scaling limits with vanishing weights and under the assumption of exchangeability~\cite{baladron:2012aa,touboul2012noise, robert:2016}.
In this light, the main theoretical challenge posed by our analysis is extending the latter exchangeability-based property to include nonzero correlations~\cite{erny:2021}, and hopefully characterize irregular synchronous state in some scaling limits.


\begin{acknowledgments}
LB, BL, NP, ES, and TT were supported by the Vision Research program of the National Institutes of Health under award number R01EY024071. LB and TT were also supported by the CRCNS program of the National Science Foundation under award number DMS-2113213.  We would like to thank Fran\c{c}ois Baccelli, David Hansel, and Nicolas Brunel for insightful discussions.
\end{acknowledgments}


\appendix


\begin{widetext}


\section{Discrete-time spiking correlation}\label{app:CorrDisc}

In this appendix, we consider first the discrete-time version of our model for possibly correlated excitatory synaptic inputs.
In this model, we consider that observing $K_\ee$ synaptic inputs during $N$ time steps specifies a $\{0,1\}$-valued matrix $\left\{ X_{k,i} \right\}_{1 \leq k \leq K_\ee, 1 \leq i \leq N}$, where $1$ indicates that an input is received and $0$ indicates an absence of inputs.
For simplicity, we further assume that the inputs are independent across time 
\begin{eqnarray*}
\Prob{\left\{ X_{k,i} \right\}_{1 \leq k \leq K_\ee, 1 \leq i \leq N} } = \prod_{i=1}^N \Prob{\left\{ X_{k,i} \right\}_{1 \leq k \leq K_\ee} } \, ,
\end{eqnarray*}
so that we can drop the time index and consider the population vector $\left\{ X_k \right\}_{1 \leq k \leq K_\ee}$.
Consequently, given the individual spiking rate $r_\ee$, we have $\Exp{X_k}=\Prob{X_k=1}=r_i \Delta t$, where $\Delta t$ is the duration of the time step where a spike may or may not occur.
Under the assumptions that $\left\{ X_k \right\}_{1 \leq k \leq K_\ee}$ belongs to an infinitely exchangeable set of random variables,
de Finetti theorem states that there exists a probability measure $F_\ee$ on $[0,1]$ such that
\begin{eqnarray*}
\Prob{\left\{ X_k \right\}_{1 \leq k \leq K_\ee} }
=
 \int \prod_{k=1}^{K_\ee} \theta_\ee^{X_k}(1-\theta_\ee)^{1-X_k} \, \dd F_\ee(\theta_\ee) \, .
\end{eqnarray*}
Assuming the directing measure $F_\ee$ known, we can compute the spiking correlation attached to our model.
To see this, first observe that specifying the above probabilistic model for $K_\ee=1$, we have
\begin{eqnarray}
\Exp{X_k} = \Exp{\Exp{ X_k \, \vert \, \theta_\ee} }=\Exp{\theta_\ee} =  \int  \theta_\ee \dd F_\ee(\theta_\ee) \, .\nonumber
\end{eqnarray}
Then, using the total law of covariance and specifying the above probabilistic model for $K=2$, we have
\begin{eqnarray}
\Cov{X_k ,X_l}
&=&
\Exp{\Cov{X_k ,X_l \, \vert \, \theta_\ee}} + \Cov{\Exp{X_k \, \vert \theta_\ee}, \Exp{ X_l \, \vert \, \theta_\ee}} \, , \nonumber\\
&=&
\one{k=l} \Exp{\Var{X_k \, \vert \, \theta_\ee}}+ \Cov{ \theta_\ee,\theta_\ee} \, , \nonumber\\
&=&
\one{k=l} \Exp{\theta_\ee(1-\theta_\ee)}+ \Var{ \theta_\ee} \, , \nonumber\\
&=&
 \one{k=l} \Exp{\theta_\ee}\left( 1- \Exp{\theta_\ee}\right) +  \one{k\neq l}\Var{\theta_\ee} \nonumber \, .
\end{eqnarray}
This directly yields that the spiking correlation reads
\begin{eqnarray}\label{eq:spCorr1}
\rho_\ee = \frac{\Cov{X_k ,X_l}}{\Var{X_k}}   = \frac{\Var{\theta_\ee}}{\Exp{\theta_\ee}(1-\Exp{\theta_\ee})}
\end{eqnarray}
The exact same calculations can be performed for the partially exchangeable case of mixed excitation and inhibition.
The assumption of partial exchangeability requires that when considered separately, the $\{0,1\}$-valued vectors $\left\{ X_1, \ldots,  X_{K_\ee} \right\}$ and $\left\{ Y_1, \ldots,  Y_{K_\ee} \right\}$ each belong to an infinitely exchangeable sequence of random variables.
Then, de Finetti's theorem states that the probability to find the full vector of inputs $\left\{ X_1, \ldots,  X_{K_\ee} , Y_1, \ldots,  Y_{K_\ee} \right\}$ in any particular configuration is given by
\begin{eqnarray}\label{eq:partExch}
\Prob{X_1, \ldots, X_{K_\ee},Y_1, \ldots, Y_{K_\ii}} 
=
 \int \prod_{k=1}^{K_\ee} \theta_\ee^{X_k} (1-\theta_\ee)^{1-X_k}  \prod_{l=1}^{K_\ii} \theta_\ii^{Y_l} (1-\theta_\ii)^{1-Y_l} \, \dd F_{\ee \ii}(\theta_\ee, \theta_\ii) \, , 
\end{eqnarray}
where the directing measure $F_{\ee\ii}$ fully parametrizes our probabilistic model.
Performing similar calculations as for the case of excitation alone within this partially exchangeable setting yields
\begin{eqnarray}\label{eq:spCorr2}
\rho_{\ee\ii} = \frac{\Cov{X_k ,Y_l}}{\sqrt{\Var{X_k} \Var{Y_l}}}  = \frac{\Cov{\theta_\ee,\theta_\ii}}{\sqrt{\Exp{\theta_\ee}(1-\Exp{\theta_\ee})\Exp{\theta_\ii}(1-\Exp{\theta_\ii})}} \, .
\end{eqnarray}


\section{Compound Poisson processes as continuous-time limits}\label{app:CorrProc1}

Let us consider the discrete-time model specified by \eref{eq:partExch}, which is obtained under the assumption of partial infinite exchangeability.
Under this assumption, the probability laws of the inputs is entirely determined by the distribution  of $(k_\ee,k_\ii)$, where $k_\ee$ denotes the number of active excitatory inputs and 
$k_\ii$ denotes the number of inhibitory inputs. 
This distribution can be computed as
\begin{eqnarray}
P_{\ee\ii,kl}
=
\Prob{k_\ee=k,k_\ii=l}
=
\binom{K_\ee}{k} \binom{K_\ii}{l} \int \theta_\ee^k (1-\theta_\ee)^{K_\ee-k}   \theta_\ii^l (1-\theta_\ii)^{K_\ii-l} \, \dd F_{\ee \ii}(\theta_\ee, \theta_\ii) \, . \nonumber
\end{eqnarray}
It is convenient to choose the directing measure as beta distributions since these are conjugate to the binomial distributions. 
Such a choice yields a class of probabilistic models referred to as beta-binomial models, which have been studied extensively~\cite{thibaux:2007,broderick2012beta}.
In this appendix, we always assume that the marginals $F_\ee$ and $F_\ii$ have the form $F_\ee \sim \mathrm{Beta}(\alpha_\ee,\beta_\ee)$
and $F_\ii \sim \mathrm{Beta}(\alpha_\ii,\beta_\ii)$.
Then, direct integrations shows that the marginal distributions for the number of excitatory inputs and inhibitory inputs are
\begin{eqnarray}
P_{\ee,k}
=
\sum_{l=0}^{K_\ii}P_{\ee\ii,kl}
=
\binom{K_\ee}{k}   \frac{B(\alpha_\ee+k,\beta_\ee+K_\ee-k)}{B(\alpha_\ee,\beta_\ee)} 
\quad \mathrm{and} \quad
P_{\ii,l}
=
\sum_{k=0}^{K_\ee}P_{\ee\ii,kl}
=
\binom{K_\ii}{l} \frac{B(\alpha_\ii+l,\beta_\ii+K_\ii-l)}{B(\alpha_\ii,\beta_\ii)} \, . \nonumber
\end{eqnarray}
Moreover, given individual spiking rates $r_\ee$ and $r_\ii$ within a time step $\Delta t$, we have
\begin{eqnarray}
r_\ee \Delta t = \Exp{X_k}=\Prob{X_k=1}=\Exp{\theta_\ee}=\frac{\alpha_\ee}{\alpha_\ee+\beta_\ee}
\quad
\mathrm{and}
\quad
r_\ii \Delta t = \Exp{Y_l}=\Prob{Y_l=1}=\Exp{\theta_\ii}=\frac{\alpha_\ii}{\alpha_\ii+\beta_\ii} \, .\nonumber
\end{eqnarray}
The continuous-time limit is obtained by taking $\Delta t \to 0^+$, which implies that the parameters $\alpha_\ee$ and $\alpha_\ii$ jointly vanish.
When $\alpha_\ee, \alpha_\ii \to 0^+$, the beta distributions $F_\ee$ and $F_\ii$ becomes deficient and we have $P_{\ee,0}, P_{\ii,0} \to 1$.
In other words, time bins of size $\Delta t$ almost surely have no active inputs in the limit $\Delta t \to 0^+$.
Actually, one can show that
\begin{eqnarray}
1-P_{\ee,0} \sim \left( \psi(K_\ee+\beta_\ee) - \psi(\beta_\ee)\right)\alpha_\ee \quad \mathrm{and} \quad 1-P_{\ii,0} \sim  \left( \psi(K_\ii+\beta_\ii) - \psi(\beta_\ii)\right)\alpha_\ii \, , \nonumber
\end{eqnarray}
where $\psi$ denotes the digamma function.
This indicates in the limit $\Delta t \to 0^+$, the times at which some excitatory inputs or some inhibitory inputs are active define a point process.
Moreover, owing to the assumption of independence across time, this point process will actually be a Poisson point process.
Specifically, consider a time $T>0$ and set $\Delta t = T/N$ for some large integer $N$.
Define the sequence of times
\begin{eqnarray}
T_{\ee,n} &=& \frac{T}{N} \cdot \inf \left\{ i > N T_{\ee,n-1}/T  \, \vert k_{\ee,i} \geq 1 \right\} \quad \mathrm{with} \quad T_{\ee,1} = \frac{T}{N} \cdot\inf \left\{ i \geq 0 \, \vert k_{\ee,i} \geq 1 \right\} \, , \nonumber\\
T_{\ii,n} &=& \frac{T}{N} \cdot \inf \left\{ i > N T_{\ii,n-1}/T   \, \vert k_{\ii, i} \geq 1 \right\} \quad \mathrm{with} \quad T_{\ii,1} = \frac{T}{N} \cdot \inf \left\{ i \geq 0 \, \vert k_{\ii, i} \geq 1 \right\} \, . \nonumber
\end{eqnarray}
Considered separately, the sequences of times $\{ T_{\ee,n}\}_{n\geq 1}$ and $\{ T_{\ii,n}\}_{n\geq 1}$ constitute binomial approximations of Poisson processes which we denote by $N_\ee$ and $N_\ii$, respectively.
It is a classical result that these limit Poisson processes are recovered exactly when $N \to \infty$ and that their rates are respectively given  by
\begin{eqnarray}
b_\ee 
&=& 
\lim_{\Delta t \to 0^+} \frac{1-P_{\ee,0}}{\Delta t}
= 
 \left( \psi(K_\ee+\beta_\ee) - \psi(\beta_\ee)\right)\left( \lim_{\Delta t \to 0^+} \frac{ \alpha_\ee}{\Delta t} \right)
 =
  \left( \psi(K_\ee+\beta_\ee) - \psi(\beta_\ee)\right) \beta_\ee r_\ee \, , \nonumber\\
  b_\ii 
&=& 
\lim_{\Delta t \to 0^+} \frac{1-P_{\ii,0}}{\Delta t}
= 
 \left( \psi(K_\ii+\beta_\ii) - \psi(\beta_\ii)\right) \left( \lim_{\Delta t \to 0^+} \frac{ \alpha_\ii}{\Delta t} \right)
 =
  \left( \psi(K_\ii+\beta_\ii) - \psi(\beta_\ii)\right) \beta_\ii r_\ii \, . \nonumber
\end{eqnarray}
For all integer $K>1$, the function $\beta \mapsto  \beta \left( \psi(K+\beta) - \psi(\beta)\right)$ is an increasing analytic functions on the domain $\mathbbm{R}^+$ with range $[1,K]$.
Thus, we always have $r_\ee \leq b_\ee \leq K_\ee r_\ee$ and $r_\ii \leq b_\ii \leq K_\ii r_\ii$ and the extreme cases are achieved for perfect or zero correlations.
Perfect correlations are achieved when $\rho_\ee=1$ or $\rho_\ii=1$, which corresponds to $\beta_\ee \to 0$ or $\beta_\ii \to 0$.
This implies that $b_\ee=r_\ee$ and $b_\ii =r_\ii$, consistent with all synapses activating simultaneously.
Zero correlations are achieved when $\rho_\ee=0$ or $\rho_\ii=0$, which corresponds to $\beta_\ee \to \infty$ or $\beta_\ii \to \infty$.
This implies that $b_\ee=K_\ee r_\ee$ and $b_\ii =K_\ii r_\ii$, consistent with all synapses activating asynchronously, so that no inputs simultaneously activate.
Observe that in all generality, the rates $b_\ee$ and $b_\ii$ are such that the mean number of spikes over the duration $T$ is conserved in the limit $\Delta t \to 0^+$.
For instance, one can check that
\begin{eqnarray}
K_\ee r_\ee T = \Exp{\sum_{T_{\ee,n} \leq T} k_{\ee, N T_{\ee,n}/T}} = \Exp{\sum_{n=1}^{N_\ee(T)} k_{\ee, n}} = \Exp{N_\ee(T)} \Exp{k_\ee}=b_\ee T \Exp{k_\ee} \nonumber
\end{eqnarray}

When excitation and inhibition are considered separately, the limit process $\Delta t \to 0^+$ specifies two compound Poisson processes 
\begin{eqnarray}
t \mapsto  \sum_{n=1}^{N_\ee(t)} k_{\ee,n} \quad \mathrm{and} \quad t \mapsto  \sum_{n=1}^{N_\ii(t)} k_{\ii,n} \, , \nonumber
\end{eqnarray}
where $N_\ee$ and $N_\ii$ are Poisson processes with rate $b_\ee$ and $b_\ii$ and where $\{k_{\ee,n}\}_{n \geq 1}$ are i.i.d according to $p_{\ee}$ and $\{k_{\ee,n} \}_{n \geq 1}$ are i.i.d according to $p_{\ii}$.
Nonzero correlations between excitation and inhibition emerge when the Poisson processes $N_\ee$ and $N_\ii$ are not independent.
This corresponds to the processes $N_\ee$ and $N_\ii$ sharing times, so excitation and inhibition occur simultaneously at these times.
To understand this point intuitively, let us consider the limit Poisson process $N$ obtained by considering synaptic events without distinguishing excitation and inhibition.
For perfect correlation, i.e., $\rho_{\ee\ii}=1$, all synapses activate synchronously and we have $N=N_\ee=N_\ii$: all times are shared.
By contrast, for zero correlation, i.e., $\rho_{\ee\ii}=0$, no synapses activate simultaneously and we have $N=N_\ee + N_\ii$: no times are shared.
For intermediary regime of correlations, a nonzero fraction of times will be shared resulting in a driving Poisson process $N$ with overall rate $b$ satisfying $\min(b_\ee,b_\ii) \leq b<b_\ee+b_\ii$. We investigate the above intuitive statements  quantitatively in Appendix \ref{app:CorrProc2} by inspecting two key examples.

Let us conclude this appendix by recapitulating the general form of the limit compound process $Z$ obtained in the continuous-time limit $\Delta t \to 0^+$ when jointly considering excitation and inhibition.
This compound Poisson process can be represented as
\begin{eqnarray*}
t \mapsto Y(t)=\left( \sum_{n}^{N(t)} W_{\ee,n}, \sum_{n}^{N(t)} W_{\ii,n}\right) \, ,
\end{eqnarray*}
where $N$ is that Poisson process registering all synaptic events without distinguishing excitation and inhibition and where the pairs $(W_{\ee,n,W_{\ii,n}})$ are i.i.d. random jumps in $\mathbbm{R}\times\mathbbm{R} \setminus \{0,0\}$.
Formally, such a process is specified by the rate of $N$, denoted by $b$, and the bivariate distribution of the jumps $(W_{\ee,n},W_{\ii,n})$, denoted by $p_{\ee\ii}$.
These are  defined as
\begin{eqnarray}\label{eq:bApp1}
b = \lim_{\Delta t \to 0^+} \frac{1-P_{\ee\ii,00}}{\Delta t} 
\quad \mathrm{and} \quad 
p_{\ee\ii,kl} =  \lim_{\Delta t \to 0^+} \frac{P_{\ee\ii,kl}}{1-P_{\ee\ii,00}} \quad \mathrm{for} \quad (k,l)\neq(0,0)\, ,
\end{eqnarray}
where $P_{\ee\ii,00}$ is the probability to register no synaptic activation during a time step $\Delta t$.
According to these definitions, $b$ is the infinitesimal likelihood that an input is active within a time bin, whereas  $p_{\ee\ii}$ is the probability that $k$ excitatory inputs and $l$ inhibitory inputs are active given that at least one input is active.
One can similarly define the excitatory and inhibitory rates of events $b_\ee$ and $b_\ii$, as well as the excitatory jump distribution $p_{\ee}$ and the inhibitory jump distribution $p_{\ii,l}$.
Specifically, we have
\begin{eqnarray}\label{eq:beApp1}
b_\ee = \lim_{\Delta t \to 0^+} \frac{1-P_{\ee,0}}{\Delta t}  
\quad &\mathrm{and}& \quad p_{\ee,k} =  \lim_{\Delta t \to 0^+} \frac{P_{\ee,k}}{1-P_{\ee,0}} \quad \mathrm{for} \quad k \neq 0\, , \\
b_\ii = \lim_{\Delta t \to 0^+} \frac{1-P_{\ii,0}}{\Delta t}  
\quad &\mathrm{and}& \quad p_{\ii,l} =  \lim_{\Delta t \to 0^+} \frac{P_{\ii,l}}{1-P_{\ii,0}} \quad \mathrm{for} \quad l\neq 0\, ,\nonumber
\end{eqnarray}
with $P_{\ee,k}=\sum_{l=0}^{K_\ii} P_{\ee\ii,k,l}$ and $P_{\ii,k}=\sum_{k=0}^{K_\ee} P_{\ee\ii,k,l}$.
Observe that thus defined, the jump distribution $p_{\ee}$ and $p_{\ii}$ are specified as conditional marginal distributions of the joint jump distribution $p_{\ee\ii}$ on the events $\{k_\ee>0\}$ and $\{k_\ii>0\}$, respectively.
These are such that $p_{\ee,k}= (b/b_\ee) \sum_{l=0}^{K_\ii}p_{\ee\ii, kl}$ and $p_{\ii,l}= (b/b_\ii) \sum_{k=0}^{K_\ee}p_{\ee\ii, kl}$.
To see why, observe for instance that
\begin{eqnarray}
p_{\ee,k} 
= 
\lim_{\Delta t \to 0^+} \frac{P_{\ee, k} }{1-P_{\ee, 0}} 
=
\lim_{\Delta t \to 0^+} \sum_{l=0}^{K_\ii} \frac{P_{\ee\ii, kl} }{1-P_{\ee\ii, 00}} \frac{1-P_{\ee\ii, 00}}{1-P_{\ee, 0}}
=
\left(\sum_{l=0}^{K_\ii}p_{\ee\ii, kl} \right) \left( \lim_{\Delta t \to 0^+} \frac{1-P_{\ee\ii, 00}}{1-P_{\ee, 0}} \right)
=
\frac{b}{b_\ee} \sum_{l=0}^{K_\ii}p_{\ee\ii, kl} 
\end{eqnarray}
where we have used the definitions of the rates $b$ and $b_\ee$ given in \eref{eq:bApp1} and \eref{eq:beApp1} to establish that
\begin{eqnarray*}
\lim_{\Delta t \to 0^+} \frac{1-P_{\ee\ii, 00}}{1-P_{\ee, 0}}
=
\frac{
\displaystyle \lim_{\Delta t \to 0^+}(1-P_{\ee\ii, 00})/\Delta t
}
{
\displaystyle \lim_{\Delta t \to 0^+} (1-P_{\ee, 0})/\Delta t
}
=\frac{b}{b_\ee} \, .
\end{eqnarray*}


\section{Continuous-time spiking correlation}\label{app:CorrCont}

\eref{eq:spCorr1} and \eref{eq:spCorr2} carry over to the continuous time limit $\Delta t \to 0^+$ by observing that for limit compound Poisson processes to emerge, one must have that
$\Exp{X_k} =\Exp{\theta_\ee} = O(\Delta t)$ and $\Exp{Y_l} =\Exp{\theta_\ii} = O(\Delta t)$.
This directly implies that when $\Delta t \to 0^+$, we have
\begin{eqnarray}\label{eq:spCorr3}
 \rho_\ee  =  \frac{\Cov{X_k ,X_l}}{\Var{X_k}}  \sim  \frac{\Exp{X_k X_l}}{\Exp{X_k^2}} = \frac{\Exp{X_k X_l}}{\Exp{X_k}} 
\quad
\mathrm{and}
\quad
\rho_{\ee\ii} = \frac{\Cov{X_k ,Y_l}}{\sqrt{\Var{X_k} \Var{Y_l}}}  \sim  \frac{\Exp{X_k Y_l}}{\sqrt{\Exp{X_k^2}\Exp{Y_l^2}}} =   \frac{\Exp{X_k Y_l}}{\sqrt{\Exp{X_k}\Exp{Y_l}}}  \, .
\end{eqnarray}
All the stationary expectations appearing above can be computed via the jump distribution of the limit point process emerging in the limit $\Delta t \to 0^+$~\cite{baccelli2003palm}.
Because this limit process is a compound Poisson process with discrete bivariate jumps, the resulting jump distribution
$p_{\ee\ii}$ is specified over $\{1,\ldots,K_\ee\} \times \{1,\ldots,K_\ii\} \setminus \{0,0\}$.
Denoting by $b$ the overall rate of synaptic events, one has $ \lim_{\Delta t \to 0^+} \Exp{X_k Y_l}/\Delta t = b \ExpPei{X_k Y_l}$.
Then by partial exchangeability of the $\{0,1\}$-valued population vectors  $\left\{ X_k \right\}_{1 \leq k \leq K_\ee}$ and  $\left\{ Y_l \right\}_{1 \leq l \leq K_\ii}$, we have
\begin{eqnarray}\label{eq:spCorr3a}
\ExpPei{X_k Y_l} =
\ExpPei{\Exp{X_k Y_l \vert (k_\ee, k_\ii)}}=
\ExpPei{\frac{k_\ee}{K_\ee} \frac{k_\ii}{K_\ii}}=
 \sum_{k=0}^{K_\ee}\sum_{l=0}^{K_\ii} \frac{k}{K_\ee}\frac{l}{K_\ii} p_{\ee\ii,kl}= \frac{\ExpPei{ k_\ee k_\ii}}{K_\ee K_\ii} \, .
\end{eqnarray}
where the bivariate jump $(k_\ee, k_\ii)$ is distributed as $p_{\ee\ii}$.

To further proceed, it is important to note the relation between the expectation $\ExpPei{\cdot}$, which is tied to the overall input process with rate $b$, and the expectation $\ExpPe{\cdot}$ which is tied to the excitatory input process with rate $b_\ee$.
This relation is best captured by remarking that $p_{\ee}$ are not defined as the marginals of $p_{\ee\ii}$, but as only as conditional marginals on $\{ k_\ee>0\}$.
In other words, we have $p_{\ee,k} = (b/b_\ee) \sum_{l=0}^{K_\ii}p_{\ee\ii, kl}$, which implies that $b \ExpPei{X_k X_l}= b_\ee \ExpPe{X_k X_l}$ and $\Exp{X_k} = b \ExpPei{X_k}= b_\ee \ExpPe{X_k}$ with 
\begin{eqnarray}
\label{eq:spCorr3b}
&\displaystyle \ExpPe{X_kX_l} = 
\ExpPe{\Exp{X_k X_l \vert k_\ee}}=
\ExpPe{\frac{k_\ee(k_\ee-1)}{K_\ee(K_\ee-1)} }=
 \sum_{k=0}^{K_\ee} \frac{k(k-1)}{K_\ee(K_\ee-1)} p_{\ee,k}= \frac{\ExpPe{ k_\ee(k_\ee-1)}}{K_\ee (K_\ee-1)} \, ,& \\
&\label{eq:spCorr3c}
\displaystyle 
\ExpPe{X_k} = 
\ExpPe{\Exp{X_k \vert k_\ee}}=
\ExpPe{\frac{k_\ee}{K_\ee} }=
 \sum_{k=0}^{K_\ee}\sum_{l=0}^{K_\ii} \frac{k}{K_\ee} p_{\ee,k}= \frac{\ExpPe{ k_\ee}}{K_\ee} \, ,&
\end{eqnarray}
with similar expressions for the inhibition-related quantities.
Injecting \eref{eq:spCorr3a}, \eref{eq:spCorr3b},  and \eref{eq:spCorr3c} in \eref{eq:spCorr3} yields 
\begin{eqnarray*}
\rho_{\ee} = \frac{\ExpPe{ k_\ee (k_\ee-1)}}{ \ExpPe{ k_\ee} (K_\ee-1)} 
\quad \mathrm{and} \quad 
\rho_{\ee\ii} = \frac{b\ExpPei{ k_\ee k_\ii}}{\sqrt{K_\ee b_\ee\ExpPe{ k_\ee} K_\ii b_\ii \ExpPi{ k_\ii}}} = \frac{\ExpPei{k_\ee k_\ii}}{\sqrt{K_\ee  \ExpPei{k_\ee}K_\ii  \ExpPei{k_\ii}}}\, .
\nonumber
\end{eqnarray*}


\section{Two examples of limit compound Poisson processes}\label{app:CorrProc2}

The probability $P_{\ee\ii,00}$ that plays a central role in Appendix \ref{app:CorrProc1} can be easily computed for zero correlation, i.e., $\rho_{\ee\ii}=0$, by considering a directing measure under product form $F_{\ee \ii}(\theta_\ee, \theta_\ii)=F_\ee(\theta_\ee) F_\ii(\theta_\ii)$.
Then integration with respect to the separable variables $\theta_\ee$ and $\theta_\ii$ yields
\begin{eqnarray}
P_{\ee\ii,kl}
=
P_{\ee, k}P_{\ii, l}
=
\binom{K_\ee}{k}   \frac{B(\alpha_\ee+k,\beta_\ee+K_\ee-k)}{B(\alpha_\ee,\beta_\ee)} \binom{K_\ii}{l} \frac{B(\alpha_\ii+l,\beta_\ii+K_\ii-l)}{B(\alpha_\ii,\beta_\ii)} \, . \nonumber
\end{eqnarray}
In turn, the limit compound Poisson process can be obtain in the limit $\Delta t \to 0^+$ by observing that
\begin{eqnarray}
1-P_{\ee, 0}
=
b_\ee \Delta t + o(\Delta t ) \, , \quad
1-P_{\ee, 0}P_{\ii, 0}
=
b_\ii \Delta t + o(\Delta t ) \, , \quad \mathrm{and} \quad
1-P_{\ee, 0}P_{\ii, 0}
=
(b_\ee + b_\ii)   \Delta t + o(\Delta t ) \, , \nonumber
\end{eqnarray}
which implies that the overall rate is determined as $b = \lim_{\Delta t \to 0^+} (1-P_{\ee, 0}P_{\ii, 0})/\Delta t = b_\ee + b_\ii$, as expected.
To characterize the limit compound Poisson process, it remains to exhibit $p_{\ee \ii, kl}$, the distribution of the jumps $k_\ee$ and $k_\ii$, 
Suppose that $k \geq 1$, then we have
\begin{eqnarray}
p_{\ee \ii, kl} 
&=&
\lim_{\Delta t \to 0^+} \frac{P_{\ee, k}P_{\ii, l}}{1-P_{\ee, 0}P_{\ii, 0}} \, , \nonumber \\
&=&
 \lim_{\Delta t \to 0^+} \left[ \left( \frac{1-P_{\ee, 0}}{1-P_{\ee, 0}P_{\ii, 0}} \right) P_{\ii, l}  \left( \frac{P_{\ee, k}}{1-P_{\ee, 0}}  \right) \right] \, , \nonumber \\
&=&
 \left( \lim_{\Delta t \to 0^+} \frac{1-P_{\ee, 0}}{1-P_{\ee, 0}P_{\ii, 0}}  \right)  \left(\lim_{\Delta t \to 0^+}  P_{\ii, l}  \right)  \left(\lim_{\Delta t \to 0^+} \frac{P_{\ee, k}}{1-P_{\ee, 0}}  \right)  \, . \nonumber 
\end{eqnarray}
Then one can use the limit behaviors
\begin{eqnarray}
 \lim_{\Delta t \to 0^+} \frac{1-P_{\ee, 0}}{1-P_{\ee, 0}P_{\ii, 0}} = \frac{b_\ee}{b_\ee+b_\ii}
 \quad \mathrm{and} \quad
\lim_{\Delta t \to 0^+}  P_{\ii, l} = \one{l=0} \, .
\nonumber
\end{eqnarray}
so that for $k \geq 1$, we have 
\begin{eqnarray}
p_{\ee \ii, kl} 
=
\frac{b_\ee}{b_\ee+b_\ii} \one{l=0} p_{\ee, k} 
\quad \mathrm{with} \quad
 p_{\ee,k} 
 = 
 \lim_{\Delta t \to 0^+} \frac{P_{\ee, k}}{1-P_{\ee, 0}} 
 =
 \binom{K_\ee}{k} \frac{B(k,\beta_\ee + K_\ee - k )}{\psi(K_\ee+\beta_\ee)-\psi(\beta)} \, . \nonumber
\end{eqnarray}
A similar calculation shows that for all $l \geq 1$, we have $p_{\ee \ii, kl} = b_\ii/ (b_\ee + b_\ii) \one{k=0} p_{\ii,l}$.
Thus $p_{\ee \ii, kl}=0$ whenever $k,l \geq 1$, so that the support of $p_{\ee \ii, kl}$ is $ \{1, \ldots, K_\ee \}\times\{0\}  \cup \{0\} \times \{1, \ldots, K_\ii \}$.
This is consistent with the intuition that excitation and inhibition happen at distinct times in the absence of correlations.

Let us now consider the case of maximum correlation for $F_\ee= F_\ii = F$, where $F$ is a beta distribution with parameters $\alpha$ and $\beta$.
Moreover, let us assume the deterministic coupling  $\theta_\ee = \theta_\ii$ such that $F_{\ee \ii}(\theta_\ee, \theta_\ii)=F(\theta_\ee) \delta(\theta_\ii-\theta_\ee)$.
Then, the joint distribution of the jumps $(k_\ee,k_\ii)$ can be evaluated via direct integration as
\begin{eqnarray}
P_{\ee \ii, kl} 
&=&
\binom{K_\ee}{k} \binom{K_\ii}{l} \int \theta_\ee^{k} (1-\theta_\ee)^{K_\ee-k}  \theta_\ee^{l} (1-\theta_\ii)^{K_\ii-l} \, dF(\theta_\ee) \delta(\theta_\ii - \theta_\ee) \, , \nonumber\\
&=&
\binom{K_\ee}{k} \binom{K_\ii}{l}  \int \theta^{k+l} (1-\theta)^{K_\ee+K_\ii-k-l}  \, dF(\theta) \, ,
\nonumber\\
&=&
\binom{K_\ee}{k} \binom{K_\ii}{l}  \frac{B(\alpha+k+l,\beta + K_\ee + K_\ii - k - l)}{B(\alpha,\beta)} \, . \nonumber
\end{eqnarray}
As excitation and inhibition are captured separately by the same marginal functions $F_\ee= F_\ii = F$, we necessarily have
$\alpha/(\alpha+\beta)=\Exp{X_k}=\Exp{Y_l}=r_\ee \Delta t=r_\ii \Delta t$ and we refer to the common spiking rate as $r$.
Then the overall rate of synaptic events is obtained as
\begin{eqnarray}\label{eq:beiDef}
b
=
\lim_{\Delta t \to 0^+} \frac{1-P_{\ee \ii, 00}}{\Delta t}
=
\lim_{\alpha \to 0^+} \frac{1-P_{\ee \ii, 00}}{\alpha}
\lim_{\Delta t \to 0^+} \frac{\alpha}{\Delta t}
=
 \left( \psi(K_\ee+K_\ii+\beta) - \psi(\beta)\right)\beta r \, ,
\end{eqnarray}
and one can check that $b$ differs from the excitatory- and inhibitory-specific rates $b_\ee$ and $b_\ii$, which satisfy
 \begin{eqnarray}\label{eq:bebiDef}
b_\ee
=
\lim_{\Delta t \to 0^+} \frac{1-P_{\ee, 0}}{\Delta t}
=
 \left( \psi(K_\ee+\beta) - \psi(\beta)\right)\beta r 
 \quad \mathrm{and} \quad
b_\ii
=
\lim_{\Delta t \to 0^+} \frac{1-P_{\ii, 0}}{\Delta t}
=
 \left( \psi(K_\ii+\beta) - \psi(\beta)\right)\beta r \, .
\end{eqnarray}
To characterize the limit compound Poisson process, it remains to exhibit $p_{\ee \ii, kl}$, the joint distribution of the jumps $(k_\ee,k_\ii)$. 
A similar calculation as for the case of excitation alone yields
\begin{eqnarray}
p_{\ee \ii, kl}
=
\lim_{\Delta t \to 0^+} \frac{P_{\ee \ii, kl} }{1-P_{\ee \ii, 00}}
=
\binom{K_\ee}{k} \binom{K_\ii}{l}  \frac{B(k+l,\beta + K_\ee + K_\ii - k - l)}{\psi(K_\ee+K_\ii + \beta)-\psi(\beta)} \, . \nonumber
\end{eqnarray}
Remember that within our model, spiking correlations do not depends on the number of neurons and that by construction we have $\rho_{\ee\ii} \leq \sqrt{\rho_{\ee}\rho_{\ii}}$.
Thus, for the symmetric case under consideration, maximum correlation corresponds to $\rho_{\ee\ii} = \rho_{\ee} = \rho_{\ii}=1/(1+\beta)$.
In particular perfect correlation between excitation and inhibition can only be attained for $\beta \to 0$.
When $\beta>0$, i.e., for partial correlations, the Poisson processes $N_\ee$ and $N_\ii$ only share a fraction of their times, yielding an aggregate Poisson process $N$ such that $\min(b_\ee,b_\ii)<b<b_\ee+b_\ii$.   
The relations between $b$, $b_\ee$, and $b_\ii$ can be directly recovered from the knowledge of $p_{\ee\ii}$ by observing that
\begin{eqnarray}
\Prob{k_\ee=0,k_\ii>0}&=&\sum_{l=1}^{K_\ii} p_{\ee \ii, 0l}  =  \frac{\psi(K_\ee+K_\ii+\beta) - \psi(K_\ee+\beta)} {\psi(K_\ee+K_\ii+\beta) - \psi(\beta)} \nonumber\\
\Prob{k_\ii=0,k_\ee>0}&=&\sum_{k=1}^{K_\ee} p_{\ee \ii, k0} =  \frac{\psi(K_\ee+K_\ii+\beta) - \psi(K_\ii+\beta)} {\psi(K_\ee+K_\ii+\beta) - \psi(\beta)} \, .
\nonumber \\
\Prob{k_\ii>0,k_\ee>0}&=&\sum_{k=1}^{K_\ee} \sum_{l=1}^{K_\ii} p_{\ee \ii, kl} =  1-\frac{2\psi(K_\ee+K_\ii+\beta) - \psi(K_\ee+\beta)-\psi(K_\ii+\beta)} {\psi(K_\ee+K_\ii+\beta) - \psi(\beta)} \, .
\nonumber 
\end{eqnarray}
This implies that the fraction of times with nonzero excitation is given by
\begin{eqnarray}
\Prob{k_\ee>0}
&=&
\Prob{k_\ee>0,k_\ii=0}+\ProbP{k_\ee>0,k_\ii>0} 
=
\frac{\psi(K_\ee+\beta) - \psi(\beta)} {\psi(K_\ee+K_\ii+\beta) - \psi(\beta)}  \, , \nonumber
\end{eqnarray}
so that we consistently recover the value of $b_\ee$ already obtained in \eref{eq:bDef} and \eref{eq:bebiDef} via
\begin{eqnarray}
b_\ee T&=& \Exp{N_\ee(T)}= \Exp{\one{k_\ee>0} N(T)}= bT\ExpPei{\one{k_\ee>0}}=bT\ProbP{k_\ee>0} \, .\nonumber
\end{eqnarray}


\section{Marcus jump rule}\label{app:Marcus}

The goal of this appendix is to justify the Marcus-type update rule given in \eref{eq:MarcusJump}.
To do so let us first remark that given a finite time interval $[0,T]$, the number of synaptic activation times  $\{T_n \}_{n \in \mathbbm{Z}}$ falling in this interval is almost surely finite.
In particular, we have $\Delta = \inf_{0\leq T_n \neq T_m \leq T} \vert T_n -T_m \vert > 0$ almost surely.
Consequently, taking $\epsilon < \Delta/ \tau_\mathrm{s}$ ensures that synaptic activation events do not overlap in time, so that it is enough to consider a single synaptic activation triggered with no lack of generality in $T_0=0$.
Let us denote the voltage just before the impulse onset as $V(T_0^-)=V_{0}$, which will serve as initial condition for the ensuing voltage dynamics.
As the dimensionless conductances remain equals to $W_e/\epsilon$ and $W_i/\epsilon$ for a duration $[0,\epsilon \tau]$, the voltage $V_\epsilon$ satisfies
\begin{eqnarray}
\tau \dot{V_\epsilon} = -V_\epsilon+(W_\ee/\epsilon) (V_\ee-V_\epsilon)+(W_\ii/\epsilon) (V_\ii-V_\epsilon) \, , \quad 0 \leq t \leq \epsilon \tau \, ,
\nonumber
\end{eqnarray}
where we assume $I=0$ for simplicity. The unique solution satisfying $V(0^-)=V_{0}$ is 
\begin{eqnarray}
V_\epsilon(t) = V_{0} e^{-t/\tau \left( 1+W_\ee/\epsilon+W_\ii/\epsilon \right)}
+ \frac{W_\ee V_\ee+ W_\ii V_\ii}{\epsilon + W_\ee + W_\ii}\left(1-e^{-t/\tau \left( 1+W_\ee/\epsilon+W_\ii/\epsilon \right)}\right)
\, , \quad 0 \leq t \leq \epsilon \tau \, .
\nonumber
\end{eqnarray}
The Marcus-type rule follows from evaluating the jump update as the limit
\begin{eqnarray}
\lim_{\epsilon \to 0^+} V_\epsilon(\epsilon \tau)-V_{0}
&=&
\lim_{\epsilon \to 0^+}
\left\{ V_{0} \left( e^{- \left( \epsilon+W_\ee+W_\ii\right)} -1 \right)
+ \frac{W_\ee V_\ee+ W_\ii V_\ii}{\epsilon + W_\ee + W_\ii}\left(1-e^{- \left( \epsilon+W_\ee+W_\ii \right)}\right) \right\} \, , \nonumber\\
&=&
\left( \frac{W_\ee V_\ee+ W_\ii V_\ii}{W_\ee + W_\ii} -V_{0} \right)\left(1-e^{- \left(W_\ee+W_\ii \right)}\right) \, , \nonumber
\end{eqnarray}
which has the same form as the rule announced in \eref{eq:MarcusJump}.
Otherwise, at fixed $\epsilon>0$, the fraction of time for which the voltage $V_\epsilon$ is exponentially relaxing toward the leak reversal potential $V_\leak=0$ is larger than $1-N\epsilon/T$, where $N$ denotes the almost surely finite number of synaptic activations, which does not depends on $\epsilon$.
Thus, the voltage $V= \lim_{\epsilon \to 0^+} V_\epsilon$ exponentially relaxes toward $V_\leak=0$, except when it has jump discontinuities in $\{T_n \}_{n \in \mathbbm{Z}}$.


\section{Stationary voltage mean}\label{app:mean}

For a positive synaptic activation time $t>0$, the classical method of the variation of the constant applies to solve \eref{eq:Vdyn}.
This yields an expression for $V_\epsilon(t)$ in terms of regular Riemann-Stieltjes integrals where the conductance traces $h_\ee(t)$ and $h_\ii(t)$ are treated as a form of  deterministic quenched disorder.
Specifically, given an initial condition $V_\epsilon(0)$, we have 
\begin{eqnarray}
V_\epsilon(t) 
= 
V_\epsilon(0) e^{ -\int_0^t \frac{1}{\tau}+h_\ee(u) +h_\ii(u) \, \dd u } 
+ \int_0^t   \left(V_\ee h_\ee(u) \! + \! V_\ii h_\ii(u) \! + \! I/C \right) e^{ -\int_u^t \frac{1}{\tau}+h_\ee(v) +h_\ii(v) \, \dd v } \, \dd u  \, .  \nonumber
\end{eqnarray}
where $V_\epsilon(t)$ depends on $\epsilon$ via the all-or-none-conductance processes $h_\ee$ and $h_\ii$.
As usual, the stationary dynamics of the voltage $V_\epsilon$ is recovered by considering the limit of arbitrary large times $t \to \infty$, for which one can neglect the influence of the initial condition $V_\epsilon(0)$.
Introducing the cumulative input processes $H=(H_\ee, H_\ii)$ defined by 
\begin{eqnarray}
\big(H_\ee(t),H_\ii(t)\big)= \left( \int_0^t h_{\ee}(u) \, du , \int_0^t h_{\ii}(u) \, du  \right) \, .
\nonumber
\end{eqnarray}
and satisfying $\tau \dd H_\ee(t) =  h_\ee(t) \, \dd t$ and $\tau \dd H_\ii(t) =  h_\ii(t) \, \dd t$,
we have
\begin{eqnarray}\label{eq:stat}
V_\epsilon = 
 && \int_{-\infty}^0   e^{\frac{t}{\tau}+H_\ee(t)+H_\ii(t)}\left( \dd [V_\ee H_\ee(t)+V_\ii H_\ii(t)]+\frac{I }{G} \frac{\dd t}{\tau}\right) \, . 
\end{eqnarray}
In turn, expanding the integrand above yields the following expression for the stationary expectation of the voltage
\begin{eqnarray}\label{eq:expmean}
\hspace{-10pt} \Exp{V_\epsilon} 
= V_\ee \int_{-\infty}^0  e^{\frac{t}{\tau}} \Exp{e^{ H_\ee(t)+H_\ii(t)}\, \dd H_\ee(t) }  
+ V_\ii \int_{-\infty}^0  e^{\frac{t}{\tau}} \Exp{ e^{ H_\ee(t)+H_\ii(t)}\, \dd H_\ii(t) }  
+ \frac{I }{G}  \int_{-\infty}^0   e^{\frac{t}{\tau}} \Exp{e^{ H_\ee(t)+H_\ii(t)} }\, \frac{\dd t}{\tau} \, .
\end{eqnarray}
Our primary task is to evaluate the various stationary expectations appearing in the above formula.
Such a goal can be achieved analytically for AONCB models.
As the involved calculations tend to be cumbersome, we only give a detailed account in Appendix.
Here we account for the key steps of the calculation, which ultimately produces an interpretable compact formula for $\Exp{V_\epsilon}$ in the limit of instantaneous synapses, i.e., when $\epsilon \to 0$.

In order to establish this compact formula, it is worth introducing the stationary bivariate function
\begin{eqnarray}\label{eq:keyQuant}
Q_\epsilon(t,s)=\Exp{e^{H_\ee(t) +H_\ii(s) }} \, ,
\end{eqnarray}
which naturally depends on $\epsilon$ via  $H_\ee(t)$ and $H_\ii(s)$.
The function $Q_\epsilon$ is of great interest because all the stationary expectations at stake in \eref{eq:expmean} can be derived from it.
Before justifying this point, an important observation is that the expectation defining $Q_\epsilon(t,s)$ only bears on the cumulative input processes $H_\ee$ and $H_\ii$, which specify bounded, piecewise continuous functions with probability one, independent of $\epsilon$.
As a result of this regular behavior, the expectation commute with the limit of instantaneous synapses allowing one to write
\begin{eqnarray}
Q(t,s)
= 
\lim_{\epsilon \to 0^+} Q_\epsilon(t,s)
=
\Exp{e^{  \lim_{\epsilon \to 0} H_\ee(t) +H_\ii(s)}}  \, ,
=
\Exp{e^{- Z_\ee(t) -  Z_\ii(s)}}  \, , \nonumber
\end{eqnarray}
where we exploit the fact that the cumulative input processes $H_\ee$ and $H_\ii$  converge toward the coupled compound Poisson processes $Z_\ee$ and $Z_\ii$ when $\epsilon \to 0^+$:
\begin{eqnarray}
Z_\ee(t) = \sum_n^{N_\ee(t)} W_{\ee,n} \quad \mathrm{and} \quad Z_\ii(t) = \sum_n^{N_\ii(t)} W_{\ii,n} \, .
\end{eqnarray}
The above remark allows one to compute the term due to current injection $I$  in \eref{eq:expmean}, where the expectation can be identified to $Q_\epsilon(t,t)$.
Indeed, utilizing the standard form for the moment-generating function for compound Poisson processes~\cite{daley2003introduction}, we find that
\begin{eqnarray}
Q(t,t)
=
e^{a_{\ee \ii,1}t/\tau} \, ,\nonumber
\end{eqnarray}
where we introduce the first-order aggregate efficacy
\begin{eqnarray}\label{eq:aei1}
a_{\ee \ii,1} =  b \tau \left(1-\ExpPei{e^{-(W_\ee+W_\ii)}} \right) \nonumber \, .
\end{eqnarray}
Remember that in the above definition, $\ExpPei{\cdot}$ denotes the expectation with respect to the joint probability of the conductance jumps, i.e., $p_{\ee\ii}$.

It remains to evaluate the expectations associated to excitation and inhibition reversal potentials  in \eref{eq:expmean}.
These terms differ from the current-associated term in that they involve expectations of stochastic integrals with respect to the cumulative input processes $H_{\eeii}$.
This is by contrast with evaluating \eref{eq:keyQuant}, which only involves expectations of functions that depends on $H_{\eeii}$.
In principle, one could still hope to adopt a similar route as for the current associated term, exploiting the compound Poisson process $Z$ obtained in the limit of instantaneous synapses.
However, such an approach would require that the operations of taking the limit of instantaneous synapses and evaluating the stationary expectation still commute.
This is a major caveat as such a commuting relation generally fails for point-process-based stochastic integrals.
Therefore, one has to analytically evaluate the expectations at stake for positive synaptic activation time $\epsilon>0$, without resorting to the simplifying limit of instantaneous synapses.
This analytical requirement is the primary motivation to consider AONCB models.

The first step in the calculation is to realize that for $\epsilon>0$, the conductance traces $h_\ee(t)=\tau \dd H_\ee(t)/\dd t$ and $h_\ii(t)=\tau \dd H_\ii(t)/\dd t$ are bounded, piecewise continuous functions with probability one.
Under these conditions, it then holds that
\begin{eqnarray}
\lim_{s \to t } \partial_t Q_\epsilon(t,s) = \Exp{\frac{\dd H_\ee(t)}{\dd t} \, e^{ H_\ee(t)+H_\ii(t) }} 
\quad \mathrm{and} \quad
\lim_{s \to t } \partial_s Q_\epsilon(t,s) = \Exp{\frac{\dd H_\ii(t)}{\dd t} \, e^{ H_\ee(t)+H_\ii(t) }} \, , \nonumber
\end{eqnarray}
so that the sought-after expectations can be deduced from the closed-form knowledge of $Q_\epsilon(t,s)$ for positive $\epsilon>0$.
The analytical expression of $Q_\epsilon(t,s)$ can be obtained via careful manipulation of the processes $H_\ee$ and $H_\ii$ featured in the exponent of \eref{eq:keyQuant} (see Appendix \ref{app:QE}).
In a nutshell, these manipulations hinge on splitting the integrals defining $H_\ee(t)$ and $H_\ii(s)$ into independent contributions arising from spiking events occurring in the five nonoverlapping, contiguous intervals bounded by the times $0 \geq -\epsilon \tau \geq t \geq s \geq t-\epsilon \tau \geq s-\epsilon \tau$.
There is no loss of generality in assuming the latter ordering and from the corresponding analytical expression, we can compute
\begin{eqnarray}
\lim_{\epsilon \to 0^+}\lim_{s \to t } \partial_t Q_\epsilon(t,s) = ba_{\ee,1} e^{ a_{\ee \ii,1} t/\tau} \quad \mathrm{and} \quad
\lim_{\epsilon \to 0^+}\lim_{s \to t } \partial_s Q_\epsilon(t,s) = ba_{\ii,1} e^{ a_{\ee \ii,1} t/\tau} \, , \nonumber
\end{eqnarray}
where the effective first-order synaptic efficacies via \eref{eq:a1} as
\begin{eqnarray}
a_{\ee,1} = b \tau \ExpPei{\frac{W_\ee}{W_\ee+W_\ii}\left(1-e^{-(W_\ee+W_\ii)} \right)}  \quad \mathrm{and} \quad
a_{\ii,1} = b \tau \ExpPei{\frac{W_\ii}{W_\ee+W_\ii}\left(1-e^{-(W_\ee+W_\ii)} \right)} \, . \nonumber
\end{eqnarray}
Observe that by definition, $a_{\ee,1}$ and $a_{\ii,1}$ satisfy $a_{\ee,1} +a_{\ii,1}=a_{\ee \ii,1}$.

Altogether,  upon evaluation of the integrals featured in \eref{eq:expmean}, these results allow one to produce the compact expression \eref{eq:statmean}  for the stationary voltage mean in the limit of instantaneous synapses:
\begin{eqnarray}
\Exp{V} 
=
\lim_{\epsilon \to 0^+}\Exp{V_\epsilon}
=
\frac{
 a_{\ee,1} V_\ee+a_{\ii,1}  V_\ii  + I/G
}{
1+ a_{\ee,1} +a_{\ii,1}
} \, . \nonumber
\end{eqnarray}



\section{Stationary voltage variance}\label{app:var}

The calculation of the stationary voltage variance is more challenging than that of the stationary voltage mean.
However, in the limit of instantaneous synapses, this calculation produces a compact, interpretable formula as well.
Adopting a similar approach as for the stationary mean calculation, we start by expressing $V^2_\epsilon$ in the stationary limit in terms of a stochastic integrals
involving the cumulative input processes $H_\ee$ and $H_\ii$.
Specifically, using \eref{eq:stat}, we have
\begin{eqnarray}\label{eq:expvar1}
V^2_\epsilon 
&=&  \left( \int_{-\infty}^0  e^{\frac{t}{\tau}+H_\ee(t)+H_\ii(t)} \left(\dd (V_\ee H_\ee(t)+V_\ii H_\ii(t))+\frac{I }{G} \frac{\dd t}{\tau}\right) \right)^2\nonumber \, ,\\
&=&  \iint_{\mathbbm{R}_-^2} e^{\frac{t+s}{\tau} + H_\ee(t)+H_\ii(t) + H_\ee(s)+H_\ii(s)} 
\left(  \dd (V_\ee H_\ee(t)+V_\ii H_\ii(t))+\frac{I }{G} \frac{\dd t}{\tau}\right) 
\left(  \dd (V_\ee H_\ee(s)+V_\ii H_\ii(s))+\frac{I }{G} \frac{ds}{\tau}\right)   \, .
\end{eqnarray}
Our main goal is to compute the stationary expectation of the above quantity.
As for the stationary voltage mean, our strategy is $(i)$ to derive the exact stationary expectation of the integrands for finite synaptic activation time, $(ii)$ to evaluate these integrands in the simplifying limit of instantaneous synapses, and $(iii)$ to rearrange the terms obtained after integration into an interpretable final form. 
Enacting the above strategy is a rather tedious task, and as for the calculation of the mean voltage, we only present the key steps of the calculation in the following.

The integrand terms at stake are obtained by expanding \eref{eq:expvar1}, which yields the following quadratic expression for the stationary second moment of the voltage
\begin{eqnarray}\label{eq:quad}
\Exp{V_\epsilon^2} 
=
 A_{\ee,\epsilon} V_\ee^2 + B_{\ee \ii,\epsilon}  V_\ee V_\ii  + A_{\ii,\epsilon}  V_\ii^2   + \left( V_\ee B_{\ee I,\epsilon} + V_\ii B_{\ii I,\epsilon} \right) (I/G)+ A_{I,\epsilon} (I/G)^2 \, , \nonumber
\end{eqnarray}
whose various coefficients need to be evaluated.
These coefficients are conveniently specified in terms of the following symmetric random function
\begin{eqnarray}
\mathcal{E}_{\ee \ii}(t,s) 
=  
e^{H_\ee(t)+H_\ii(t)+H_\ee(s)+H_\ii(s)} \,  . \nonumber
\end{eqnarray}
which features prominently in \eref{eq:expvar1}.
Moreover, drawing on the calculation of the stationary mean voltage, we anticipate that the quadrivariate version of $\mathcal{E}_{\ee \ii}(t,s)$ will play a central role in the calculation via its stationary expectation.
Owing to this central role, we denote this expectation as
\begin{eqnarray}
R_\epsilon(t,u,s,v) = \Exp{e^{H_\ee(t)+H_\ii(u)+H_\ee(s)+H_\ii(v)}} \, . \nonumber
\end{eqnarray}
where we make the $\epsilon$-dependence explicit. 
As a mere expectation with respect to the cumulative input processes $(H_\ee,H_\ii)$, the expectation can be evaluated in closed form for AONCB models.
This again requires careful manipulations of the processes $H_\ee$ and $H_\ii$, which need to split into independent contributions arising from spiking events occurring in nonoverlapping intervals.
By contrast with the bivariate case, the quadrivariate case requires to consider nine contiguous intervals.
There is no loss of generality to consider these interval bounds to be determined by the two following time orderings:
\begin{description}
\item[$O$-order] $0 \geq -\epsilon \tau \geq t \geq u \geq t-\epsilon \tau \geq u-\epsilon \tau \geq s \geq v \geq s-\epsilon \tau \geq v-\epsilon \tau$ ,
\item[$D$-order] $0 \geq -\epsilon \tau \geq t \geq u \geq s \geq v \geq t-\epsilon \tau \geq u-\epsilon \tau  \geq s-\epsilon \tau \geq v-\epsilon \tau$ ,
\end{description}
where $O$ stands for off-diagonal ordering and $D$ for diagonal ordering.

The reason to only consider the $O/D $-orders is that all the relevant calculations will be made in the limit $(u,v) \to (t,s)$. By symmetry of $R_\epsilon(t,u,s,v)$, it is then enough to restrict our consideration to the limit $(u,v) \to (t^-,s^-)$, which leaves the choice of $t,s \leq 0$ to be determined.
By symmetry, one can always choose $t>s$, so that the only remaining alternative is to decide wether $(t,s)$ belong to the diagonal region $\mathcal{D}_\epsilon= \{ t,s \leq 0 \, \vert \, \epsilon \tau \geq  \vert t-s \vert  \}$ or the off-diagonal region $\mathcal{O}_\epsilon = \{ t,s \leq 0 \, \vert \,  \epsilon \tau <\vert t-s \vert  \}$.
For the sake of completeness, we give the two expressions of $R_\epsilon(t,u,s,v)$ on the regions $\mathcal{O}_\epsilon$ and $\mathcal{D}_\epsilon$ in Appendix \ref{app:RO}.
Owing to their tediousness, we do not give the detailed calculations leading to these expressions, which are lengthy but straightforward elaborations on those used in Appendix \ref{app:QE}. 
Here we stress that for $\epsilon>0$, these expressions reveal that $R_\epsilon(t,u,s,v)$ is defined as a twice-differentiable quadrivariate function.\\

With these remarks in mind, the coefficients featured in \eref{eq:quad} can be categorized in three classes:\\

$1.$ There is a single current-dependent inhomogeneous coefficient 
\begin{eqnarray}
A_{I,\epsilon} &=& \iint_{\mathbbm{R}_-^2}  e^{\frac{t+s}{\tau}} \Exp{\mathcal{E}_{\ee \ii}(t,s)}  \, \frac{\dd t \, ds}{\tau^2} \, , \nonumber
\end{eqnarray}
where we recognize that $\Exp{\mathcal{E}_{\ee \ii}(t,s)}=R_\epsilon(t,t,s,s)\stackrel{\mathrm{def}}{=}R_\epsilon(t,s)$.
As $R_\epsilon(t,s)$ is merely a stationary expectation with respect to the cumulative input processes $(H_\ee,H_\ii)$, it  can be directly evaluated in the limit of instantaneous synapses. 
In other word, step $(ii)$ can be performed before step $(i)$, similarly as for the stationary voltage mean calculation.
However, having a general analytical expression for $R_\epsilon(t,u,s,v)$ on $\mathcal{O}_\epsilon$ (see Appendix \ref{app:RO}), we can directly evaluate for all $t \neq s$ that
\begin{eqnarray}\label{eq:Rts}
R(t,s) 
= \lim_{\epsilon \to 0^+} R_\epsilon(t,s) 
= e^{\left( 2a_{\ee \ii,2} \max(t,s) - a_{\ee \ii,1} \vert t-s \vert \right)/\tau }  \, ,
\end{eqnarray}
where we define the second-order aggregate efficacy 
\begin{eqnarray}
a_{\ee \ii,2} = \frac{b \tau}{2} \left( 1 - \ExpPei{e^{-2(W_\ee+W_\ii)}} \right) \, . \nonumber
\end{eqnarray}
It is clear that the continuous function $R(t,s)$ is smooth everywhere except on the diagonal where it admits a slope discontinuity.
As we shall see, this slope discontinuity is the reason why one needs to consider the $\mathcal{D}_\epsilon$ region carefully, even when only concerned with the  limit $\epsilon \to 0^+$.
That being said, the diagonal behavior plays no role here and straightforward integration of $R(t,s)$ on the negative orthant gives
\begin{eqnarray}
A_I 
=
\lim_{\epsilon \to 0^+}
A_{I,\epsilon}
= 
\frac{1}{(1 + a_{\ee \ii,1})(1+ a_{\ee \ii,2})} \, . \nonumber
\end{eqnarray}

$2.$ There are two current-dependent linear coefficients
\begin{eqnarray}
B_{\ee I,\epsilon} =  2\iint_{\mathbbm{R}_-^2}  e^{\frac{t+s}{\tau}} \Exp{\mathcal{E}_{\ee \ii}(t,s) \dd H_\ee(t)} \frac{\dd s}{\tau} \quad \mathrm{and} \quad
B_{\ii I,\epsilon} =  2\iint_{\mathbbm{R}_-^2}  e^{\frac{t+s}{\tau}} \Exp{\mathcal{E}_{\ee \ii}(t,s) \dd H_\ii(t)} \frac{\dd s}{\tau} \, , \nonumber
\end{eqnarray}
where the coefficient $2$ above comes from the fact that $B_{\ee I,\epsilon}$ and $B_{\ii I,\epsilon}$ are actually resulting from the contributions of two symmetric terms in the expansion of \eref{eq:expvar1}.
Both $B_{\ee I,\epsilon}$ and $B_{\ii I,\epsilon}$ involve expectations of stochastic integrals akin to those evaluated for the stationary mean calculation.
Therefore, these terms can be treated similarly by implementing step $(i)$ and $(ii)$ sequentially.
The trick is to realize that for positive $\epsilon$ and $t \neq s \leq 0$, it holds that 
\begin{eqnarray}
\Exp{\mathcal{E}_{\ee \ii}(t,s) \, \frac{\dd H_\ee(t)}{\dd t} }  = \lim_{u \to t } \partial_t R_\epsilon(t,u, s,s)   \quad \mathrm{and} \quad
\Exp{\mathcal{E}_{\ee \ii}(t,s) \, \frac{\dd H_\ii(t)}{\dd t} } = \lim_{v \to s } \partial_s R_\epsilon(t,t,s,v)  \, . \nonumber
\end{eqnarray}
Thus for any $(t, s)$ in the off-diagonal region $O_\epsilon$, the analytical knowledge of $R_\epsilon(t,u,s,v)$  (see Appendix \ref{app:RO})
allows one to evaluate 
\begin{eqnarray} \label{eq:kernel}
 \lim_{u \to t^- } \tau \frac{\partial_t R_\epsilon(t,u, s,s)}{R_\epsilon(t, s)} = 
\left\{
\begin{array}{ccc}
a_{\ee,1}  &  \mathrm{if} &  t>s  \, , \\
a_{\ee,2}-a_{\ee,1} & \mathrm{if} &   t<s  \, ,
\end{array}
\right. 
\quad \mathrm{and} \quad
 \lim_{v \to s^- } \tau \frac{\partial_s R_\epsilon(t,u, s,s)}{R_\epsilon(t, s)} = 
\left\{
\begin{array}{ccc}
 a_{\ii,1}   &  \mathrm{if} &  t>s  \, , \\
 a_{\ii,2}-a_{\ii,1}  & \mathrm{if} &   t<s  \, ,
\end{array}
\right.
\end{eqnarray}
where the second-order synaptic efficacies are defined as
\begin{eqnarray}
\label{eq:a2}
a_{\ee,2} = \frac{b \tau}{2} \ExpPei{ \frac{W_\ee}{W_\ee+W_\ii} \left(1- e^{-2(W_\ee+W_\ii)}\right)} 
\quad \mathrm{and} \quad
a_{\ii,2} &=& \frac{b \tau}{2} \ExpPei{ \frac{W_\ii}{W_\ee+W_\ii} \left(1- e^{-2(W_\ee+W_\ii)}\right)} \, .
\end{eqnarray}
Observe that these efficacies satisfy the familiar relation $a_{\ee,2}+a_{\ii,2}=a_{\ee \ii,2}$.
Taking the limits of \eref{eq:kernel} when $\epsilon \to 0^+$ specify two bivariate functions that are continuous everywhere, except on the diagonal $t=s$, where these functions present a jump discontinuity.
This behavior is still regular enough to discard any potential contributions from diagonal terms, so that we can restrict ourselves to the region $O_\epsilon$.
Then, taking the limit $\epsilon \to 0^+$ after integration of over $O_\epsilon$, we find that 
\begin{eqnarray}
B_{\ee I} 
= 
\lim_{\epsilon \to 0^+}
B_{\ee I,\epsilon}
= 
\frac{a_{\ee,2}}{(1+a_{\ee \ii,1})(1+a_{\ee \ii,2})}
\quad \mathrm{and} \quad
B_{\ii I} 
=
\lim_{\epsilon \to 0^+}
B_{\ii I,\epsilon}
=
\frac{b\tau a_{\ii,2}}{(1+a_{\ee \ii,1})(1+a_{\ee \ii,2})} \, . \nonumber
\end{eqnarray}

$3.$ There are four quadratic coefficients associated to the reversal-potential $V_\ee$ and $V_\ii$, including two diagonal terms
\begin{eqnarray}
A_{\ee,\epsilon}
=
\iint_{\mathbbm{R}_-^2} e^{\frac{t+s}{\tau}} \Exp{\mathcal{E}_{\ee \ii}(t,s) \dd H_\ee(t) \dd H_\ee(s) } \quad \mathrm{and} \quad
A_{\ii,\epsilon}
=  
\iint_{\mathbbm{R}_-^2} e^{\frac{t+s}{\tau}} \Exp{\mathcal{E}_{\ee \ii}(t,s) \dd H_\ii(t) \dd H_\ii(s) } \, , \nonumber
\end{eqnarray}
and two symmetric cross terms contributing
\begin{eqnarray}
B_{\ee \ii,\epsilon}&=&  2 \iint_{\mathbbm{R}_-^2}  e^{\frac{t+s}{\tau}}  \Exp{\mathcal{E}_{\ee \ii}(t,s)   \dd H_\ee(t) \dd H_\ii(s) } \, . \nonumber
\end{eqnarray}
Notice that it is enough to compute only one diagonal term as the other term can be deduced by symmetry.
Following the same method as for the linear terms, we start by remarking that for all $(t,s)$ in the off-diagonal region $\mathcal{O}_\epsilon$, it holds that 
\begin{eqnarray}
\Exp{\mathcal{E}_{\ee \ii}(t,s) \, \frac{\dd H_\ee(t)}{\dd t} \frac{\dd H_\ee(s)}{\dd s}}  &=& \lim_{(u,v) \to (t,s) } \partial_t \partial_s R_\epsilon(t,u, s,v)   \, ,
\, \nonumber\\
\Exp{\mathcal{E}_{\ee \ii}(t,s) \, \frac{\dd H_\ee(t)}{\dd t} \frac{\dd H_\ii(s)}{\dd s}} &=&  \lim_{(u,v) \to (t,s)} \partial_t \partial_v R_\epsilon(t,u,s,v)  \, , \nonumber
\end{eqnarray}
As before,  the analytical knowledge of $R_\epsilon(t,u,s,v)$ on the $O_\epsilon$ region  (see Appendix \ref{app:RO})
allows one to evaluate 
\begin{eqnarray}
\lim_{(u,v) \to (t,s)^-}  \tau^2 \frac{ \partial_t \partial_u R_\epsilon(t,u,s,s)}{R_\epsilon(t, s)} &=& a_{\ee,1}(2a_{\ee,2}-a_{\ee,1})  \, ,  \nonumber\\
\lim_{(u,v) \to (t,s)^-}  \tau^2 \frac{\partial_t \partial_s R_\epsilon(t,u,s,v)}{R_\epsilon(t, s)}  &=& \frac{1}{2}( a_{\ee,1}(2a_{\ii,2}-a_{\ii,1}) +  a_{\ii,1}(2a_{\ee,2}-a_{\ee,1})) \, .  \nonumber
\end{eqnarray}
The above closed-form expressions allow one to compute $A_{\ee,\epsilon}'$ and $B_{\ee \ii,\epsilon}'$, the part of the coefficients $A_{\ee,\epsilon}$ and $B_{\ee \ii,\epsilon}$ resulting from integration over the off-diagonal region $O_\epsilon$, which admit  well-defined limit values $A_{\ee}' =\lim_{\epsilon \to 0^+} A_{\ee,\epsilon}'$ and $B_{\ee\ii}' =\lim_{\epsilon \to 0^+} B_{\ee\ii,\epsilon}'$ with:
\begin{eqnarray}
A_{\ee}' 
&=&
\lim_{\epsilon \to 0^+}  \iint_{\mathcal{O}_\epsilon} e^{\frac{t+s}{\tau}} \Exp{\mathcal{E}_{\ee \ii}(t,s) \dd H_\ee(t) \dd H_\ee(s) } 
= \frac{ a_{\ee,1}(2a_{\ee,2}-a_{\ee,1}) }{(1+ a_{\ee \ii,1})(1+b\tau a_{\ee \ii,2})} \, , \nonumber\\
B_{\ee \ii}'
&=& 2 \lim_{\epsilon \to 0^+} \iint_{\mathcal{O}_\epsilon} e^{\frac{t+s}{\tau}} \Exp{\mathcal{E}_{\ee \ii}(t,s) \dd H_\ee(t) \dd H_\ii(s) } 
= \frac{ a_{\ee,1}(2a_{\ii,2}-a_{\ii,1}) +  a_{\ii,1}(2a_{\ee,2}-a_{\ee,1}) }{(1+ a_{\ee \ii,1})(1+ a_{\ee \ii,2})} \, .  \nonumber
\end{eqnarray}
However, for quadratic terms, one also needs to include the contributions arising from the diagonal region $\mathcal{D}_\epsilon$, as suggested be the first-order jump discontinuity of $R(t,s)=\lim_{\epsilon \to 0^+}R_\epsilon(t,s)$ on the diagonal $t=s$.
To confirm this point, one can show from the analytical expression of $R_\epsilon(t,u,s,v)$ on $\mathcal{D}_\epsilon$ (see Appendix \ref{app:RO}), that all relevant second-order derivative terms scale as $1/\epsilon$ over $\mathcal{D}_\epsilon$.
This scaling leads to the nonzero contributions $A_{\ee,\epsilon}''$ and $B_{\ee \ii,\epsilon}''$ resulting form the integration of these second-order derivative terms over the diagonal region $\mathcal{D}_\epsilon$, even in the limit $\epsilon \to 0^+$.
Actually, we find that these contributions also admit  well-defined limit values $A_{\ee}'' =\lim_{\epsilon \to 0^+} A_{\ee,\epsilon}''$ and $B_{\ee\ii}'' =\lim_{\epsilon \to 0^+} B_{\ee\ii,\epsilon}''$ with: (see Appendix \ref{app:ID})
\begin{eqnarray}\label{eq:Ae2}
A_\ee''
&=& \lim_{\epsilon \to 0^+} \iint_{\mathcal{D}_\epsilon} e^{\frac{t+s}{\tau}} \Exp{\mathcal{E}_{\ee \ii}(t,s) \dd H_\ee(t) \dd H_\ee(s) } 
= \frac{a_{\ee,12}-c_{\ee \ii} }{1+ a_{\ee \ii,2}} \, , \\
B_{\ee \ii}'' \label{eq:Bei2}
&=&2 \lim_{\epsilon \to 0^+} \iint_{\mathcal{D}_\epsilon} e^{\frac{t+s}{\tau}} \Exp{\mathcal{E}_{\ee \ii}(t,s) \dd H_\ee(t) \dd H_\ii(s) } 
=  \frac{2 c_{\ee \ii} }{1+ a_{\ee \ii,2}} \, .  
\end{eqnarray}
Remembering that the expression of $A_\ii''$ can be deduced from that of $A_\ee''$ by symmetry,
\eref{eq:Ae2} defines $A_\ee''$,  and thus $A_\ii''$, in terms of the useful auxiliary second-order efficacies $a_{\ee,12} = a_{\ee,1} - a_{\ee,2}$ and $a_{\ii,12} = a_{\ii,1} - a_{\ii,2}$.
These efficacies will feature prominently in the final variance expression and it is worth mentioning their explicit definitions as 
\begin{eqnarray}
\label{eq:a12}
a_{\ee,12} = \frac{b\tau}{2}  \ExpPei{\frac{W_\ee}{W_\ee+W_\ii}\left(1- e^{-(W_\ee+W_\ii)}\right)^2} 
\quad \mathrm{and} \quad
a_{\ii,12} = \frac{b\tau}{2}  \ExpPei{\frac{W_\ii}{W_\ee+W_\ii}\left(1- e^{-(W_\ee+W_\ii)}\right)^2} \, .
\end{eqnarray}
The other quantity of interest is the coefficient $c_{\ee \ii}$, which appears in both \eref{eq:Ae2} and \eref{eq:Bei2}. 
This nonnegative coefficient, defined as
\begin{eqnarray}
\label{eq:cei}
c_{\ee \ii} = \frac{b \tau}{2}  \ExpPei{\frac{W_\ee W_\ii}{(W_\ee +W_\ii)^2}\left(1- e^{-(W_\ee+W_\ii)}\right)^2} \, ,
\end{eqnarray}
entirely captures the (nonnegative) correlation between excitatory and inhibitory inputs and shall be seen as an efficacy as well.
Keeping these definitions in mind, the full quadratic coefficients are finally obtained as $A_\ee=A_\ee'+A_\ee''$, $A_\ii=A_\ii'+A_\ii''$, and $B_{\ee \ii}=B_{\ee \ii}'+B_{\ee \ii}''$.\\

 From there, injecting the analytical expressions of the various coefficients in the quadratic form \eref{eq:quad} leads to an explicit formula for the stationary voltage variance in the limit of instantaneous synapses.
Then, one is only left with step $(iii)$, which aims at  exhibiting a compact, interpretable form for this formula. 
We show in Appendix \ref{app:VE1} that lengthy but straightforward algebraic manipulations lead to the simplified form given in \eref{eq:statvar}:
\begin{eqnarray}
\Var{V} = \lim_{\epsilon \to 0^+}\Var{V_\epsilon} == \frac{1}{1+ a_{\ee \ii,2}} \left( a_{\ee,12} (V_\ee \!-\! \Exp{V})^2 + a_{\ii,12} (V_\ii \!-\! \Exp{V})^2 - c_{\ee \ii} (V_\ee \!-\! V_\ii)^2 \right) \nonumber \, .
\end{eqnarray}


\section{Evaluation of $Q_\epsilon(t,s)$ for $\epsilon>0$} \label{app:QE}

The goal here is to justify the closed-form expression of $Q_\epsilon(t,s) = \Exp{e^{H_\ee(t)+H_\ii(s)}}$ via standard manipulation of exponential functionals of Poisson processes.
By definition, assuming with no loss of generality the order $0 \geq t \geq s$, we have 
\begin{eqnarray}
H_\ee(t)+H_\ii(s)
&=&
-\frac{1}{\tau} \left( \int_t^0 h_\ee(u) \, \dd u +  \int_s^0 h_\ii(u) \, \dd u \right) \, ,\nonumber \\
&=&
-\frac{1}{\epsilon \tau}  \left( \int_t^0   \dd u \sum_{N(u-\epsilon \tau)+1}^{N(u)} W_{\ee,k} +  \int_s^0  \dd u  \sum_{N(u-\epsilon \tau)+1}^{N(u)} W_{\ii,k}  \right) \, , \nonumber \\
&=&
-\frac{1}{\epsilon \tau}  \left( \int_t^0  \dd u \sum_{N(u-\epsilon \tau)+1}^{N(u)} (W_{\ee,k} +W_{\ii,k})  +  \int_s^t  \dd u \sum_{N(u-\epsilon \tau)+1}^{N(u)} W_{\ii,k}  \right) \, . \label{eq:intForm1}
\end{eqnarray}
We will evaluate $Q_\epsilon(t,s) = \Exp{e^{H_\ee(t)+H_\ii(s)}}$ as a product of independent integral contributions. 

Isolating these independent contributions from \eref{eq:intForm1} requires to establish two preliminary results about the quantity 
\begin{eqnarray}
I(t,s) = \int_s^t \sum_{k=N(u-\Delta)+1}^{N(u)} X_k \, \dd u \, ,
\end{eqnarray}
where $N$ denotes a Poisson process, $X_k$ denotes i.i.d. nonnegative random variables, and where $\Delta$ is positive activation time.
Assume $t-s \geq \Delta$, then given some real $w<u-\Delta$, we have
\begin{eqnarray}
I(t,s) 
&=& 
\int_s^t  \dd u \sum_{k=N(v)+1}^{N(u)} X_k  - \int_s^t  \dd u \sum_{k=N(v)+1}^{N(u-\Delta)} X_k  \, , \nonumber\\
&=& 
\int_s^t  \dd u \sum_{k=N(v)+1}^{N(u)} X_k  - \int^{t-\Delta}_{s-\Delta}  \dd u \sum_{k=N(v)+1}^{N(u)} X_k  \, , \nonumber\\
&=& 
\int_{t-\Delta}^t  \dd u \sum_{k=N(v)+1}^{N(u)} X_k  - \int_{s-\Delta}^s  \dd u \sum_{k=N(v)+1}^{N(u)} X_k  \, , \nonumber\\
&=& 
\left( \int_{t-\Delta}^t  \dd u \sum_{k=N(v)+1}^{N(t-\Delta)} X_k  +  \int_{t-\Delta}^t  \dd u \sum_{k=N(t-\Delta)+1}^{N(u)} X_k  \right) 
- 
\left( \int_{s-\Delta}^s  \dd u \sum_{k=N(v)+1}^{N(s)} X_k   - \int_{s-\Delta}^s  \dd u \sum_{k=N(u)+1}^{N(s)} X_k  \right)\, , \nonumber\\
&=& 
\int_{t-\Delta}^t   \dd u \sum_{k=N(t-\Delta)+1}^{N(u)} X_k  + \Delta \sum_{k=N(s)+1}^{N(t-\Delta)} X_k + \int_{s-\Delta}^s  \dd u \sum_{k=N(u)+1}^{N(s)} X_k \, .\label{eq:intForm2}
\end{eqnarray}
One can check that the three terms in \eref{eq:intForm2} above are independent for involving independent numbers of i.i.d. draws over the intervals $(t-\Delta,t]$, $(s,t-\Delta]$, and $(s-\Delta,s]$, respectively.
Similar manipulations for the order for $t-s \leq \Delta$ yields
\begin{eqnarray}
I(t,s) 
&=& 
\int_s^t   \dd u \sum_{k=N(s)+1}^{N(u)} X_k  + (t-s) \sum_{k=N(t-\Delta)+1}^{N(s)} X_k + \int_{s-\Delta}^{t-\Delta}  \dd u \sum_{k=N(u)+1}^{N(t-\Delta)} X_k  \, ,
\label{eq:intForm3}
\end{eqnarray}
where that three independent contributions corresponds to independent numbers of i.i.d. draws over the intervals $(s,t]$, $(t-\Delta,s]$, and $(s-\Delta,t-\Delta]$, respectively.

As evaluating $Q_\epsilon$ only  involves taking the limit $s \to t^-$ at fixed $\epsilon>0$, it is enough to consider the order $0 \geq -\epsilon \tau \geq t \geq s \geq t - \epsilon \tau$ 
With that in mind, we can apply \eref{eq:intForm2}  \eref{eq:intForm3} with $\Delta=\epsilon \tau$ and $X_k=W_{\ee,k}+W_{\ii,k}$ or $X_k=W_{\ii,k}$, to decompose the two terms of \eref{eq:intForm1} in six contributions
\begin{eqnarray}
I(t,s) 
&=& 
\int_{-\epsilon \tau}^0   \dd u \sum_{k=N(t-\epsilon \tau)+1}^{N(u)} (W_{\ee,k}+W_{\ii,k})  + \epsilon \tau \sum_{k=N(t)+1}^{N(-\epsilon \tau)} (W_{\ee,k}+W_{\ii,k}) + \int_{t-\epsilon \tau}^{t}   \dd u \sum_{k=N(u)+1}^{N(t)} (W_{\ee,k}+W_{\ii,k})  \nonumber\\
&&+
\int_s^t  \dd u \sum_{k=N(s)+1}^{N(u)} W_{\ii,k}   + (t-s) \sum_{k=N(t-\epsilon \tau)+1}^{N(s)} W_{\ii,k}  + \int_{s-\epsilon \tau}^{t-\epsilon \tau}   \dd u \sum_{k=N(u)+1}^{N(t-\epsilon \tau)} W_{\ii,k} \, . \nonumber
\end{eqnarray}
It turns out that the contribution of the third term overlaps with that of the fourth and fifth terms.
Further splitting of that third term produces the following expression
\begin{eqnarray}
I(t,s) 
&=& 
\underbrace{\int_{-\epsilon \tau}^0  \dd u \sum_{k=N(t-\epsilon \tau)+1}^{N(u)} (W_{\ee,k}+W_{\ii,k})}_{\displaystyle I_1} 
+ 
\underbrace{\epsilon \tau \sum_{k=N(t)+1}^{N(-\epsilon \tau)} (W_{\ee,k}+W_{\ii,k})}_{\displaystyle I_2(t)}    \nonumber\\
&&
+  
\underbrace{\int_{s}^{t}  \dd u \left(  \sum_{k=N(u)+1}^{N(t)} (W_{\ee,k}+W_{\ii,k}) + \sum_{k=N(s)+1}^{N(u)} W_{\ii,k}  \right)  +  (s-t+\epsilon \tau) \sum_{k=N(s)+1}^{N(t)} (W_{\ee,k}+W_{\ii,k})}_{\displaystyle I_3(t,s)} \nonumber\\
&&
+ 
\underbrace{\left( \int_{t-\epsilon \tau}^{s}  \dd u  \sum_{k=N(u)+1}^{N(s)} (W_{\ee,k}+W_{\ii,k})  + (t-s) \sum_{k=N(t-\epsilon \tau)+1}^{N(s)} W_{\ii,k}  \right)}_{\displaystyle I_4(s,t)}
+ 
\underbrace{\int_{s-\epsilon \tau}^{t-\epsilon \tau}  \dd u \sum_{k=N(u)+1}^{N(t-\epsilon \tau)} W_{\ii,k}}_{\displaystyle I_5(t,s)} \, , \nonumber
\end{eqnarray}
where all five terms correspond to independent numbers of i.i.d. draws over the intervals $(-\epsilon \tau,0]$, $(t,-\epsilon \tau]$, $(s,t]$, $(t-\epsilon \tau,s]$,  and $(s-\epsilon \tau,t-\epsilon \tau]$.
Then, we have
\begin{eqnarray}
Q_\epsilon(t,s) 
= 
\Exp{e^{H_\ee(t)+H_\ii(s)}}
= \Exp{e^{-I_1/(\epsilon \tau)}} \Exp{e^{-I_2(t)/(\epsilon \tau)}} \Exp{e^{-I_3(t,s)/(\epsilon \tau)}} \Exp{e^{-I_4(s,t)/(\epsilon \tau)}} \Exp{e^{-I_5(t,s)/(\epsilon \tau)}} \, , \nonumber
\end{eqnarray}
where all expectation terms can be computed via standard manipulation of the moment-generating function of Poisson processes~\cite{daley2003introduction}.
The trick is to remember that for all $t \geq s$, given that a Poisson process admits $K=N(t)-N(s)$ points in $(s,t]$, all these $K$ points are uniformly i.i.d. over $(s,t]$.
This trick allows one to simply represent all integral terms in terms of uniform random variables, whose expectations are easily computable.
To see this, let us consider $I_3(t,s)$ for instance.
We have
\begin{eqnarray}
I_3(t,s)
&=&
(t-s)\sum_{k=N(s)+1}^{N(t)}\big[(1-U_k)(W_{\ee,k}+W_{\ii,k}) + U_k W_{\ii,k} \big] + (s-t+\epsilon \tau) \sum_{k=N(s)+1}^{N(t)} (W_{\ee,k}+W_{\ii,k}) \, , \nonumber\\
&=&
(t-s) \sum_{k=N(s)+1}^{N(t)} U_k W_{\ee,k} +\epsilon \tau \sum_{k=N(s)+1}^{N(t)} (W_{\ee,k}+W_{\ii,k}) \, ,  \nonumber
\end{eqnarray}
where $\{ U_k \}_{N(s)+1 \leq k \leq N(t)}$ are uniformly i.i.d. on $[0,1]$.
From the knowledge of the moment-generating function of Poisson random variables~\cite{daley2003introduction}, one can evaluate
\begin{eqnarray}
\Exp{e^{-I_3(t,s)/(\epsilon \tau)}}
&=&
\Exp{e^{-\frac{t-s}{\epsilon \tau} \sum_{k=N(s)+1}^{N(t)} U_k W_{\ee,k} - \sum_{k=N(s)+1}^{N(t)} (W_{\ee,k}+W_{\ii,k}) } }  \, , \nonumber\\
&=&
\Exp{\Exp{e^{-\frac{t-s}{\epsilon \tau} U W_{\ee} - (W_{\ee}+W_{\ii})}}^{N(t)-N(s)} \,\bigg \vert \, N(t)-N(s) } \, , \nonumber\\
&=&
\exp{ \left(b(t-s) \left( \Exp{e^{-\frac{t-s}{\epsilon \tau} U W_{\ee} - (W_{\ee}+W_{\ii}) } }- 1 \right)\right)} \, , \nonumber
\end{eqnarray}
where $(W_{\ee},W_{\ii})$ denotes exemplary conductance jumps and $U$ denotes an independent uniform random variable.
Furthermore we have
\begin{eqnarray}
\Exp{e^{-\frac{t-s}{\epsilon \tau} U W_{\ee} - (W_{\ee}+W_{\ii}) } }
&=& 
\Exp{\Exp{e^{-\frac{t-s}{\epsilon \tau} U W_{\ee} - (W_{\ee}+W_{\ii}) } } \,\Big \vert \, W_\ee, W_\ii} \, , \nonumber\\
&=& 
\ExpPei{ e^{ - (W_{\ee}+W_{\ii}) } \Exp{e^{-\frac{t-s}{\epsilon \tau} U W_{\ee}  } } } \, , \nonumber\\
&=& 
\ExpPei{ e^{ - (W_{\ee}+W_{\ii}) } \frac{\left(1-e^{-\frac{t-s}{\epsilon \tau} W_{\ee} }\right)}{\frac{t-s}{\epsilon \tau}W_\ee} } \, , \nonumber
\end{eqnarray}
so that we finally obtain
\begin{eqnarray}
\ln \Exp{e^{-I_3(t,s)/(\epsilon \tau)}} 
=
\epsilon b \tau \left(  \ExpPei{ e^{ - (W_{\ee}+W_{\ii}) } \frac{\left(1-e^{-\frac{t-s}{\epsilon \tau} W_{\ee} }\right)}{W_\ee}}- \frac{t-s}{\epsilon \tau} \right) \, . \nonumber
\end{eqnarray}
Similar calculations show that we have
\begin{eqnarray}
\ln \Exp{e^{-I_1/(\epsilon \tau)}}
=
\epsilon b  \tau \left( \ExpPei{\frac{1-e^{-(W_\ee+W_\ii)}}{W_\ee+W_\ii}} -1 \right) \, , \nonumber
\end{eqnarray}
\begin{eqnarray}
\ln \Exp{e^{-I_2(t)/(\epsilon \tau)}}
=
 b (\epsilon \tau +t) \left(1- \ExpPei{e^{-(W_\ee+W_\ii)}}  \right)  \, , \nonumber
\end{eqnarray}
\begin{eqnarray}
\ln \Exp{e^{-I_4(s,t)/(\epsilon \tau)}}
=
\epsilon b \tau \left( \ExpPei{ e^{ - \frac{t-s}{\epsilon \tau}W_{\ii} } \frac{\left(1-e^{-\left(1+\frac{s-t}{\epsilon \tau}\right) ( W_{\ee}+W_\ii )}\right)}{W_\ee+ W_\ii} }- \left(1+\frac{s-t}{\epsilon \tau}\right) \right)  \, , \nonumber
\end{eqnarray}
\begin{eqnarray}
\ln \Exp{e^{-I_5(t,s)/(\epsilon \tau)}}
=
\epsilon b \tau \left( \ExpPei{\frac{1-e^{-\frac{t-s}{\epsilon \tau} W_\ii}}{W_\ii}} - \frac{t-s}{\epsilon \tau}\right) \, . \nonumber
\end{eqnarray}


\section{Expression of $R_\epsilon(t,u,s,v)$ on $\mathcal{O}_\epsilon$ and $\mathcal{D}_\epsilon$}\label{app:RO}

Using the similar calculations as in Appendix \ref{app:QE}, we can evaluate the quadrivariate expectation $R_\epsilon(t,u,s,v)$ on the region $\mathcal{O}_\epsilon$, for which the $O$-order holds: $0 \geq -\epsilon \tau \geq t \geq u \geq t-\epsilon \tau \geq u-\epsilon \tau \geq s \geq v \geq s-\epsilon \tau \geq v-\epsilon \tau$.
This requires to isolate consider $9$ independent contributions, corresponding to the $9$ contiguous intervals specified by the $O$-order.
We find
\begin{eqnarray}
\ln R_\epsilon(t,u,s,v) = A_1+A_2(t)+A_3(t,u)+A_4(u,t)+A_5(t,u)+A_6(u,s)+A_7(s,v)+A_8(v,s)+A_9(s,v) \, , \nonumber
\end{eqnarray}
where the nonnegative terms making up the above sum are defined as
\begin{eqnarray}
A_1
=
\epsilon b \tau \left( \ExpPei{\frac{1-e^{-2(W_\ee+W_\ii)}}{2(W_\ee+W_\ii)}} -1 \right) \, , \nonumber
\end{eqnarray}
\begin{eqnarray}
A_2(t)
=
 b (\epsilon \tau +t) \left(1- \ExpPei{e^{-2(W_\ee+W_\ii)}}  \right) \, , \nonumber
\end{eqnarray}
\begin{eqnarray}
A_3(t,u)
=
\epsilon b \tau  \left(  \ExpPei{e^{-2(W_\ee+W_\ii)}\frac{\left(1-e^{-\frac{t-u}{\epsilon \tau}W_\ee}\right)}{W_\ee}} -\frac{t-u}{ \epsilon\tau} \right) \, , \nonumber
\end{eqnarray}
\begin{eqnarray}
A_4(u,t)
=
\epsilon b \tau  \left(  \ExpPei{e^{-W_\ee-\left(1+\frac{t-u}{\epsilon \tau} \right)W_\ii} \frac{\left(1-e^{- \left(1+\frac{u-t}{\epsilon \tau} \right)(W_\ee+W_\ii)}\right)}{W_\ee+W_\ii}} -\left(1+\frac{u-t}{ \epsilon\tau} \right)\right) \, , \nonumber
\end{eqnarray}
\begin{eqnarray}
A_5(t,u)
=
\epsilon b \tau \left(  \ExpPei{e^{-(W_\ee+W_\ii)} \frac{\left(1-e^{- \frac{t-u}{\epsilon \tau} W_\ii}\right)}{W_\ii}} -\frac{t-u}{ \epsilon\tau} \right) \, , \nonumber
\end{eqnarray}
\begin{eqnarray}
A_6(u,s)
=
 b (s+\epsilon \tau -u) \left(1- \ExpPei{e^{-(W_\ee+W_\ii)}}  \right) \nonumber
\end{eqnarray}
\begin{eqnarray}
A_7(s,v)
=
\epsilon b \tau  \left(  \ExpPei{e^{-(W_\ee+W_\ii)}\frac{\left(1-e^{-\frac{s-v}{\epsilon \tau}W_\ee}\right)}{W_\ee}} -\frac{s-v}{ \epsilon\tau} \right) \, , \nonumber
\end{eqnarray}
\begin{eqnarray}
A_8(v,s)
=
\epsilon b \tau  \left(  \ExpPei{e^{-\frac{s-v}{\epsilon \tau}W_\ii} \frac{\left(1-e^{- \left(1-\frac{s-v}{\epsilon \tau} \right)(W_\ee+W_\ii)}\right)}{W_\ee+W_\ii}} -\left(1-\frac{s-v}{ \epsilon\tau} \right)\right) \, , \nonumber
\end{eqnarray}
\begin{eqnarray}
A_9(s,v)
=
\epsilon b \tau  \left(  \ExpPei{ \frac{\left(1-e^{- \frac{s-v}{\epsilon \tau} W_\ii}\right)}{W_\ii}} -\frac{s-v}{ \epsilon\tau} \right) \, .\nonumber
\end{eqnarray}
One can check that $A_3(t,t)=A_5(t,t)=0$ and $A_7(s,s)=A_9(s,s)=0$ and that $A_1$, $A_4(u,t)$, and $A_8(v,s)$ are all uniformly $O(\epsilon)$ on the region $\mathcal{O}_\epsilon$.
This implies that for all $(t,s)$ in $\mathcal{O}_\epsilon$, we have
\begin{eqnarray}
R(t,s) 
=
\lim_{\epsilon \to 0^+} R_\epsilon(t,t,s,s)  
= 
\lim_{\epsilon \to 0^+} e^{A_2(t) + A_6(t,s)} 
=
e^{2bt a_{\ee \ii,2} - b \vert t-s\vert a_{\ee \ii,1}} \, . \nonumber
\end{eqnarray}

Using the similar calculations as in Appendix \ref{app:QE}, we can evaluate the quadrivariate expectation $R_\epsilon(t,u,s,v)$ on the region $\mathcal{D}_\epsilon$, for which the $D$-order holds: $0 \geq -\epsilon \tau \geq t \geq u \geq s \geq v \geq t-\epsilon \tau \geq u-\epsilon \tau  \geq s-\epsilon \tau \geq v-\epsilon \tau$.
This requires to isolate consider $9$ independent contributions, corresponding to the $9$ contiguous intervals specified by the $O$-order.
We find
\begin{eqnarray}\label{eq:RD}
\lefteqn{ \ln R_\epsilon(t,u,s,v) =} \\
&& B_1+B_2(t)+B_3(t,u)+B_4(t,u,s)+B_5(t,u,s,v)+B_6(t,u,s,v)+B_7(t,u,s,v)+B_8(u,s,v)+B_9(s,v) \nonumber
\end{eqnarray}
where the nonnegative terms making up the above sum are defined as
\begin{eqnarray}
B_1
=
\epsilon b \tau \left( \ExpPei{\frac{1-e^{-2(W_\ee+W_\ii)}}{2(W_\ee+W_\ii)}} -1 \right)  \, , \nonumber
\end{eqnarray}
\begin{eqnarray}
B_2(t)
=
 b (\epsilon \tau +t) \left(1- \ExpPei{e^{-2(W_\ee+W_\ii)}}  \right)  \, , \nonumber
\end{eqnarray}
\begin{eqnarray}
B_3(t,u)
=
\epsilon b \tau  \left(  \ExpPei{e^{-2(W_\ee+W_\ii)}\frac{\left(1-e^{-\frac{t-u}{\epsilon \tau}W_\ee}\right)}{W_\ee}} -\frac{t-u}{ \epsilon\tau} \right)  \, , \nonumber
\end{eqnarray}
\begin{eqnarray}
B_4(t,u,s)
=
\epsilon b \tau  \left(  \ExpPei{e^{-\left(2-\frac{t-s}{\epsilon \tau} \right)W_\ee - \left(2-\frac{u-s}{\epsilon \tau} \right)W_\ii } \frac{\left(1-e^{- \frac{u-s}{\epsilon \tau} (W_\ee+W_\ii)}\right)}{W_\ee+W_\ii}} -\frac{u-s}{ \epsilon\tau} \right)  \, , \nonumber
\end{eqnarray}
\begin{eqnarray}
B_5(t,u,s,v)
=
\epsilon b \tau  \left(  \ExpPei{e^{-\left(2-\frac{t-v}{\epsilon \tau} \right)W_\ee - \left(2-\frac{u-v}{\epsilon \tau} \right)W_\ii } \frac{\left(1-e^{- \frac{s-v}{\epsilon \tau} (2W_\ee+W_\ii)}\right)}{2W_\ee+W_\ii}} -\frac{s-v}{ \epsilon\tau} \right)  \, , \nonumber
\end{eqnarray}
\begin{eqnarray}
B_6(t,u,s,v)
=
\epsilon b \tau \left(  \ExpPei{e^{-\left(\frac{t-s}{\epsilon \tau} \right)W_\ee - \left(\frac{2t-(u+v)}{\epsilon \tau} \right)W_\ii } \frac{\left(1-e^{-\left(1- \frac{t-v}{\epsilon \tau} \right) 2(W_\ee+W_\ii)}\right)}{2(W_\ee+W_\ii)}} -\left(1- \frac{t-v}{ \epsilon\tau} \right)\right)  \, , \nonumber
\end{eqnarray}
\begin{eqnarray}
B_7(t,u,s,v)
=
\epsilon b \tau \left(  \ExpPei{e^{-\left(\frac{u-s}{\epsilon \tau} \right)W_\ee - \left(\frac{u-v}{\epsilon \tau} \right)W_\ii } \frac{\left(1-e^{-\frac{t-u}{\epsilon \tau}  (W_\ee+2W_\ii)}\right)}{W_\ee+2W_\ii}} -\frac{t-u}{ \epsilon\tau} \right)  \, , \nonumber
\end{eqnarray}
\begin{eqnarray}
B_8(u,s,v)
=
\epsilon b \tau  \left(  \ExpPei{e^{- \left(\frac{s-v}{\epsilon \tau} \right)W_\ii } \frac{\left(1-e^{-\frac{u-s}{\epsilon \tau}  (W_\ee+W_\ii)}\right)}{W_\ee+W_\ii}} -\frac{u-s}{ \epsilon\tau} \right)  \, , \nonumber
\end{eqnarray}
\begin{eqnarray}
B_9(s,v)
=
b  \epsilon\tau  \left(  \ExpPei{e^{- \left(\frac{s-v}{\epsilon \tau} \right)W_\ii } \frac{\left(1-e^{-\frac{s-v}{\epsilon \tau}  W_\ii}\right)}{W_\ii}} -\frac{s-v}{ \epsilon\tau} \right)  \, . \nonumber
\end{eqnarray}
Observe that $B_1=A_1$ and $B_2(t)=A_2(t)$ and that $B_3(t,t)=B_7(t,t,s,v)=0$ and $B_5(t,u,s,s)=B_9(s,s)=0$.
Moreover, one can see that $R(t,s)$ is continuous over the whole negative orthant by checking that:
\begin{eqnarray}
\lim_{s \to (t-\epsilon \tau)^-} B_4(t,t,s) = \lim_{s \to (t-\epsilon \tau)^+} A_4(t,s) \, , \nonumber\\
\lim_{s \to (t-\epsilon \tau)^-} B_6(t,t,s,s) = \lim_{s \to (t-\epsilon \tau)^+} A_6(t,s) \, , \nonumber\\\
\lim_{s \to (t-\epsilon \tau)^-} B_8(t,s,s) = \lim_{s \to (t-\epsilon \tau)^+} A_8(t,s) \, . \nonumber\
\end{eqnarray}
Actually, by computing the appropriate limit values of the relevant first- and second-order derivatives of $R_\epsilon(t,u,s,v)$, one can check that for $\epsilon>0$, all the integrands involved in specifying the coefficients of the quadratic form \eref{eq:quad} define continuous functions.


\section{Integrals of the quadratic terms on  $\mathcal{D}_\epsilon$}\label{app:ID}

Here, we only treat the quadratic term $A_\ee$ as the other quadratic terms $A_\ii$ and $B_{\ee \ii}$ involve a similar treatment.
The goal is to compute $A_\ee''$, which is defined as the contribution to $A_\ee$ resulting from integrating $\lim_{(u,v) \to (t,s)^-}\partial_t \partial_s R_\epsilon(t,u,s,v)$ over the diagonal region $\mathcal{D}_\epsilon=\{ t,s \leq 0 \, \vert \,  \tau \epsilon \geq \vert t-s \vert \}$, in the limit $\epsilon \to 0^+$. 
To this end we first remark that
\begin{eqnarray}
\frac{\partial_t \partial_s R_\epsilon(t,u,s,v)}{R_\epsilon(t,u,s,v)}  = \partial_t \partial_s \ln R_\epsilon(t,u,s,v) + \big(\partial_t \ln R_\epsilon(t,u,s,v)\big) \big( \partial_s \ln R_\epsilon(t,u,s,v)\big) \, . \nonumber
\end{eqnarray}
Injecting the analytical expression \eref{eq:RD} into the above relation and evaluating $I_\epsilon(t,s)=\lim_{(u,v \to (t,s)^-)}\partial_t \partial_s R_\epsilon(t,u,s,v)$ reveals that $I_\epsilon(t,s)$ scales as $1/\epsilon$, so that one expects that 
\begin{eqnarray}
A_\ee'' = \lim_{\epsilon \to 0^+} \iint_{\mathcal{D}_\epsilon} e^{\frac{t+s}{\tau}} I_\epsilon(t,s) \, \dd t \dd s  > 0 \, .  \nonumber
\end{eqnarray}
To compute the exact value of $A_\ee''$, we perform the change of variable $x=(t-s)/(\epsilon \tau) \Leftrightarrow s=t-\epsilon \tau x$ to write
\begin{eqnarray}
\iint_{\mathcal{D}_\epsilon}  e^{\frac{t+s}{\tau}} I_\epsilon(t,s) \, \dd t \dd s = 2 \int_{-\infty}^0  \left( \int_0^1  \epsilon \tau e^{-\epsilon x}I_\epsilon(t,t+\epsilon \tau x) \, \dd x \right) e^{\frac{2t}{\tau}}  \dd t \, ,  \nonumber
\end{eqnarray}
where the function $\epsilon e^{-\frac{\epsilon x}{\tau}} I_\epsilon(t,t+\epsilon x)$ remains of order one on $\mathcal{D}_\epsilon$ in the limit of instantaneous synapses.
Actually, one can compute that
\begin{eqnarray}
\lim_{\epsilon \to 0^+} \epsilon e^{-\epsilon x} I_\epsilon(t,t+\epsilon \tau x) = \frac{b}{2\tau} \ExpPei{\frac{W_\ee^2}{W_\ee+W_\ii} e^{-x(W_\ee+W_\ii)} \left( 1-e^{-2\left(1-x\right)(W_\ee+W_\ii)}\right)} e^{2bt a_{\ee \ii,2}} \, .  \nonumber
\end{eqnarray}
Then, for dealing with positive, continuous, uniformly bounded functions, one can safely exchange the integral and limit operations to get
\begin{eqnarray}
A_\ee'' 
&=& 
2 \int_{-\infty}^0  \left( \int_0^1 \lim_{\epsilon \to 0^+}  \epsilon \tau e^{-\epsilon x}I_\epsilon(t,t+\epsilon \tau x) \, \dd x \right) e^{\frac{2t}{\tau}} \dd t  \, , \nonumber\\
&=& 
\left( \int_{-\infty}^0  e^{2\frac{t}{\tau}\left(1+ a_{\ee \ii,2} \right)} \, \dd t  \right) 
\left( \int_0^1 b \ExpPei{\frac{W_\ee^2}{W_\ee+W_\ii} e^{-\frac{x}{\tau}(W_\ee+W_\ii)} \left( 1-e^{-2\left(1-\frac{x}{\tau}\right)(W_\ee+W_\ii)}\right)} \, \dd x \right) 
\, , \nonumber\\
&=&
\frac{b \tau}{2(1 + a_{\ee \ii,2})} \, \ExpPei{\frac{W_\ee^2}{(W_\ee+W_\ii)^2} \left( 1-e^{-(W_\ee+W_\ii)^2}\right)} \, .  \nonumber
\end{eqnarray}
A similar calculation for the quadratic cross term $B''_{\ee \ii}$ yields
\begin{eqnarray}
B_{\ee \ii}'' 
&=& 
\frac{2 c_{\ee \ii}}{1 +  a_{\ee \ii,2}}  \quad \mathrm{with} \quad c_{\ee \ii}=\frac{ b \tau}{2} \ExpPei{\frac{W_\ee W_\ii}{(W_\ee+W_\ii)^2} \left( 1-e^{-(W_\ee+W_\ii)^2}\right)} \, .  \nonumber
\end{eqnarray}
In order to express $A_\ee''$ in term of  $c_{\ee \ii}$, we need to introduce the quantity $a_{\ee,12} = a_{\ee,1}-a_{\ee,2}$ which satisfies
\begin{eqnarray}
a_{\ee,12} 
&=& 
b \tau \ExpPei{\frac{W_\ee}{(W_\ee+W_\ii)} \left( 1-e^{-(W_\ee+W_\ii)}\right)}-\frac{1}{2}\ExpPei{\frac{W_\ee}{(W_\ee+W_\ii)} \left( 1-e^{-(W_\ee+W_\ii)}\right)^2}  \, ,  \nonumber\\
&=& 
b \tau \ExpPei{\frac{W_\ee}{(W_\ee+W_\ii)} \left( 1-e^{-(W_\ee+W_\ii)}\right)\left( 1-\frac{1}{2} \left( 1-e^{-(W_\ee+W_\ii)}\right) \right)} \, ,  \nonumber\\
&=& 
b \tau \ExpPei{\frac{W_\ee}{(W_\ee+W_\ii)} \left( 1-e^{-(W_\ee+W_\ii)}\right)\left(\frac{\left( 1+e^{-(W_\ee+W_\ii)}\right)}{2} \right)} \, ,  \nonumber\\
&=& 
\frac{b \tau}{2}\ExpPei{\frac{W_\ee}{(W_\ee+W_\ii)} \left( 1+e^{-(W_\ee+W_\ii)}\right)^2 } \, ,  \nonumber
\end{eqnarray}
With the above observation, we remark that
\begin{eqnarray}
(1 +  a_{\ee \ii,2})A_\ee'' -  a_{\ee,12} 
&=&
\frac{b \tau}{2} \left( \ExpPei{\frac{W_\ee^2}{(W_\ee+W_\ii)^2} \left( 1-e^{-(W_\ee+W_\ii)}\right)^2}-\ExpPei{\frac{W_\ee}{(W_\ee+W_\ii)} \left( 1-e^{-(W_\ee+W_\ii)}\right)^2}\right) \, , \nonumber\\
&=&
\frac{b \tau}{2} \ExpPei{\frac{W_\ee^2-W_\ee(W_\ee+W_\ii)}{(W_\ee+W_\ii)^2} \left( 1-e^{-(W_\ee+W_\ii)}\right)^2} \, ,  \nonumber\\
&=&
-\frac{b \tau}{2} \ExpPei{\frac{W_\ee W_\ii}{(W_\ee+W_\ii)^2} \left( 1-e^{-(W_\ee+W_\ii)}\right)^2} \, ,  \nonumber\\
&=&- c_{\ee \ii} \nonumber
\end{eqnarray}
so that we have the following compact expression for the quadratic diagonal term
\begin{eqnarray}
A_\ee'' 
=
\frac{a_{\ee,12}-c_{\ee \ii}}{1 +  a_{\ee \ii,2}} \, . \nonumber
\end{eqnarray}


\section{Compact variance expression}\label{app:VE1}

Our goal is to find a compact, interpretable formula for the stationary variance $\Var{V}$ from the knowledge of the quadratic form
\begin{eqnarray}\label{eq:quadApp}
\Exp{V^2} 
=
 A_{\ee} V_\ee^2 + B_{\ee \ii}  V_\ee V_\ii  + A_{\ii}  V_\ii^2  + \left( V_\ee B_{\ee I} + V_\ii B_{\ii I} \right) (I/G)+ A_{I} (I/G)^2 \, . \nonumber
\end{eqnarray}
Let us first assume no current injection, $I=0$, so that one only has to keep track of the quadratic terms.
Specifying the quadratic coefficient $A_\ee=A_\ee'+A_\ee''$, $A_\ii =A_\ii'+A_\ii''$ and $B_{\ee \ii} =B_{\ee \ii}'+B_{\ee \ii}''$ in \eref{eq:quadApp}, we get
\begin{eqnarray}
\Exp{V^2} 
&=&
\left(  \frac{ a_{\ee,1}(2a_{\ee,2}-a_{\ee,1}) }{(1+ a_{\ee \ii,1})(1+ a_{\ee \ii,2})} + \frac{a_{\ee,12}-c_{\ee \ii}}{1 +  a_{\ee \ii,2}} \right) V_\ee^2   \nonumber\\
&&+ 
\left(\frac{ a_{\ee,1} (2a_{\ii,2}-a_{\ii,1}) +  a_{\ii,1}(2a_{\ee,2}-a_{\ee,1}) }{(1+ a_{\ee \ii,1})(1+ a_{\ee \ii,2})}   + \frac{2  c_{\ee \ii}}{1 + a_{\ee \ii,2}} \right) V_\ee V_\ii  \nonumber\\
&&+ 
\left(\frac{ a_{\ii,1}(2a_{\ii,2}-a_{\ii,1}) }{(1+ a_{\ee \ii,1})(1+ a_{\ee \ii,2})} + \frac{a_{\ii,12}-c_{\ee \ii}}{1 +  a_{\ee \ii,2}}\right) V_\ii^2   \, , \nonumber\\
&=&
\left(  \frac{ a_{\ee,1}(2a_{\ee,2}-a_{\ee,1}) +(1+ a_{\ee,1}+a_{\ii,1})(a_{\ee,1}- a_{\ee2})}{(1+ a_{\ee \ii,1})(1+ a_{\ee \ii,2})}  \right) V_\ee^2   \nonumber\\
&&+ 
\left(\frac{a_{\ee,1}(2a_{\ii,2}-a_{\ii,1}) +  a_{\ii,1}(2a_{\ee,2}-a_{\ee,1}) }{(1+ a_{\ee \ii,1})(1+ a_{\ee \ii,2})}  \right) V_\ee V_\ii \nonumber\\
&&+ 
\left(  \frac{ a_{\ii,1}(2a_{\ii,2}-a_{\ii,1}) +(1+ a_{\ee,1}+a_{\ii,1})(a_{\ii,1}- a_{\ii2})}{(1+  a_{\ee \ii,1})(1+ a_{\ee \ii,2})}  \right) V_\ii^2 - \frac{ c_{\ee \ii}}{1 + a_{\ee \ii,2}} (V_\ee-V_\ii)^2  \, , \nonumber
\end{eqnarray}
where we collect separately all the terms containing the coefficient $c_{\ee \ii}$ and where we use the facts that by definition $a_{\ee,12}=a_{\ee,1}-a_{\ee,2}$, $a_{\ii,12}=a_{\ii,1}-a_{\ii,2}$, and  $a_{\ee \ii,1}=a_{\ee,1}+a_{\ii,1}$.
Expanding and simplifying the coefficients of $V_\ee^2$ and $V_\ii^2$ above yield
\begin{eqnarray}
\Exp{V^2} 
&=&
\left(  \frac{a_{\ee,1}a_{\ee,2}+(1+ a_{\ii,1})(a_{\ee,1}- a_{\ee2})}{(1+ a_{\ee \ii,1})(1+ a_{\ee \ii,2})}  \right) V_\ee^2   \nonumber\\
&&+ 
\left(\frac{ a_{\ee,1}(2a_{\ii,2}-a_{\ii,1}) +  a_{\ii,1}(2a_{\ee,2}-a_{\ee,1})}{(1+ a_{\ee \ii,1})(1+ a_{\ee \ii,2})}  \right) V_\ee V_\ii  \nonumber\\
&&+ 
\left(  \frac{ a_{\ii,1} a_{\ii,2} +(1+ a_{\ee,1})(a_{\ii,1}- a_{\ii2})}{(1+ a_{\ee \ii,1})(1+ a_{\ee \ii,2})}  \right)   V_\ii^2 - \frac{ c_{\ee \ii}}{1 +  a_{\ee \ii,2}} (V_\ee-V_\ii)^2  \, . \nonumber
\end{eqnarray}
Then, we can utilize the expression above for $\Exp{V^2}$ together with the stationary mean formula
\begin{eqnarray}\label{eq:meanApp}
\Exp{V}= \frac{ a_{\ee,1} V_\ee+  a_{\ii,1} V_\ii}{1+  a_{\ee \ii,1}} \, ,
\end{eqnarray}
to write the variance $\Var{V}=\Exp{V^2} -\Exp{V}^2$ as
\begin{eqnarray}
\Var{V} 
&=&
\left(  \frac{(a_{\ee,1}-a_{\ee,2})(1+ a_{\ii,1})^2+(a_{\ii,1}-a_{\ii,2})a_{\ee,1}^2}{(1+ a_{\ee \ii,1})^2(1+ a_{\ee \ii,2})}  \right) V_\ee^2   \nonumber\\
&&
- \left(\frac{
(a_{\ee,1}-a_{\ee,2})a_{\ee,1}(1+ a_{\ee,1})+a_{\ii,1}(a_{\ii,1}-a_{\ii,2})(1+ a_{\ii,1})
 }{(1+ a_{\ee \ii,1})^2(1+ a_{\ee \ii,2})}  \right) V_\ee V_\ii  \nonumber\\
&&
+\left(   \frac{(a_{\ii,1}-a_{\ii,2})(1+ a_{\ee,1})^2+(a_{\ee,1}-a_{\ee,2})a_{\ii,1}^2}{(1+ a_{\ee \ii,1})^2(1+ a_{\ee \ii,2})}  \right) V_\ii^2 - \frac{c_{\ee \ii}}{1 +  a_{\ee \ii,2}} (V_\ee-V_\ii)^2  \, . \nonumber
\end{eqnarray}
To factorize the above expression, let us reintroduce $a_{\ee,12}=a_{\ee,1}-a_{\ee,2}$ and $a_{\ii,12}=a_{\ii,1}-a_{\ii,2}$ and collect the terms where these two coefficients occur.
This yields
\begin{eqnarray}
\Var{V} 
&=&
 \frac{ a_{\ee,12}}{(1+ a_{\ee \ii,1})^2(1+ a_{\ee \ii,2})}
\left( (1+ a_{\ii,1})^2V_\ee^2-  a_{\ii,1}(1+ a_{\ee,1})^2 V_\ee V_\ii + ( a_{\ee,1})^2 V_\ii^2  \right)   \nonumber\\
&&+ \frac{ a_{\ii,12}}{(1+ a_{\ee \ii,1})^2(1+ a_{\ee \ii,2})}
\left( (1+ a_{\ee,1})^2V_\ii^2-  a_{\ee,1}(1+ a_{\ii,1})^2 V_\ee V_\ii + ( a_{\ii,1})^2 V_\ee^2  \right)  \ \nonumber\\
&&- \frac{ c_{\ee \ii}}{1 +  a_{\ee \ii,2}} (V_\ee-V_\ii)^2  \, , \nonumber\\
&=&
 \frac{ a_{\ee,12}}{1+ a_{\ee \ii,2}} 
\left( \frac{(1+ a_{\ii,1})V_\ee - a_{\ee,1} V_\ii }{1+ a_{\ee \ii,1}} \right)^2  
+
 \frac{ a_{\ii,12}}{1+a_{\ee \ii,2}} 
\left( \frac{(1+ a_{\ii,e})V_\ii -  a_{\ii,1} V_\ee }{1+ a_{\ee \ii,1}} \right)^2   \nonumber\\
&&- \frac{ c_{\ee \ii}}{1 + a_{\ee \ii,2}} (V_\ee-V_\ii)^2  \nonumber \, .
\end{eqnarray}
Finally, injecting the expression of stationary mean \eref{eq:meanApp} in both parentheses above produces the compact formula
\begin{eqnarray} \label{eq:statvarapp}
\Var{V} 
=
 \frac{a_{\ee,12}}{1+a_{\ee \ii,2}} 
\left( V_\ee-\Exp{V} \right)^2  
+
 \frac{ a_{\ii,12}}{1+ a_{\ee \ii,2}} 
\left( V_\ii-\Exp{V} \right)^2  
- 
\frac{c_{\ee \ii}}{1 + a_{\ee \ii,2}} (V_\ee-V_\ii)^2   \, ,
\end{eqnarray}
which is the same as  the one given in \eref{eq:statvar}.


\section{Factorized variance expression}\label{app:VE2}

In this appendix, we reshape the variance expression given in \eref{eq:statvarapp} under a form that is clearly nonnegative.
To this end, let us first remark that the calculation in Appendix \ref{app:ID} shows that
\begin{eqnarray}
a_{\ee,12} - c_{\ee \ii} = \frac{b \tau}{2}\ExpPei{\frac{W_\ee^2}{(W_\ee+W_\ii)^2} \left( 1+e^{-(W_\ee+W_\ii)}\right)^2}  \, . \nonumber
\end{eqnarray}
Then, setting $ (V_\ee-V_\ii)^2=  ((V_\ee-\Exp{V}) -(V_\ii-\Exp{V}))^2=(V_\ee-\Exp{V})^2 - 2 (V_\ee-\Exp{V})(V_\ii-\Exp{V})+ (V_\ii-\Exp{V})^2$ in  \eref{eq:statvarapp}, we obtain
\begin{eqnarray}
\Var{V} 
&=&  
\frac{1}{1+ a_{\ee \ii,2}}  \left( a_{\ee,12} (V_\ee-\Exp{V})^2 + a_{\ii,12} (V_\ii-\Exp{V})^2 - c_{\ee \ii} (V_\ee-V_\ii)^2 \right) \, , \nonumber\\
&=&  
\frac{1}{1+ a_{\ee \ii,2}}  \left( (a_{\ee,12} - c_{\ee \ii} ) (V_\ee-\Exp{V})^2 + 2 c_{\ee \ii} (V_\ee-\Exp{V})(V_\ii-\Exp{V}) + (a_{\ii,12} - c_{\ee \ii} )(V_\ii-\Exp{V})^2  \right) \, , \nonumber\\
&=&  
\frac{b \tau}{1+ a_{\ee \ii,2}} \nonumber\\
&& \hspace{20pt}
\ExpPei{ \left( \frac{W_\ee^2(V_\ee-\Exp{V})^2}{2(W_\ee+W_\ii)^2}  +  \frac{2 W_\ee(V_\ee-\Exp{V})W_\ii(V_\ii-\Exp{V})}{2(W_\ee+W_\ii)^2} + \frac{W_\ii^2(V_\ii-\Exp{V})^2}{2(W_\ee+W_\ii)^2} \right) \left( 1- e^{- (W_\ee+W_\ii)}\right)^2 } \, , \nonumber\\
&=&  
\frac{b \tau}{2(1+ a_{\ee \ii,2})} 
\ExpPei{ \left( \frac{\big[ W_\ee(V_\ee-\Exp{V})+W_\ii(V_\ii-\Exp{V}) \big]^2}{(W_\ee+W_\ii)^2}  \right) \left( 1- e^{-(W_\ee+W_\ii)}\right)^2 } \, . \label{eq:nonIvar}
\end{eqnarray}
Note that the above quantity is clearly non negative as any variance shall be. 
From there, one can include the impact of the injected current $I$ by further considering all the terms in \eref{eq:quadApp}, including the linear and inhomogeneous current-dependent terms.
Similar algebraic manipulations confirm that  \eref{eq:nonIvar} remains valid so that the only impact of $I$ is via  altering the expression $\Exp{V}$, so that we ultimately obtain the following explicit compact form:
\begin{eqnarray}
\Var{V} = 
\frac{
\ExpPei{ 
 \left(\frac{W_\ee V_\ee+W_\ii V_\ii}{W_\ee+W_\ii}-\Exp{V}\right)^2 \left(1- e^{-(W_\ee+W_\ii)}\right)^2
}
}
{
2/(b \tau)
+ 
\ExpPei{ 
\left(1- e^{-2(W_\ee+W_\ii)}\right)
}
} 
\quad \mathrm{with} \quad
\Exp{V} = 
\frac{
b \tau \ExpPei{ 
 \left(\frac{W_\ee V_\ee+W_\ii V_\ii}{W_\ee+W_\ii}\right) \left(1- e^{-(W_\ee+W_\ii)}\right) 
}
+ I/G
}
{
1
+ 
b \tau \ExpPei{ 
\left(1- e^{-(W_\ee+W_\ii)}\right)
}
} 
\, . \nonumber
\end{eqnarray}
The above expression shows that as expected $\Var{V} \geq 0$ and that the variability vanishes if and only if $W_\ee/W_\ii =(\Exp{V}-V_\ii)/ (V_\ee-\Exp{V})$
with probability one.
In turn plugging this relation into the mean voltage expression and solving for  $\Exp{V}$ reveals that we necessarily have $\Exp{V}=I/G$.
This is consistent with the intuition that variability can only vanish if excitation and inhibition perfectly cancel one another.


\section{Variance in the small-weight approximation}\label{app:smallApprox1}

In this appendix, we compute the simplified expression for the variance $\Var{V}$ obtained via the small-weight approximation.
Second, let us compute the small-weight approximation of the second-order efficacy
\begin{eqnarray}
c_{\ee\ii}
=
\frac{b\tau}{2} \ExpPei{\frac{W_\ee W_\ii}{(W_\ee+W_\ii)^2}\left( 1-e^{-(W_\ee+W_\ii)}\right)^2}
\simeq
\frac{b\tau}{2}  \ExpPei{W_\ee W_\ii}
=
\frac{b \tau}{2} w_\ee w_\ii \ExpPei{k_\ee k_\ii} \, , \nonumber
\end{eqnarray}
which amounts to compute the expectation of the crossproduct of the jumps $k_\ee$ and $k_\ii$.
To estimate the above approximation, it is important to remember that first that $p_{\ee}$ and $p_{\ii}$ are not defined as the marginals of $p_{\ee\ii}$, but as conditional marginals, for which 
we have $p_{\ee,k} = (b/b_\ee) \sum_{l=0}^{K_\ii}p_{\ee\ii, kl}$ and $p_{\ii,l} = (b/b_\ii) \sum_{k=0}^{K_\ee}p_{\ee\ii, kl}$.
Then by the definition of the correlation coefficient $\rho_{\ee\ii}$ in \eref{eq:corrEI}, we have 
\begin{eqnarray}
\rho_{\ee\ii}
=
\frac{b\ExpPei{k_\ee k_\ii}}{\sqrt{K_\ee b\ExpPei{k_\ee}K_\ii b\ExpPei{k_\ii}}} 
=
\frac{b\ExpPei{k_\ee k_\ii}}{\sqrt{K_\ee b_\ee \ExpPe{k_\ee}K_\ii b_\ii \ExpPi{k_\ii}}} 
=
\frac{b\ExpPei{k_\ee k_\ii}}{K_\ee K_\ii \sqrt{r_\ee r_\ii}} \, , \nonumber
\end{eqnarray}
as the rates $b_\ee$ and $b_\ii$ are such that $b_\ee \ExpPe{k_\ee}=K_\ee r_\ee$ and $b_\ii \ExpPe{k_\ii}=K_\ii r_\ii$.
As a result, we obtain a simplified expression for the cross-correlation coefficient:
\begin{eqnarray}
c_{\ee\ii}
=
(\rho_{\ee\ii} \sqrt{r_\ee r_\ii }\tau/ 2) (K_\ee w_\ee) (K_\ii w_\ii) \, . \nonumber
\end{eqnarray}
Observe that as expected, $c_{\ee\ii}$ vanishes when $\rho_{\ee\ii}=0$.
Second, let us compute the small-weight approximation of the second-order efficacy
\begin{eqnarray}
a_{\ee,12}
=
\frac{b\tau}{2} \ExpPei{\frac{W_\ee}{W_\ee+W_\ii}\left( 1-e^{-(W_\ee+W_\ii)}\right)^2}
\simeq
\frac{b\tau}{2}  \ExpPei{W_\ee(W_\ee + W_\ii)}
=
\frac{b\tau}{2} \left(w_\ee^2 \ExpPei{k_\ee^2} + w_\ee w_\ii \ExpPei{k_\ee k_\ii} \right) \, . \nonumber
\end{eqnarray}
To estimate the above approximation, we use the definition of the correlation coefficient $\rho_{\ee}$ in \eref{eq:CorrRhoe},
\begin{eqnarray}
\rho_{\ee}
=
\frac{b_\ee \ExpPe{k_\ee(k_\ee-1)}}{b_\ee \ExpPe{k_\ee}(K_\ee -1)}
=
\frac{b \ExpPei{ k_\ee(k_\ee-1)}}{K_\ee(K_\ee -1)r_\ee}
\, , \nonumber
\end{eqnarray}
as the rate $b_\ee$ is such that $b_\ee \ExpPe{k_\ee}=K_\ee r_\ee$.
This directly implies that
\begin{eqnarray}
b\ExpPei{k_\ee^2}
=
b\ExpPei{k_\ee(k_\ee-1)} + b\ExpPei{k_\ee}
=
\rho_\ee K_\ee(K_\ee -1)r_\ee +K_\ee r_\ee
=
K_\ee r_\ee (1 +\rho_\ee (K_\ee -1) ) \, . \nonumber
\end{eqnarray}
so that  we evaluate  
\begin{eqnarray}
a_{\ee,12}
=
\frac{b\tau}{2} \left(w_\ee^2 \ExpPei{k_\ee^2} + w_\ee w_\ii \ExpPei{k_\ee k_\ii} \right) 
=
\frac{r_\ee\tau}{2} K_\ee(1+\rho_\ee(K_\ee-1))w_\ee^2 +  \rho_{\ee\ii} \frac{ \sqrt{r_\ii r_\ee}\tau}{2}(K_\ee w_\ee) (K_\ii w_\ii)  \, , \nonumber
\end{eqnarray}
which simplifies to $a_{\ee,12}=(r_\ee \tau/2) K_\ee(1+\rho_\ee(K_\ee-1))w_\ee^2$ when excitation and inhibition act independently.
A symmetric expression holds for the inhibitory efficacy $a_{\ii,12}$.
Plugging the above expressions for synaptic efficacies in the variance expression \eref{eq:statvar} yields the small-weight approximation
\begin{eqnarray}
\Var{V} 
&\simeq&
\frac{
(1+\rho_\ee(K_\ee-1))K_\ee r_\ee w_\ee^2 (V_\ee - \Exp{V})^2 
+
(1+\rho_\ii(K_\ii-1))K_\ii r_\ii w_\ii^2 (V_\ii - \Exp{V})^2 
}{ 
2(1/\tau+ K_\ee r_\ee w_\ee + K_\ii r_\ii w_\ii) 
}   \nonumber\\
&&
\hspace{20pt}
+
\frac{
\rho_{\ee\ii} \sqrt{r_\ee r_\ii}( K_\ee  w_\ee) (K_\ii w_\ii)  \big[ (V_\ee - \Exp{V})^2 +  (V_\ii - \Exp{V})^2- (V_\ee - V_\ii)^2\big]
}{ 
2(1/\tau+ K_\ee r_\ee w_\ee + K_\ii r_\ii w_\ii)
}  \nonumber \, . 
\end{eqnarray}
Let us note that the first-term in the right-hand side above represents the small-weight approximation of the voltage variance  in the absence of correlation between excitation and inhibition, i.e., for $\rho_{\ee\ii}=0$.
Denoting the latter approximation by $\Var{V}\vert_{\rho_{\ee\ii}=0}$ and using the fact that the small-weight expression for the mean voltage
\begin{eqnarray*}
\Exp{V} 
=
\frac{K_\ee r_\ee w_\ee V_\ee + K_\ii r_\ii  w_\ii V_\ii}{1/\tau + K_\ee r_\ee w_\ee + K_\ii r_\ii w_\ii} \, ,
\end{eqnarray*}
is independent of correlations, we observe that as intuition suggests, synchrony-based correlation between excitation and inhibition results in a decrease of the neural variability:
\begin{eqnarray}
\Delta \Var{V}_{\rho_{\ee\ii}}
=
\Var{V} - \Var{V}\vert_{\rho_{\ee\ii}=0}
&\simeq&
-\frac{
\rho_{\ee\ii}\sqrt{r_\ee r_\ii}  ( K_\ee w_\ee) (K_\ii w_\ii)  (V_\ee - \Exp{V})(\Exp{V} -V_\ii)
}{ 
1/\tau+ K_\ee r_\ee w_\ee + K_\ii r_\ii w_\ii
} \leq 0 \nonumber \, .
\end{eqnarray}
However, the overall contribution of correlation is to increase variability in the small-weight approximation.
This can be shown under the assumptions that $K_\ee \gg 1$ and $K_\ii \gg 1$, by observing that
\begin{eqnarray}
\Delta \Var{V}_{\rho_{\ee\ii},\rho_{\eeii}}
=
\Var{V} - \Var{V}\vert_{\rho_\eeii=\rho_{\ee\ii}=0}
&\simeq&
\frac{
 \big( \sqrt{\rho_\ee r_\ee}  K_\ee w_\ee (V_\ee - \Exp{V}) - \sqrt{\rho_\ii r_\ii}  K_\ii w_\ii (V_\ii - \Exp{V}) \big)^2
}{ 
2(1/\tau+ K_\ee r_\ee w_\ee + K_\ii r_\ii w_\ii)
} 
 \nonumber\\
&&
+
( \sqrt{\rho_\ee\rho_\ii}-\rho_{\ee\ii})
\frac{
\sqrt{r_\ee r_\ii}  ( K_\ee w_\ee) (K_\ii w_\ii)  (V_\ee - \Exp{V})(\Exp{V} -V_\ii)
}{ 
1/\tau+ K_\ee r_\ee w_\ee + K_\ii r_\ii w_\ii
}
\geq 0 \nonumber \, ,
\end{eqnarray}
where both terms are positive since we always have $0 \leq \rho_{\ee\ii} \leq \sqrt{\rho_\ee\rho_\ii}$.


\section{Validity of the small-weight approximation}\label{app:smallApprox2}

Biophysical estimates of the synaptic weights $w_\ee<0.01$, $w_\ii<0.04$ and the synaptic input numbers $K_\ee<10000$, $K_\ii<2500$, suggest that neurons operates in the small-weight regime.
In this regime, we claim that exponential corrections due to finite-size effect can be neglected in the evaluation of synaptic efficacies, as long as the spiking correlations remains weak.
Here, we make this latter statement quantitative by focusing on the first-order efficacies in the case of excitation alone.
The relative error due to neglecting exponential corrections can be quantified as
\begin{eqnarray}
\mathcal{E}= \frac{\ExpPe{W_\ee} - \ExpPe{1-e^{-W_\ee}}}{\ExpPe{1-e^{-W_\ee}}} \ge 0 \, .
\nonumber
\end{eqnarray}
Let us evaluate this relative error, assumed to be small, when correlations are parametrized via beta distributions with parameter $\beta_\ee=1/\rho_\ee-1$.
Assuming correlations to be weak,  $\rho_\ee \ll 1$, amounts to assuming large, $\beta_\ee \gg 1$
Under the assumptions of small error, we can compute
\begin{eqnarray}
 \ExpPe{1-e^{-W_\ee}}
\simeq
 \ExpPe{W_\ee}
 =
 w_\ee \ExpPe{k_\ee}
 \quad \mathrm{and} \quad
  \ExpPe{W_\ee -1+e^{-W_\ee}}
 \simeq  
 \ExpPe{W_\ee^2}/2
 =
 w_\ee^2 \ExpPe{k_\ee^2} /2\, , \nonumber
\end{eqnarray}
By the calculations carried out in Appendix \ref{app:smallApprox2}, we have
\begin{eqnarray}
b_\ee \ExpPe{k_\ee} = K_\ee r_\ee \quad \mathrm{and} \quad  b_\ee \ExpPe{k_\ee^2} =  K_\ee r_\ee (1+\rho_\ee(K_\ee-1)) \, . \nonumber
\end{eqnarray}
Remembering that $\beta_\ee=1/\rho_\ee-1$, this implies that  we have 
\begin{eqnarray}
\mathcal{E} 
\simeq
\frac{\ExpPe{W_\ee^2}/2}{\ExpPe{W_\ee}-\ExpPe{W_\ee^2}/2}
\simeq
\frac{w_\ee(1+\rho_\ee (K_\ee-1))/2}{1-w_\ee(1+\rho_\ee (K_\ee-1))/2}\, , \nonumber
\end{eqnarray}
For a correlation coefficient $\rho_\ee \leq 0.05$, this means that neglecting exponential corrections incurs less than a $e=3\%$ error if the number of inputs is smaller than $K_\ee \leq 1000$ for moderate synaptic weight $w_\ee=0.001$ or than $K_\ee \leq 100$ for large synaptic weight $w_\ee=0.01$.


\section{Infinite-size limit with spiking correlations}\label{app:InfSize}

The computation of the first two moments $\Exp{V}$ and $\Exp{V^2}$ requires to evaluate various efficacies as expectations.
Upon inspection,  these expectations are all of the form $b\ExpPei{f(W_\ee,W_\ii)}$, where $f$ is a smooth positive function that is bounded on $\mathbbm{R}^+\times \mathbbm{R}^+$ with $f(0,0)=0$.
Just as for the L\'evy-Khintchine decomposition of stable jump processes~\cite{khintchine:1934,levy:1954}, this observation allows one to generalize our results to processes that exhibit and countable infinity of jumps over finite, nonzero time intervals.
For our parametric forms based on beta distributions, such processes emerge in the limit of an arbitrary large number of inputs, i.e., for $K_\ee, K_\ii \to \infty$.
Let us consider the case of excitation alone for simplicity.
Then, we need to make sure that all expectations of the form $b_\ee \ExpPei{f(W_\ee)}$ remain well-posed in the limit $K_\ee \to \infty$ for smooth, bounded test function $f$ with $f(0)=0$.
To check this, observe that for all $0 < k \leq K_\ee$, we have by \eref{eq:betaParam} and \eref{eq:bDef} that
\begin{eqnarray}
b_\ee p_{\ee,k} = \beta r_\ee \binom{K_e}{k} B(k, \beta+ K_\ee-k)= \beta r_\ee \frac{\Gamma(K_\ee+1)}{\Gamma(k+1)\Gamma(K_\ee-k+1)} \frac{\Gamma(k)\Gamma(\beta+K_\ee-k+1)}{\Gamma(\beta+K_\ee)} \, , \nonumber
\end{eqnarray}
where we have introduce the Gamma function $\Gamma$.
Rearranging terms and using the fact that $\Gamma(z+1)=z\Gamma(z)$ for all $z>0$, we obtain
\begin{eqnarray}
b_\ee p_{\ee,k} 
= 
\frac{\beta r_\ee}{k} \frac{K_\ee \Gamma(K_\ee)}{\Gamma(\beta+K_\ee)}  \frac{\Gamma(\beta+K_\ee-k)}{(K_\ee-k)\Gamma(K_\ee-k)} 
= 
\frac{\beta  r_\ee}{k} \left(1-\frac{k}{K_\ee} \right)^{\beta-1} + o\left(\frac{1}{K_\ee}\right) \, , \nonumber
\end{eqnarray}
where the last equality is uniform in $k$ and follows from the fact that for all $x>0$, we have
\begin{eqnarray*}
\lim_{z \to \infty} \frac{\Gamma(z+x)}{\Gamma(z)}=z^x \left( 1 + \binom{x}{2} \frac{1}{z} + o\left(\frac{1}{z}\right)\right)
\end{eqnarray*}
From there, given a test function $f$, let us consider
\begin{eqnarray}
b_\ee \ExpPe{f(W_\ee)} 
&=&  \int \sum_{k=1}^{K_\ee} b_\ee p_{\ee,k} \delta \left(W_\ee-\frac{k\Omega_\ee}{K_\ee}\right) f(W_\ee)\, \dd W_\ee \, , \nonumber\\
&=&  \sum_{k=1}^{K_\ee} b_\ee p_{\ee,k} f\left(\frac{k\Omega_\ee}{K_\ee}\right)\, , \nonumber\\
&=& r_\ee \sum_{k=1}^{K_\ee} \frac{\beta}{k} \left(1-\frac{k}{K_\ee} \right)^{\beta-1}f\left(\frac{k\Omega_\ee}{K_\ee}\right) + o(1) \, . \nonumber
\end{eqnarray}
The order zero term above can be interpreted as a Riemann sum so that one has
\begin{eqnarray}
\lim_{K_\ee \to \infty} b_\ee \ExpPe{f(W_\ee)} 
&=&
r_\ee \lim_{K_\ee \to \infty} \frac{1}{K_\ee}\sum_{k=1}^{K_\ee} \frac{\beta K_\ee}{k} \left(1-\frac{k}{K_\ee}\right)f\left(\frac{k\Omega_\ee}{K_\ee} \right) \, , \nonumber\\
&=&
r_\ee \int_0^1 \beta \theta^{-1}(1-\theta)^{\beta-1}f(\theta \Omega_\ee) \, \dd \theta \, , \nonumber\\
&=&
r_\ee \int_0^{\Omega_\ee} \frac{\beta}{w}\left(1-\frac{w}{\Omega_\ee}\right)^{\beta-1}f(w) \, \dd w \, . \nonumber
\end{eqnarray}
Thus, the jump densities is specified via the L\'evy-Khintchine measure
\begin{eqnarray}
\nu_\ee(w)= \frac{\beta}{w}\left(1-\frac{w}{\Omega_\ee}\right)^{\beta-1} \, , \nonumber
\end{eqnarray}
which is a deficient measure for admitting a pole in zero.
This singular behavior indicates that the limit jump process obtained when $K_\ee \to \infty$ has a countable infinity of jumps within any finite, nonempty time interval.
Generic stationary jump processes with independent increments, as is the case here, are entirely specified by their  L\'evy-Khintchine measure $\nu_\ee$~\cite{khintchine:1934,levy:1954}.
Moreover, one can check that given knowledge of $\nu_\ee$, one can consistently estimate the corresponding pairwise spiking correlation as
\begin{eqnarray} 
 \rho_\ee
=
\lim_{K_\ee \to \infty} \frac{\ExpPe{k_\ee(k_\ee-1)}}{\ExpPe{k_\ee}(K_\ee-1)} 
=
\lim_{K_\ee \to \infty} \frac{b_\ee \ExpPe{(k_\ee/K_\ee )^2}}{b_\ee \ExpPe{k_\ee/K_\ee}} 
=
\frac{
\int_0^{\Omega_\ee} w^2 \nu_\ee(w) \, \dd w 
}
{
\Omega_\ee \int_0^{\Omega_\ee} w \nu_\ee(w) \, \dd w 
} \, . \nonumber
\end{eqnarray}
Performing integral with respect to the L\'evy-Khintchine measure $\nu_\ee$ instead of the evaluating the expectation $\ExpPe{\cdot}$ in \eref{eq:statmean} and \eref{eq:statvar} yields
\begin{eqnarray}
\Exp{V} = \frac{V_\ee \int_0^{\Omega_\ee} (1-e^{-w}) \nu_\ee(\dd w)}{1/\tau +\int_0^{\Omega_\ee} (1-e^{-w}) \nu_\ee(\dd w)} 
\quad \mathrm{and} \quad
\Var{V} = \frac{(V_\ee -\Exp{V})^2 \int_0^{\Omega_\ee} (1-e^{-w})^2 \nu_\ee(\dd w)}{2/\tau +\int_0^{\Omega_\ee} (1-e^{-2w}) \nu_\ee(\dd w)}  \, . \nonumber
\end{eqnarray}
Observe that as $(1-e^{-w})^2 \leq w^2$ for all $w \geq 0$, the definition of the spiking correlation and voltage variance implies that we have $ \Var{V}=O(\rho_\ee)$ so that neural variability consistently vanishes in the absence of correlations.

\end{widetext}

\bibliography{prx_references}

\end{document}